\documentclass[11pt,xcolor=dvipsnames,fleqn]{article}
\DeclareMathAlphabet{\scr}{U}{rsfs}{m}{n}

\pdfoutput=1
\usepackage{scalerel,stackengine}
\usepackage{subfigure}
\usepackage{latexsym}
\usepackage{epsfig}
\usepackage[mathscr]{eucal}
\usepackage{amsfonts}
\usepackage{amscd}
\usepackage{multirow}
\usepackage{cite}
\usepackage{array}
\usepackage{amssymb}
\usepackage{amsthm}
\usepackage{colordvi}
\usepackage[centertags]{amsmath}
\usepackage{indentfirst}
\usepackage{geometry}
\usepackage{float} 
\usepackage{enumerate}
\usepackage{graphicx}
\usepackage{booktabs}
\usepackage{theorem}
\usepackage[footnotesize]{caption}
\usepackage{soul}
\usepackage{mcite}
\usepackage{slashed}
\usepackage{braket}
\usepackage{color,xcolor}
\usepackage{bbm}

\usepackage[utf8]{inputenc}
\usepackage{fancyvrb}
\usepackage{framed}
\usepackage{xspace}
\usepackage{todonotes}
\usepackage{siunitx}

\usepackage{psfrag,rotating}
\usepackage{epstopdf}
\numberwithin{figure}{section}
\numberwithin{table}{section}
\def\zz{\mathbb{Z}}

\allowdisplaybreaks

\usepackage{cleveref}
\crefname{chapter}{Chapter}{Chapter}
\crefname{section}{Sec.}{Secs.}
\crefname{table}{Tab.}{Tabs.}
\crefname{figure}{Fig.}{Figs.}
\crefname{equation}{Eq.}{Eqs.}
\crefname{appendix}{Appendix\ }{Appendix\ }

\begin{document}
\pdfoutput=1

\title{
    \vspace*{-3.7cm}
    \vspace*{2.7cm}
    \textbf{One-loop radiative corrections to $e^+ e^-\to Zh^0/H^0A^0$ in the Inert Higgs Doublet Model \\[4mm]}}

\date{}
\author{
Hamza Abouabid$^{1\,}$\footnote{E-mail:
    \texttt{hamza.abouabid@gmail.com}} ,
Abdesslam Arhrib$^{1\,}$\footnote{E-mail:
    \texttt{aarhrib@gmail.com}} ,
Rachid Benbrik$^{2\,}$\footnote{E-mail:
    \texttt{r.benbrik@uca.ma}} ,
\\
Jaouad El Falaki$^{3\,}$\footnote{E-mail: \texttt{jaouad.elfalaki@gmail.com}},
Bin Gong$^{4,5\,}$\footnote{{E-mail: \texttt{twain@ihep.ac.cn}}},
Wenhai Xie$^{4,5\,}$\footnote{E-mail:
    \texttt{xiewh@ihep.ac.cn}},
Qi-Shu Yan$^{5,6\,}$\footnote{E-mail: \texttt{yanqishu@ucas.ac.cn}},
\\[5mm]
{\small\it
$^1$Universit\'{e} Abdelmalek Essaadi, FSTT, B. 416, Tangier, Morocco.} \\[3mm]
{\small\it $^2$Laboratoire de Physique fondamentale et Appliq\'{e}e Safi, Facult\'{e} Polydisciplinaire de Safi,} \\
{\small \it Sidi Bouzid, BP 4162, Safi, Morocco.} \\[3mm]
%
{\small\it $^3$ EPTHE, Physics Department, Faculty of Science, Ibn Zohr University, P.O.B. 8106 Agadir, Morocco.} \\[3mm]
{\small\it $^4$ Theory Division, Institute of High Energy Physics, Chinese Academy of Sciences, Beijing 100049, China.} \\[3mm]
{\small\it $^5$School of Physics Sciences, University of Chinese Academy of Sciences, Beijing 100049, China.} \\[3mm]
{\small\it $^6$ Center for Future High Energy Physics, Chinese Academy of Sciences,  Beijing 100049, China.} \\[3mm]
}

\maketitle

\begin{abstract}
We compute the full one-loop radiative corrections
(including both weak and QED corrections) for two 
processes $e^{+}e^{-}\to Z h^0,H^0 A^{0}$ in the Inert Higgs
Doublet model (IHDM). Up to $O(\alpha_{w})$ and $O(\alpha_{em})$ order, we
use FeynArts/FormCalc to compute the one-loop virtual corrections and Feynman
Diagram Calculation (FDC) to evaluate the real emission, respectively. Being equipped 
with these computing tools, we investigate radiative corrections of new physics for five scenarios 
with three typical collision energies of future electron-positron
colliders: 250 GeV, 500 GeV, and 1000 GeV. By scanning the parameter space of IHDM, we identify the allowed 
regions which are consistent with constraints and bounds, from both theoretical and experimental sides. We find 
that the radiative corrections of the IHDM to $e^+ e^- \to Z h^0$ can be sizeable and are within the detection potentials 
of future Higgs factories. We also find that the new physics of IHDM could also be directly detected by observing 
the process $e^{+}e^{-}\to H^0 A^{0} $ which could have large enough production rate. We propose six benchmark 
points and examine their salient features which can serve as physics targets for future electron-positron 
colliders, such as CEPC/CLIC/FCC-ee/ILC as well as for LHC.
\end{abstract}
\thispagestyle{empty}
\vfill
\newpage
\setcounter{page}{1}
\section{Introduction}
\label{sec:introduction}
The first LHC run with $7\oplus 8$ TeV and the second one with 13 TeV were successful operations which led to the discovery of a new scalar particle \cite{Aad:2012tfa, Chatrchyan:2012xdj} including the recent observation of its production in association with $t\bar{t}$ \cite{Aaboud:2018urx, Sirunyan:2018hoz} and its decay to $b\bar{b}$ \cite{Aaboud:2018zhk,Sirunyan:2018kst} among other achievements. The LHC program has already performed several precise measurements in term of production cross sections and branching fractions. These measurements demonstrate that the Standard Model (SM) works well to explain these observed phenomena at the electroweak scale.

One of the main goals of the future run of the LHC with 14 TeV and its High Luminosity option (HL-LHC) is to improve 
the aforementioned measurements  and pin down the uncertainties to few percent level \cite{Dawson:2013bba,Zeppenfeld:2000td,Gianotti:2000tz,Cepeda:2019klc,deBlas:2019rxi}. On the other hand, it is also expected from the future LHC run to establish a new measurement such as the triple Higgs coupling~\cite{DiMicco:2019ngk}, and Higgs decay into $\gamma Z$ and $\mu^+\mu^-$. Moreover, it is well known that a precise measurement program which already began at the LHC,  
is expected to be performed at the $e^+e^-$ 
machines \cite{Moortgat-Picka:2015yla, Fujii:2015jha, Fujii:2017vwa} such as the Circular Electron 
Positron Collider (CEPC) \cite{CEPC-SPPCStudyGroup:2015csa, An:2018dwb}, the Compact Linear Collider (CLIC) \cite{Battaglia:2004mw, Aicheler:2012bya, Linssen:2012hp,Charles:2018vfv,deBlas:2018mhx}, 
the Future Circular Collider (FCC-ee) \cite{Gomez-Ceballos:2013zzn,Abada:2019zxq}, and the International 
Linear Collider (ILC)\cite{Moortgat-Picka:2015yla,Fujii:2019zll}.
The $e^+e^-$  machines which are expected to deliver rather high luminosity and  possess a very clean environment, 
would be able to improve  the Higgs couplings and production cross section measurements below the percent level \cite{Moortgat-Picka:2015yla,Fujii:2015jha,Fujii:2017vwa}. Such a precision, if achieved, 
will be very useful to discover  the evidence of new physics beyond the SM.

In Table \ref{tab:HFs}, we provide the projected experimental precisions for $\delta g_{Z Z h^0}$ at future Higgs factories, which is taken from Table 4.2 of reference \cite{Abada:2019zxq}. It should be declared that the numbers quoted here are just for illustration on the potential precision which could be reached, but they are not necessarily exact what those Higgs factories are planned to achieve since the actual precisions depend on more detailed run programs. 

\begin{table}[!htb]
\centering
\begin{tabular}{|c|c|c|c|c|c|}
\hline
& CEPC & CLIC  & FCC-ee  & ILC & LEP3  \\
\hline
Collision energy (GeV) & 250 & 380  & 240+365 & 250 & 240\\
\hline
No. of collision points & 2 & 1 & 2 & 1 & 3\\
\hline
Run duration (years) & 7  & 8  & $3_{240} + 4_{365}$ & 15  & 6\\
\hline
Integrated Lum (1/ab) & $5$ & $1$ & $5_{240}+1.5_{365}$ & $2$ & $3$ \\
\hline
Projected precision $\delta g_{ZZh^0}/ g_{ZZh^0}$ & $0.25\%$ & $0.6 \%$& $0.2\%$  & $0.3 \%$ & $0.32\%$ \\
\hline
\end{tabular}
\caption{The projected precisions of $\delta g_{Z Z h^0}$ are tabulated. }
\label{tab:HFs}
\end{table}

However, there are several evidences both theoretical and experimental which indicate that the SM could not be the 
ultimate theory. Instead, the SM should be 
viewed as a low energy effective theory of some more complete and fundamental  one yet to be discovered. 
It is believed that a precise measurement of Higgs boson productions and decays can be a promising probe both to test the prediction of the SM as well as to search for new physics beyond the SM.  After the discovery of the new scalar particle, 
there have been many theoretical and phenomenological studies devoted to non-minimal Higgs sector models that can explain 
such discovery and address some of the weakness of the SM.
One of the simplest non-minimal Higgs model is the popular Two Higgs Doublet Model (2HDM) where both Higgs doublets possess a vacuum expectation value (VEV) and participate to electroweak symmetry breaking.  A subclass of the 2HDM, 
that can provide a dark matter (DM) candidate, 
is the Inert Higgs Doublet Model (IHDM) where one Higgs doublet 
does not  develop a VEV and may act as a dark matter candidate while the other one plays the role of the SM Higgs doublet \cite{Deshpande:1977rw}. 
We notice that the IHDM possesses an exact discrete $\zz_2$ symmetry where the new Higgs doublet is odd under $\zz_2$ while all the SM fields are even under this symmetry. Therefore, the Lightest Odd Particle (LOP) under $\zz_2$ is stable and can be a viable candidate for DM \cite{Gustafsson:2007pc, Hambye:2007vf, Agrawal:2008xz, Dolle:2009fn, Andreas:2009hj, Barbieri:2006dq}.
The spectrum of the IHDM contains one CP-even Higgs $h^0$ which is identified with the 125 GeV SM Higgs  and four $\zz_2$  odd Higgses: 
one CP-even $H^0$, one CP-odd $A^0$ and a pair of charged Higgs $H^\pm$. 
As a simple extension of the SM, recently, the IHDM has been under intensive phenomenological investigation. For example, there are works dealing with loop corrections \cite{Arhrib:2015hoa,Banerjee:2019luv,Banerjee:2021oxc,Banerjee:2021anv,Banerjee:2021xdp,Banerjee:2021hal,Braathen:2019pxr,Braathen:2019zoh,Senaha:2018xek}, works on collider signal searches \cite{Dolle:2009ft,Aoki:2013lhm,Datta:2016nfz,Dutta:2017lny,Kalinowski:2018ylg,Kalinowski:2018kdn,Guo-He:2020nok,Yang:2021hcu,Kalinowski:2020rmb} including the analysis of LHC constraints and indirect searches \cite{Melfo:2011ie,Abercrombie:2015wmb,Ilnicka:2015jba,Blinov:2015qva,Poulose:2016lvz,Hashemi:2016wup,Wan:2018eaz,Belyaev:2018ext,Dercks:2018wch,Lu:2019lok}, works on the vacuum metastability and the contributions of the inert Higgs bosons to electroweak phase transition and gravitational waves \cite{Gil:2012ya,Swiezewska:2015paa,Blinov:2015vma,Huang:2017rzf}, etc. It is noteworthy that dedicated phenomenological investigations for the IHDM which address DM \cite{Keus:2014jha,Arcadi:2019lka}, astrophysics 
as well as collider constraints \cite{Kalinowski:2018ylg,Belyaev:2016lok,Arhrib:2013ela,Eiteneuer:2017hoh}, concluded that the IHDM is still consistent with all theoretical and experimental bounds.

At $e^+e^-$ machines,  using the recoil mass technique 
one can measure $\sigma(e^+e^-\to Zh^0)$ independently of the decay modes of the Higgs boson. This measurement is expected to be at the percent level and would be promising for precision analysis 
\cite{Moortgat-Picka:2015yla,Fujii:2015jha,Fujii:2017vwa}. Therefore, with such planed precision, this process would be sensitive to higher order effects at loop level and could be used to disentangle between various models beyond the SM. 

Radiative corrections for Higgs production at $e^+e^-$ machines 
have been performed in many models beyond the Standard Model. In the SM,
full one loop radiative corrections to $e^+e^- \to Zh^0$ have been evaluated long time ago
in Refs.~\cite{Fleischer:1982af, Kniehl:1991hk, Denner:1992bc}. These corrections could be of the order of 
several percent and  become large and negative for high center of mass energy \cite{Denner:1992bc}. 
The next-to-next-to-leading order (NNLO) corrections to Higgsstrahlung process have 
been presented in Refs.~\cite{Sun:2016bel, Gong:2016jys}
where NNLO effects are integrated into the total cross sections via the mixed QCD-electroweak corrections 
$\mathcal{O}(\alpha\alpha_s)$. These corrections turn out to reach $1.3\%$  (resp $0.7\%$) of the leading-order results for a center-of-mass energy around 240 GeV (resp 500 GeV).
 However,
in the IHDM, one loop radiative corrections to SM Higgs decays 
$h^0\to b\bar{b}$, $h^0\to ZZ, WW$ and also $h^0\to \gamma \gamma$ and $h^0\to \gamma Z$ have been considered 
in \cite{Kanemura:2015mxa,Kanemura:2016sos,Arhrib:2014pva,Arhrib:2012ia,Swiezewska:2012eh,Krawczyk:2013pea,Arhrib:2015hoa}.
We notice that to our best knowledge radiative corrections to $e^+e^- \to Zh^0$ and $e^+e^- \to H^0A^0$ 
in the IHDM  are still missing in the literature. The aim of this paper is to calculate in the framework of  IHDM the full one-loop radiative corrections to: 
$e^+e^-\to Zh^0$ and $e^+e^-\to H^0 A^0$.
We will include not only the full weak corrections,
but also the QED corrections including both soft 
 and hard photon emissions. In our analysis, we will take into account theoretical constraints on the IHDM as well as experimental constraints from LHC, like the Higgs decaying into two photons, the invisible Higgs dacay and the electroweak precision tests. We also take into account constraints from dark matter and monojet searches.
 
The paper is organized  as follows: In section \ref{sec:model}, we briefly describe the IHDM, its mass spectra and key trilinear and quartic Higgs couplings, list various theoretical and experimental constraints that we will take into account in this work, and show our results of the scan over the parameter space both for the degenerate and non-degenerate IHDM spectrum. In section \ref{sec:LO}, we provide the leading order formula for differential and total cross sections for $e^+ e^- \to  Zh^0/H^0 A^0$ processes. In section \ref{sec:renormalization}, we introduce the on-shell renormalization scheme for the IHDM and set up basic notations and conventions. Then we study the one-loop contributions to $e^+e^-\to Zh^0/H^0 A^0$ processes and examine the importance of soft and hard photon emission in order to guarantee the cancellation of the infrared (IR) as well as the soft collinear divergences  at the next leading order calculation. We present our numerical results for $e^+e^-\to Zh^0$ and $e^+e^- \to H^0A^0 $ in Section \ref{sec:results}. In Section \ref{bps5}, we propose six benchmark points (BPs) and examine the radiative corrections of them for future $e^+ e^-$ colliders.  We end this work with concluding remarks and brief discussions in section \ref{sec:conclusions}.
\section{Review of IHDM, its theoretical and experimental constraints}
\label{sec:model}
\subsection{A brief introduction to IHDM}
The IHDM is a simple extension of the SM which can also provide a viable dark matter candidate. It is a version of the 2HDM with an exact discrete $\zz_2$ symmetry. The SM scalar sector parametrized by $H_1$  is extended by an inert scalar doublet $H_2$ which can provide a stable dark matter candidate. Under $\zz_2$ symmetry all the  SM particles are even while $H_2$ is odd.
We shall use the following parameterization of the two doublets :
\begin{eqnarray}
H_1 = \left (\begin{array}{c}
G^\pm \\
\frac{1}{\sqrt{2}}(v + h^0 + i G^0) \\
\end{array} \right)
\qquad , \qquad
 H_2 = \left( \begin{array}{c}
H^\pm\\
\frac{1}{\sqrt{2}}(H^0 + i A^0) \\
\end{array} \right)
\end{eqnarray}
with $G^0$ and $G^\pm$ are the Nambu-Goldstone bosons absorbed, after electroweak symmetry breaking,  by the longitudinal component of $W^\pm$ and $Z^0$, respectively. 
$v$ is the vacuum expectation value (VEV) of the SM Higgs $H_1$. 
The most general renormalizable, gauge invariant and
CP invariant potential is given by :
\begin{eqnarray}
V  &=&  \mu_1^2 |H_1|^2 + \mu_2^2 |H_2|^2  + \lambda_1 |H_1|^4
+ \lambda_2 |H_2|^4 +  \lambda_3 |H_1|^2 |H_2|^2 + \lambda_4
|H_1^\dagger H_2|^2 \nonumber \\
&+&\frac{\lambda_5}{2} \left\{ (H_1^\dagger H_2)^2 + {\rm h.c} \right\}
\label{potential}
\end{eqnarray}
In the above potential, because of $\zz_2$ symmetry, there is no mixing terms like $\mu_{12}^2 (H_1^\dagger H_2 +
h.c)$.  In addition, by hermicity of the potential, all $\lambda_i, i = 1, \cdots, 4$ are real valued. The phase of
$\lambda_5$ can be absorbed by a suitable redefinition
 of the fields $H_1$ and $H_2$, therefore the scalar sector of the IHDM  is CP conserving. After spontaneous symmetry breaking of $SU(2)_L \otimes U(1)_Y$
 down to electromagnetic $U(1)_{em}$, the spectrum of
the above potential will have five scalar particles: two CP even $H^0$ and $h^0$ which will be identified as the SM Higgs boson with 125 GeV mass, one CP odd $A^0$ and
a pair of charged scalars $H^\pm$.
Their masses are given by:
\begin{eqnarray}
&& m_{h^0}^2 = - 2 \mu_1^2 = 2 \lambda_1 v^2 \nonumber \\
&& m_{H^0}^2 = \mu_2^2 + \lambda_L v^2 \nonumber \\
&&  m_{A^0}^2 = \mu_2^2 + \lambda_S v^2 \nonumber \\
&&  m_{H^{\pm}}^2 = \mu_2^2 + \frac{1}{2} \lambda_3 v^2
  \label{spect.IHDM}
\end{eqnarray}
where $\lambda_{L,S}$ are defined as:
\begin{eqnarray}
\lambda_{L,S} &=& \frac{1}{2} (\lambda_3 + \lambda_4 \pm \lambda_5)
\end{eqnarray}

From above relations, one can easily express $\lambda_i$ as a function of physical masses: \footnote{The value of the self-coupling $\lambda_1$ is fixed by 
	$m_{h^0}$ and $v$. Hence, the experimentally measured Higgs mass, $m_{h^0}= 125$ GeV, implies that $\lambda_1 \simeq 0.13$}
\begin{eqnarray}
\lambda_1&=&\dfrac{m_{h^0}^2}{2v^2} \nonumber\\
\lambda_3&=&\dfrac{2(m_{H^\pm}^2-\mu_{2}^2)}{v^2} \nonumber\\
\lambda_4&=&\dfrac{(m_{H^0}^2+m_{A^0}^2 -2 m_{H^\pm} ^2)}{v^2} \nonumber\\
\lambda_5&=&\dfrac{(m_{H^0}^2-m_{A^0}^2)}{v^2} 
\label{eq:lams}
\end{eqnarray}
The IHDM involves $8$ independent parameters: five $\lambda_{1,...,5}$, two $\mu_{1,2}$
and $v$. 
One parameter is eliminated by the minimization condition and the VEV is fixed by the $Z$ boson mass, fine-structure constant  and Fermi constant  $G_F$. Finally, we are left with six independent  parameters which we choose as follow :
\begin{eqnarray}
 \{ \mu_2^2, \lambda_2, m_{h^0}, m_{H^\pm}, m_{H^0}, m_{A^0} \}
 \label{param.IHDM}
\end{eqnarray}
For completeness we list here the triple and quartic Higgs couplings that are needed in our analysis:
\begin{eqnarray}
&&h^0H^0H^0 = -2 v \lambda_L = -v (\lambda_3+\lambda_4+\lambda_5)\equiv v\lambda_{h^0H^0H^0}\nonumber\\
&&h^0A^0A^0 = -2 v \lambda_S = -v (\lambda_3+\lambda_4-\lambda_5)\equiv v\lambda_{h^0A^0A^0}\nonumber\\
&&h^0 H^\pm H^\mp = -v \lambda_3 \equiv v\lambda_{h^0H^\pm H^\mp}\nonumber\\
&&H^0H^0A^0A^0=-2 \lambda_2
\label{scalar-coup}
\end{eqnarray}
As we will see later, these triple Higgs couplings are either 
directly involved in the processes 
$e^+e^-\to Zh^0$ and $e^+e^-\to H^0A^0$ under investigation or in the experimental constraints that have to be fulfilled, 
while the quartic Higgs coupling $H^0H^0A^0A^0$ enters only in $e^+e^-\to H^0A^0$ production.

\subsection{Theoretical constraints}
\label{sec:thexcon}
The parameter space of the scalar potential of the IHDM should be consistent with theoretical requirements. 
The important theoretical requirements in our consideration include perturbativity of the scalar quartic couplings, vacuum stability and tree-level perturbative unitarity conditions for various scattering amplitudes of all scalar bosons.
\begin{itemize}
\item \underline{\bf Perturbativity:}

To guarantee the perturbation expansion, it is required that each of  the quartic couplings of the scalar potential in Eq.~(\ref{potential}) should obey the following conditions:
\begin{eqnarray}
|\lambda_i| \le 8 \pi
\end{eqnarray}

\item \underline{\bf Vacuum Stability:}

The vacuum stability requires the potential $V$ should remain positive when the values of scalar fields 
become extremely large \cite{Deshpande:1977rw}. From this condition, we have the following constraints
on the IHDM parameters (for a review see \cite{Branco:2011iw}):
\begin{eqnarray}
\lambda_{1,2} > 0 \quad \rm{and} \quad \lambda_3 + \lambda_4 -|\lambda_5| +
2\sqrt{\lambda_1 \lambda_2} >0 \quad\rm{and} \quad\lambda_3+2\sqrt{\lambda_1
  \lambda_2} > 0
\end{eqnarray}

\item \underline{\bf  Charge-breaking minima:}

Likewise, a neutral, charge-conserving vacuum can be guaranteed by demanding that \cite{Ginzburg:2010wa}
\begin{equation}
\lambda_4-|\lambda_5|\leq 0,
\label{chargebreaking}
\end{equation}

which is a sufficient but not necessary condition for the vacuum to be neutral. 
Because, a neutral vacuum can also be achieved for positive $ \lambda_4-|\lambda_5|$ with suitable $\mu_1^2$ and $\mu_2^2$,
but in this case  the dark matter (DM) particle can not be neutral. Condition (\ref{chargebreaking}) avoids this scenario.
\footnote{If $H^0$ is the DM particle i.e $m_{H^0}\leq m_{A^0}$, then 
$\lambda_4-|\lambda_5|=\frac{2}{v^2}(m_{H^0}^2-m_{H^\pm}^2)$. So condition (\ref{chargebreaking}) implies
that $m_{H^0}\leq m_{H^\pm}$. In the other case when $A^0$ is the DM candidate i.e $m_{A^0}\leq m_{H^0}$,  
$\lambda_4-|\lambda_5|=\frac{2}{v^2}(m_{A^0}^2-m_{H^\pm}^2)$. Consequently, condition (\ref{chargebreaking}) gives $m_{A^0}\leq m_{H^\pm}$. 
In  both cases, $m_{H^0}\leq m_{A^0},m_{H^\pm}$ or $m_{A^0} \leq m_{H^0},m_{H^\pm}$ is satisfied.}
\item \underline{{\bf Inert Vacuum:}}

In order to insure that the CP-conserving minimum described earlier is the global one, we need to impose the following conditions \cite{Ginzburg:2010wa}:
\begin{eqnarray}
m_{h^0}^2, m_{H^0}^2, m_{A^0}^2, m_{H^\pm}^2 >0 \qquad {\rm and} \qquad
\mu_1^2/\sqrt{\lambda_1}<
\mu_2^2/\sqrt{\lambda_2}
\label{eq:inertvac}
\end{eqnarray}
\item\underline{\bf Unitarity:}

To constrain the scalar potential parameters of the IHDM, the tree-level perturbative unitarity is imposed to the various scattering amplitudes of scalar bosons at high energy. From the technique developed in \cite{Lee:1977eg}, we get the following set of eigenvalues:
\begin{eqnarray}
&&e_{1,2}=\lambda_3 \pm \lambda_4 \quad , \quad
e_{3,4}= \lambda_3 \pm \lambda_5\\
&&e_{5,6}= \lambda_3+ 2 \lambda_4 \pm 3\lambda_5\quad , \quad
e_{7,8}=-\lambda_1 - \lambda_2 \pm \sqrt{(\lambda_1 - \lambda_2)^2 + \lambda_4^2}
\\
&&
e_{9,10}= -3\lambda_1 - 3\lambda_2 \pm \sqrt{9(\lambda_1 - \lambda_2)^2 + (2\lambda_3 +
   \lambda_4)^2}
\\
&&
e_{11,12}=
 -\lambda_1 - \lambda_2 \pm \sqrt{(\lambda_1 - \lambda_2)^2 + \lambda_5^2}
\end{eqnarray}

We impose perturbative unitarity constraint on all $e_i$'s. $e_i \le 8 \pi ~, \forall ~ i = 1,...,12$.
\end{itemize}

\subsection{Experimental constraints}
\label{sec:expxcon}
The parameter space of the scalar potential of the IHDM should also satisfy experimental search constraints. We will consider the following experimental constraints: 1) from Higgs data at the LHC, 2) the direct collider searches from the LEP, 3) the indirect searches from electroweak precision tests, and 4) the data from dark matter searches. Below we elaborate more of  these constraints. 
\begin{enumerate}

\item Constraints from Higgs data at the LHC (HD)

In the IHDM, because of the exact $\zz_2$ symmetry, 
$H_2$ does not couple to SM fermions which lead to natural flavor conservation. Only SM Higgs doublet couples to fermions, 
therefore all SM Higgs couplings to fermions and gauge bosons W and Z 
are the same as in the SM. Therefore, Higgs production cross section through conventional channels 
at the LHC such as gluon fusion, vector boson fusion, Higgsstrahlung and $t\bar{t}h^0$ are  exactly the 
same as in the SM. Similarly, all tree level Higgs decays $h^0\to b\bar{b}, \tau^+ \tau^-, ZZ^*, WW^*$ are 
identical to SM. Nevertheless, Higgs data provides important constraints to the IHDM.

\begin{enumerate}
\item[(HD.1)] The one-loop decay channels $h^0\to \gamma \gamma$ and $h^0\to \gamma Z$ can receive contributions from charged Higgs boson via loop processes which may modify the SM predictions \cite{Arhrib:2012ia}. Therefore, in our analysis we will take into account the existing constraints on $h^0\to \gamma \gamma$ (see Sec.~\ref{sec:hpp} for details). 

\item[(HD.2)] Besides, when a DM candidate is lighter than $m_{h^0}/2$, the Higgs decay into invisible channel can be open. To be precise, in the IHDM, $h^0$ can decay into a pair of dark Higgs $\chi=H^0, A^0$:  $h^0\to \chi \chi$ if kinematically allowed where $\chi$ is the LOP.  Such invisible decay of the SM Higgs has been investigated experimentally  by both ATLAS and CMS and set an upper limit on the branching ratio: $Br(h^0\to \mathrm{invisible})$ to 26\% (resp 19\% ) at 95\% confidence level (CL) for ATLAS (resp for CMS)  \cite{Sirunyan:2018owy,Aaboud:2019rtt,Aaboud:2018sfi}. Moreover,  global fit studies performed on LHC data can in turn put limits on the invisible decay of the SM Higgs  which  is constrained to be less than about 8.4\% \cite{Cheung:2018ave}. Note that the upper limit on the invisible decay of the SM Higgs can be inverted  into a limit on the coupling $h^0\chi\chi$ \cite{Arhrib:2013ela}. A more recent global fit \cite{Kraml:2019sis}  found that a tighter constraint on $BR(h^0\to\mathrm{invisible})\lesssim 5\%$ at 95\% CL is achievable. In this work, we adopt the most recent result from ATLAS~\cite{ATLAS:2020kdi} for which we have $BR(h^0\to\mathrm{invisible})< 11\%$ at 95\% CL.
\end{enumerate}

\item Direct search from LEP  (DS)

The direct search bounds from collider signatures
of the inert Higgs bosons at hadron collider or at lepton collider are rather similar to charginos and neutralinos production of the Minimal Supersymetric Standard Model (MSSM) \cite{Belanger:2015kga,Lundstrom:2008ai}. 
We will follow the strategy adopted in \cite{Arhrib:2012ia,Swiezewska:2012eh, Lundstrom:2008ai, Belanger:2015kga}.
These constraints can be roughly summarized as follows:

\begin{itemize}
\item[(DS.1)] $m_{H^\pm}>80$ GeV (adapted from charginos search at LEP-II),

\item[(DS.2)] max$(m_{A^0},m_{H^0})>100$ GeV (adapted from neutralinos search at LEP-II),

\item[(DS.3)] $m_{A^0}+m_{H^0}>m_Z$ from the $Z$ width and 
$m_{A^0}+m_{H^\pm}>m_W$ from the $W$ width.
\end{itemize}

\item Electroweak Precision Tests (EWPT)

The EWPT is very sensitive to extra electroweak multiplets, which can contribute to the vacuum polarization processes. Therefore it is reasonable to constrain the Higgs spectrum of the IHDM by  using the global electroweak fit 
through the oblique parameters S, T and U  \cite{Peskin:1991sw}.
The contribution to  S and T parameters of an extra weak Higgs doublet ~\cite{Barbieri:2006dq} can be written as
\begin{equation}
S = \frac{1}{2\pi} \int_0^1 x (1-x) \log\left[\dfrac{x m_{H^0}^2+ (1-x) m_{A^0}^2}{m_{H^+}^2}\right] dx
\end{equation}
%
\begin{equation}
T = \frac{1}{32\pi^2\alpha v^2}\left[F(m_{H^{+}}^2,m_{A^0}^2) + F(m_{H^{+}}^2,m_{H^0}^2) - F(m_{A^0}^2,m_{H^0}^2)\right]
\end{equation}
where the function $F(x,y)$ is defined by
\begin{equation*}
F(x,y) = 
\begin{cases}
\frac{x+y}{2}-\frac{xy}{x-y}\log{\left(\frac{x}{y}\right)}, & x\neq y\\
0, & x = y
\end{cases}
\end{equation*}
From the above expressions, one can easily check that T parameter vanishes in the following 2 limits: $m_{H^+}\to m_{A^0}$ or $m_{H^+}\to m_{H^0}$ while S parameter vanishes only in the degenerate case $m_{H^+}\approx m_{A^0}\approx m_{H^0}$.

To study the correlation between S and T, we define the $\chi^2$ test as follows:
\begin{equation}
\chi^2 = \frac{(T - \widehat{T})^2}{\sigma_T^2(1-\rho^2)}+\frac{(S - \widehat{S})^2}{\sigma_S^2(1-\rho^2)}-\frac{2\rho(T-\widehat{T})(S-\widehat{S})}{\sigma_T \sigma_S(1-\rho^2)}
\end{equation}
Where S and T are the computed quantities in the IHDM given above, while $\widehat{S}$  and $\widehat{T}$ are the experimental central measured values of $\Delta S $ and $\Delta T$,  $\sigma_{S,T}$ are their one-sigma errors and $\rho$ is their correlation. 
 Using the PDG values of S and T with U fixed to be zero, we allow S and T parameters, 
 to be as follows \cite{Tanabashi:2018oca} :
\begin{eqnarray}
T&=& 0.06 \pm 0.06 \quad , \quad S = 0.02 \pm  0.07 
\end{eqnarray}
with the correlation coefficient $\rho_{S,T}=0.92$.

\item DM relic density,  direct, indirect and collider searches (DM)

\begin{enumerate}
\item [(DM.1)] From Plank data, the dark matter relic density $\Omega h^2 $ has been determined as $0.1200\pm0.0012$ \cite{Zyla:2020zbs}.  In the current work, we set this value as an upper bound since it is possible that some particles like the right-handed neutrinos or the axion \cite{Abbott:1982af,Dine:1982ah,Preskill:1982cy} can contribute to the relic density of our universe. 

\item [(DM.2)] Generally speaking, dark matter can be detected by direct search (via the dark matter particle scattering with nuclei), indirect search (via a pair of dark matter particles annihilating into cosmic rays), and collider search (via mono-jet, mono-photon, mono-$W^\pm /Z^0/h^0$ boson processes to measure the missing energy). In micrOMEGAs 5.2\cite{Belanger:2020gnr}, experimental constraints from XENON1T\cite{Aprile:2018dbl}, PICO-60\cite{Amole:2019fdf}, CRESST-III \cite{Abdelhameed:2019hmk} and DarkSide-50 \cite{Agnes:2018ves} have been implemented. From 6 GeV to 1000 GeV, XENON1T provided the most stringent bounds. We will use these bounds in this work. These bounds are obtained by assuming equal proton and neutron spin-independent cross section and assuming a specific choice of astrophysical parameters (i.e the DM velocity distribution is Maxwellian). In the IHDM, dark matter candidates can be either $A^0$ or $H^0$, they can scatter with nuclei via t-channel by exchanging the SM-like Higgs boson (so-called Higgs portal). At leading order, the scattering cross section is isospin-symmetric and spin-independent. Therefore, these bounds will be taken into account.

\item[(DM.3)] At the LHC, one can search for dark matter candidate directly by looking 
for events with high $p_T$ monojet balanced by a large missing transverse energy
\cite{Aaboud:2017phn,Sirunyan:2017hci}. 
The mono-jet final states can  arise from one of the following processes:  $gg \to \chi \chi + g$, $qg\to \chi \chi  + q$ and $q\bar{q} \to \chi \chi  + g$ where $ \chi=H^0, A^0$ is the dark matter candidate. In the IHDM, most of these processes can proceed by producing SM Higgs in association with monojet: $\{gg,qq\}\to h^0 g$ or $h^0 q$ followed by the decay $h^0\to \chi \chi$. Therefore these monojet processes are proportional to $\lambda_{L/S}$.
An attempt to set a limit on IHDM parameters using the ATLAS and CMS data was done 
in \cite{Belyaev:2018ext} where  a projection for future LHC run including the high luminosity option is also provided. Ref. \cite{Lu:2019lok} did a recasting for ATLAS analysis and set a limit on $\lambda_{L/S}$. It is found that for $m_\chi<m_h/2$, $\lambda_{L/S}>3\times 10^{-2}$ is excluded \cite{Belyaev:2018ext,Lu:2019lok} while for $m_\chi>m_h/2$ the limit is weaker and $\lambda_{L/S}>5$ apply only to the case where $62.5<m_\chi<100$ GeV 
\cite{Lu:2019lok}.   For higher dark matter mass, the cross section is suppressed, therefore the constraints on   $\lambda_{L/S}$ would be weaker.
\end{enumerate}

\end{enumerate}

\subsection{More about the constraints from $h^0\rightarrow \gamma\gamma$}
\label{sec:hpp}
We now discuss the impact of the LHC experimental searches on the IHDM. Taking into account the latest measurement of the  di-photon signal strength, we study the constraint on the charged Higgs mass and  $h^0 H^+ H^-=-v \lambda_3$ coupling that are involved in $h^0\to \gamma \gamma$ with $h^0$ being the SM Higgs. Since in the IHDM, the Higgs boson production cross section is identical to the SM one.
Therefore, the di-photon signal strength reduces to the ratio
of branching fractions of $Br(h^0\to \gamma \gamma)$
in the IHDM and in the SM:
\begin{eqnarray}
\mu_{\gamma\gamma} \approx R_{\gamma \gamma}\equiv \frac{Br(h^0\to \gamma \gamma)^{IHDM}}{Br(h^0\to \gamma \gamma)^{SM}}
\end{eqnarray}
Moreover, in case where the decay $h^0\to invisible$ is not open,
the above ratio reduces to:
\begin{eqnarray}
R_{\gamma \gamma}\approx \frac{\Gamma(h^0\to \gamma \gamma)^{IHDM}}{\Gamma(h^0\to \gamma \gamma)^{SM}}
\end{eqnarray}
where $\Gamma(h^0\to \gamma \gamma)$ is the partial decay width of $h^0$ decay into two photons. 
The measured signal strength relative 
to the SM expectation  from ATLAS and CMS are given respectively by:
$\mu_{\gamma\gamma}^{ATLAS}=0.99_{-0.14}^{+0.15}$
\cite{Aaboud:2018xdt} and  $\mu_{\gamma\gamma}^{CMS}=1.18_{-0.14}^{+0.17}$\cite{Sirunyan:2018ouh}.
The combined ATLAS and CMS result at the 2 $\sigma$ level is given by:
 \begin{eqnarray}
 \mu_{\gamma\gamma}=1.04\pm 0.10
 \end{eqnarray}
\begin{figure}[H]\centering
\includegraphics[width=0.35\textwidth]{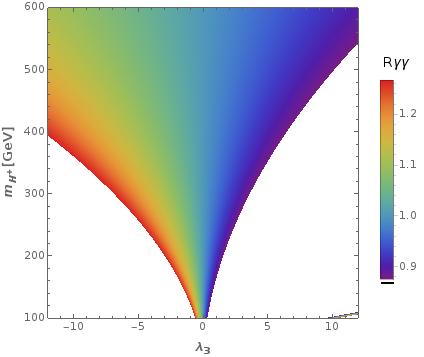}
\includegraphics[width=0.35\textwidth]{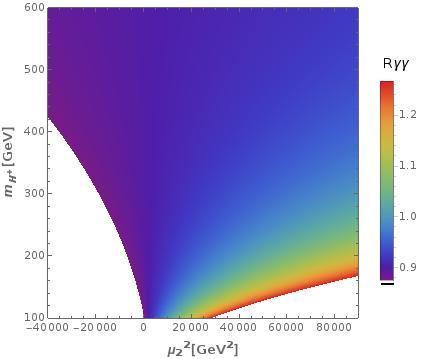}
\caption{Allowed regions in ($m_{H^\pm}, \lambda_3$) plane (left) and ($m_{H^\pm}, \mu_2^2$) plane (right). Only $\mu_{\gamma\gamma}$ constraint is taken into account.}
\label{fig:Rgama}
\end{figure}

In the SM, it is well known that $
\Gamma(h^0\to \gamma \gamma)$ is dominated by the W loops. In the IHDM, the charged Higgs loops can interfere
 constructively (respectively destructively) with the W loops for $\lambda_3<0$ (resp $\lambda_3>0$). In Fig.~(\ref{fig:Rgama}) (left),
 we present $R_{\gamma \gamma}$ in the $(m_{H^\pm}, \lambda_3)$ plane. It is clear that for light charged Higgs boson, its contribution is rather important and could violate $R_{\gamma \gamma}$ measurement. This is translated into a severe constraint
  on $\lambda_3$. Namely, for $m_{H^\pm}=200$ GeV, 
  $\lambda_3$ is forced to be in the range $[-3, 2]$. The allowed range for $\lambda_3$ becomes larger as far as the charged Higgs mass increase. When the charged Higgs is heavier than 400 GeV, the contribution of charged Higgs boson loops will be suppressed, i.e. in the decoupling limit, there is no limit at all on $\lambda_3$.

It is remarkable that there exists a small parameter region where a light charged Higgs boson is allowed, i.e. the light charged Higgs boson can be around 100 GeV but with a large and positive $\lambda_3$ (say $\lambda_3 > 10$). For such a large $\lambda_3$, the contribution of charged Higgs boson is around twice larger than that of W bosons with opposite signs. While in the SM case, the contribution of W bosons is the dominant one while the top contribution is subleading.
In Fig.~(\ref{fig:Rgama}) (right) we show the allowed region in ($m_{H^\pm}, \mu_2^2$) plane.
From this plot one can see that values of $-40\times10^3 \,\mathrm{GeV}^2\leq\mu_2^2\leq 0$ are excluded for charged Higgs mass
$100\,\mathrm{GeV}\leq m_{H^\pm} \leq 420\, \mathrm{GeV}$. While, values of $28\times 10^3 \,\mathrm{GeV}^2\leq\mu_2^2\leq 90\times 10^3 \,\mathrm{GeV}^2$ are not allowed
in a mass range of $100\,\mathrm{GeV}\leq m_{H^\pm} \leq 170 \,\mathrm{GeV}$.

\subsection{Allowed parameter space and selected Scenarios}
\label{sec:allowed_space}
 Before ending this section, we present the effect of the various aforementioned theoretical and experimental constraints on 
the IHDM parameter space. To be precise, the degenerate spectra are defined as the case where all Hidden Higgs bosons are degenerate, i.e.
\begin{eqnarray}
m_{H^0}=m_{A^0}=m_{H^\pm}=m_S\,. \label{eq:deg}
\end{eqnarray}
According to Eq.~(\ref{eq:lams}), we have $\lambda_{4}=\lambda_5=0$ while $\lambda_3$ could be either positive or negative depending on the splitting between $m_{H^\pm}$ and  $\mu_{2}^2$. Then IHDM is fully described by 
three parameters which are:
\begin{eqnarray}
\{m_S, \mu_2^2, \lambda_2\}\,.
\label{eq:deg-scen}
\end{eqnarray}
 We perform a systematic scan over these three parameters taking into account all the above theoretical constraints.
We notice that from vacuum stability constraints $\lambda_2$ must be positive. Using unitarity constraints $e_{9,10}$ 
one gets the strongest limit on $\lambda_{1,2}$ 
which gives $\lambda_{1,2}\le \frac{4 \pi}{3}$.

For non-degenerate spectra, there will be no relation of Eq. (\ref{eq:deg}) for Higgs boson masses. Moreover, due to the direct search bounds on mass from LEP listed in (DS), no degenerate scenarios can be defined if new invisible decay is open.

\begin{figure}[H]\centering
\includegraphics[width=0.32\textwidth]{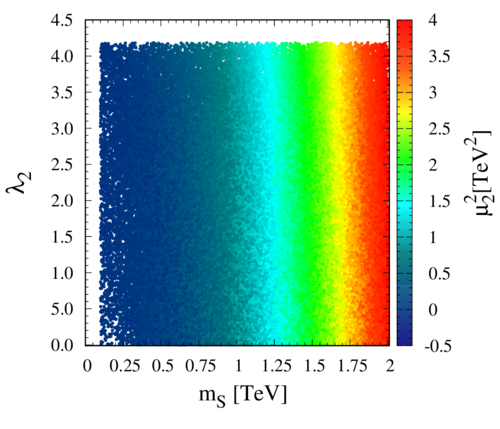}
\includegraphics[width=0.32\textwidth]{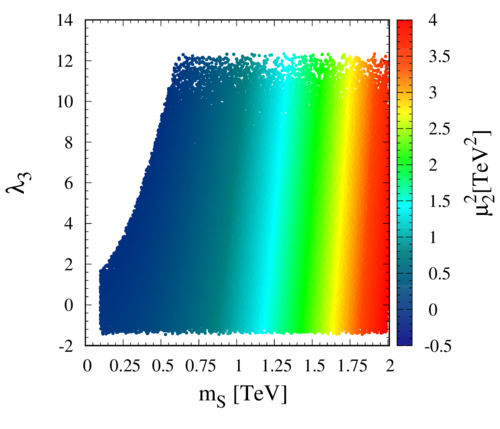}
\includegraphics[width=0.32\textwidth]{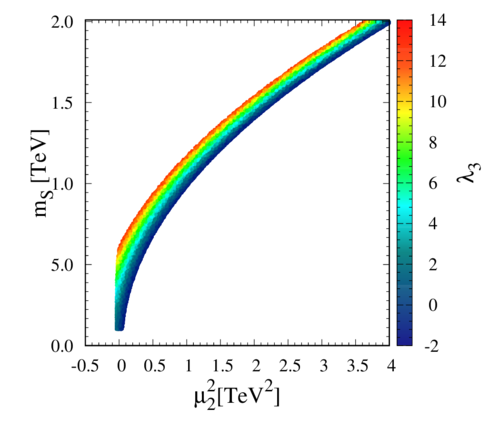}\\
\caption{Allowed parameter space  in the degenerate IHDM spectra are shown, where the various theoretical 
constraints and experimental bounds from the DS and EWPT are taken into account: a) $(\lambda_2,m_S)$ plane with $\mu_2^2$ colored on the vertical axis; b) $(\lambda_3,m_S)$ plane with $\mu_2^2$ colored on the vertical axis and c) $(m_{S}, \mu_2^2)$ plane with $\lambda_3$ colored on the vertical axis.}
\label{fig:deg}
\end{figure}

In  Fig. (\ref{fig:deg}), there are only three free parameters and we present the allowed parameter 
space in $(\lambda_2, m_S)$ plane with $\mu^2_2$ colored in Fig.~(\ref{fig:deg}a).
It should be pointed out that large and negative value of $\mu_2^2$ is excluded both by unitarity constraints as well as by  
the inert vacuum constraints Eq.~(\ref{eq:inertvac}). Only small region with negative $\mu_2^2$ survives.
As one can see, the decoupling limit $m_S\gg m_Z$ is achieved for very large $\mu_2^2$, namely $\mu_2^2>10^6\, GeV^2$.
On the other hand, as discussed before, the size of $h^0H^+H^-$ triple coupling is directly related to the value of $\lambda_3$. 
Therefore, it is interesting to know the allowed space for $\lambda_3$ and its correlation with other parameters.\\
In Figs.~(\ref{fig:deg}b) and (\ref{fig:deg}c), we show the $(\lambda_3,m_S)$ plane with $\mu_2^2$
colored on the vertical axis and $(m_{S}, \mu_2^2)$ plane
where the values of $\lambda_3$ are color coded as indicated  on the right of the plot, respectively.
It can be seen from the figures that above decoupling limit could be reached for quite a wide range of $\lambda_3$.  
It is also clear that $\lambda_3$ could be either positive or negative depending on the splitting between $m_S$ and $\mu_2^2$. 
However, vacuum stability constraints, request that 
$\lambda_3$ could not take large negative values. 
This is clearly seen in Fig.~(\ref{fig:deg}b) where $\lambda_3\in [-2,12]$.

\begin{figure}[!htb]
\centering
\includegraphics[width=0.32\textwidth]{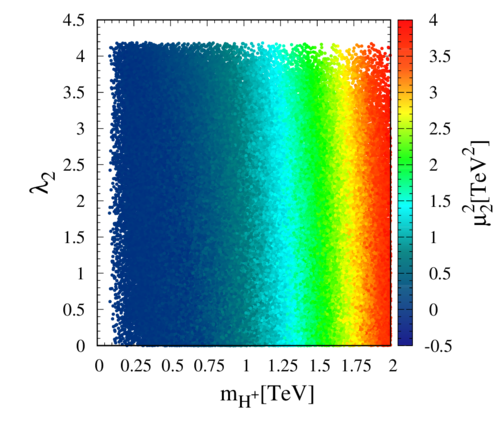}
\includegraphics[width=0.32\textwidth]{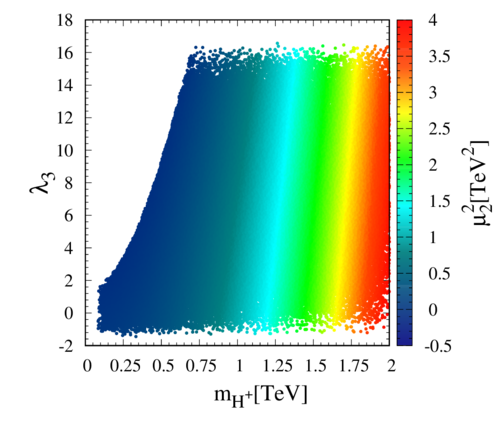}
\includegraphics[width=0.32\textwidth]{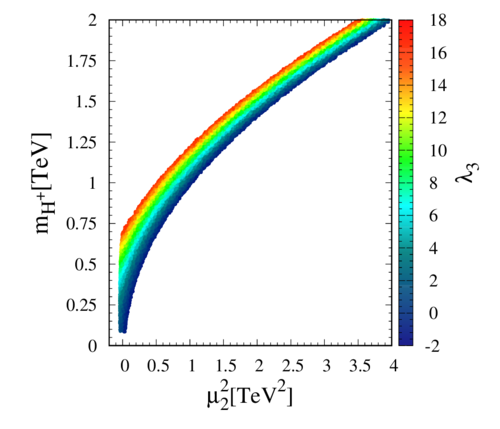}\\
\includegraphics[width=0.32\textwidth]{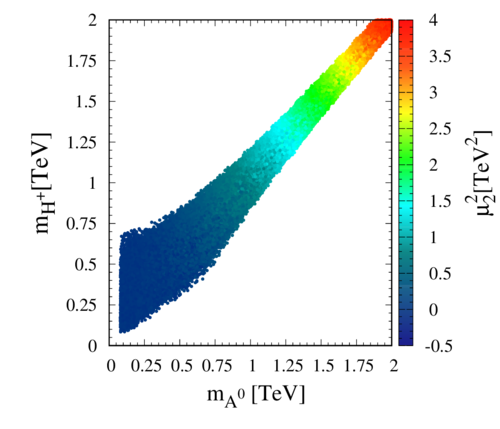}
\includegraphics[width=0.32\textwidth]{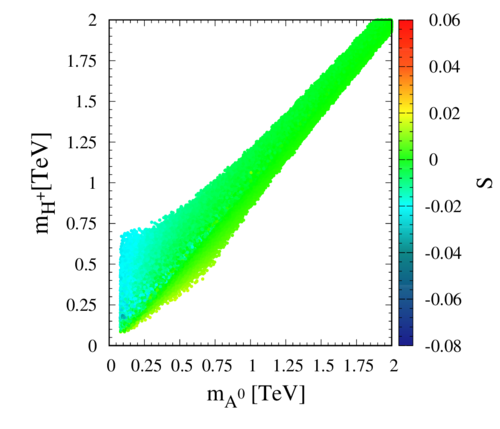}
\includegraphics[width=0.32\textwidth]{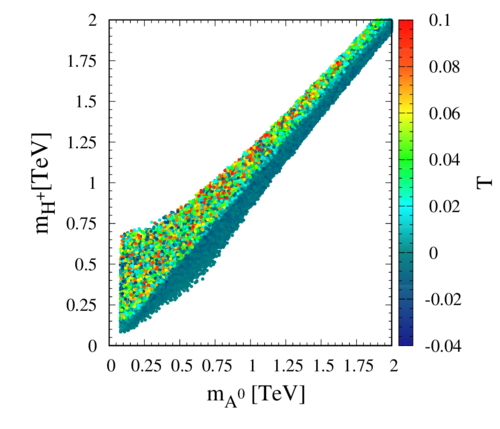}
\caption{Allowed parameter space in the  non-degenerate IHDM spectra are shown, where various theoretical constraints and experimental bounds like the DS and EWPT are taken into account: a) $(\lambda_2,m_S)$ plane with $\mu_2^2$ colored on the vertical axis; b) $(\lambda_3,m_S)$ plane with $\mu_2^2$ colored on the vertical axis; c) $(m_{S}, \mu_2^2)$ plane with
$\lambda_3$ colored on the vertical axis; d,e,f) $(m_{H^\pm}, m_A)$ plane with $\mu_2^2$, S and T parameters colored on the vertical axis, respectively.}
\label{fig:nondeg}
\end{figure}

In  Fig. (\ref{fig:nondeg}), all five parameters of IHDM are free. In Figs.~(\ref{fig:nondeg}a-\ref{fig:nondeg}c), we illustrate our scan  in a similar way to Fig. (\ref{fig:deg}). Fig.~(\ref{fig:nondeg}a) seems almost the same as Fig.~(\ref{fig:deg}a), except that there are less points near the right up corner in  Fig.~(\ref{fig:nondeg}a) and the difference is rather tiny.  Figs.~(\ref{fig:nondeg}b) and (\ref{fig:nondeg}c) have the same shapes with Figs.~(\ref{fig:deg}b) and (\ref{fig:deg}c), but we can see that the upper bound of $\lambda_3$ is changed from 12 to 16, which indicates that larger $h^0H^+H^-$ coupling is allowed in this case. 
It is well known that the electroweak precision observables S and T put a strong constraint on the splitting between the masses that contribute to these parameters. This is clearly illustrated in Figs.~(\ref{fig:nondeg}d-\ref{fig:nondeg}f). 
From Fig.~(\ref{fig:nondeg}d), one can see that the splitting between $A^0$ and $H^\pm$ can not exceed 600 GeV. One can also see from this panel that the decoupling limit with large CP-odd and large charged Higgs can be reached for large $\mu_2^2$. 
It is also visible that 
in the decoupling limit, the splitting between $A^0$ and $H^\pm$ becomes small and does not exceed 100-200 GeV.
In Figs.~(\ref{fig:nondeg}e) and (\ref{fig:nondeg}f), we illustrate the values of S and T parameters on the vertical axis as a function of $m_{A^0}$ and $m_{H^\pm}$.
One can see that in the degenerate case, as expected,
both S and T parameters vanish along the diagonal line $m_{A^0}=m_{H^\pm}$. Away from the diagonal line, S and T get 
non vanishing values but remain within the allowed range.

\begin{figure}[H]
\centering
\includegraphics[width=0.32\textwidth]{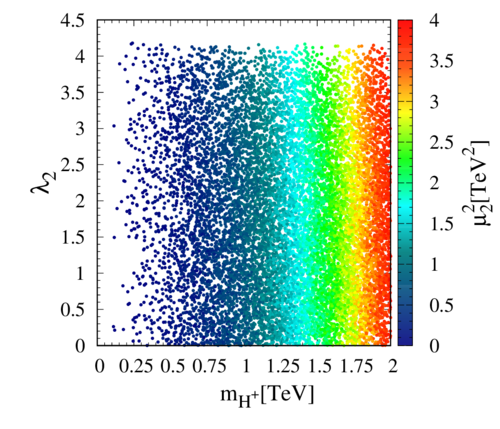}
\includegraphics[width=0.32\textwidth]{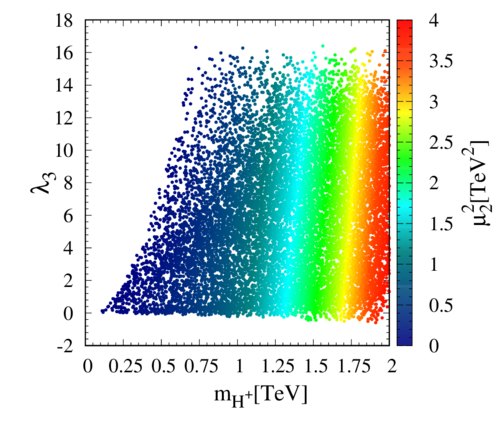}
\includegraphics[width=0.32\textwidth]{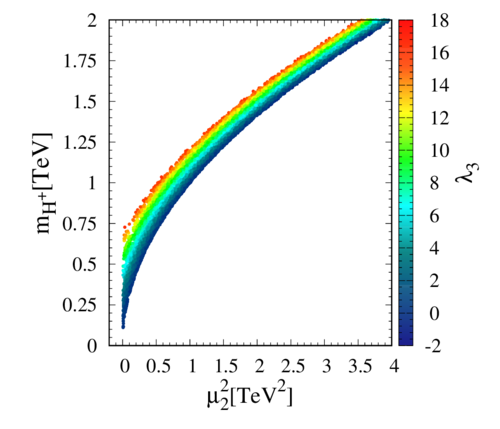}\\
\includegraphics[width=0.32\textwidth]{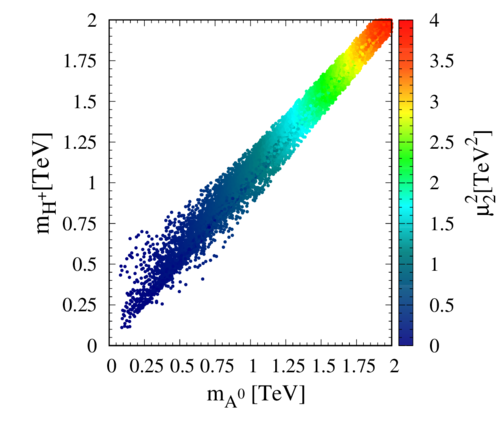}
\includegraphics[width=0.32\textwidth]{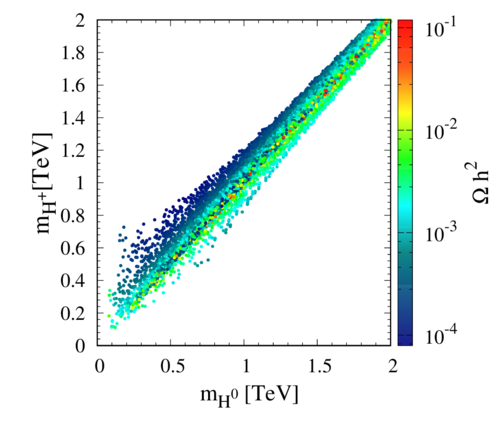}
\includegraphics[width=0.32\textwidth]{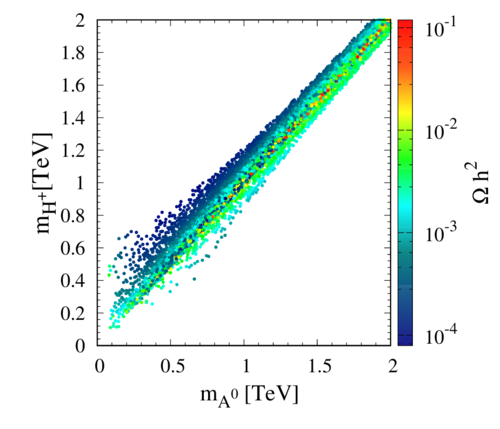}
\caption{Allowed parameter space in  the non-degenerate IHDM spectra using all theoretical and experimental constraints are shown, which correspond to Scenario III defined in Table \ref{tab:scenarios}.}
\label{fig:nondeg-dark}
\end{figure}

In Fig. (\ref{fig:nondeg-dark}), the allowed parameter space in Scenario III is shown. It is observed that the constraints from dark matter conditions can kill $99\%$ of points in Fig. (\ref{fig:nondeg}). It is remarkable that from last two plots in Fig.~(\ref{fig:nondeg-dark}), it can be seen that they are the same despite the permutation of $m_{H^0}$ to $m_{A^0}$ in th x-axis. As mentioned before, the independent parameters in this work are chosen as in Eq.~(\ref{param.IHDM}). Actually it is found that  not only all the allowed parameter space, but also all the results of the processes studied in this work, are totally symmetric under the exchange $m_{H^0}\leftrightarrow m_{A^0}$. 
In the following sections, we will consider 5 Scenarios, which are tabulated in Table \ref{tab:scenarios} and categorized in terms of degenerate and non-degenerate in Higgs boson masses, without/with new invisible decay open, and without/with DM constraints.

\begin{table}[!htb]
\centering
\begin{tabular}{|c|c|c|c|c|c|}
\hline
& Scenario I & Scenario II & Scenario III & Scenario IV & Scenario V  \\
\hline
Theoretical constraints & $\checkmark$ & $\checkmark$& $\checkmark$& $\checkmark$& $\checkmark$\\
\hline 
Degenerate spectrum &$\checkmark$&&&&\\
\hline
\hline
Higgs Data &$\checkmark$&$\checkmark$&$\checkmark$&$\checkmark$&$\checkmark$\\
\hline
Higgs Invisible decay open &&&&$\checkmark$&$\checkmark$\\
\hline
Direct searches from LEP &$\checkmark$&$\checkmark$&$\checkmark$&$\checkmark$&$\checkmark$ \\
\hline
Electroweak precision tests & $\checkmark$&$\checkmark$&$\checkmark$&$\checkmark$&$\checkmark$\\
\hline
Dark matter constraints & & &$\checkmark$&&$\checkmark$\\
\hline
\end{tabular}
\caption{Scenarios and their conditions are tabulated.}
\label{tab:scenarios}
\end{table}
We will propose benchmark points from Scenario III and V and examine their radiative corrections, since both of them can pass all current experimental bounds.

\section{Radiative corrections to: $e^+ e^- \to Zh^0/H^0A^0$}
\subsection{Lowest order results}
\label{sec:LO}
\begin{figure}[!htb]\centering
\includegraphics[width=0.3\textwidth]{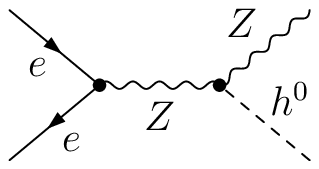}
~~~~
\includegraphics[width=0.3\textwidth]{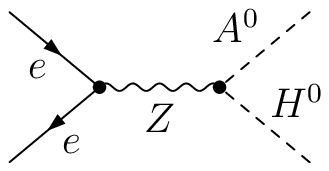}
\caption{Tree level Feynman diagrams for $e^+ e^- \to Zh^0$ and $e^+e^-\to  H^0A^{0}$. }
\label{fig:eeha}
\end{figure}

In our calculation, due to the tininess of electron mass and the corresponding Yukawa couplings, it is justified numerically to neglect the contributions of the Feynman diagrams which involve $e^{+}e^{-}h^0$, $e^{+}e^{-}G^{0}$, 
$e^{-}\overline{\nu_e}G^{+}$ and $e^{+}\nu_eG^{-}$ vertices.
For this reason, at the tree-level, the only one contributes in these two processes is the s-channel Z-exchange diagram, as shown in Fig.~(\ref{fig:eeha}).

From the covariant derivative of the Higgs doublet, one can derive the Higgs coupling to gauge bosons. We list hereafter a part of the Lagrangian needed for our study 
\begin{eqnarray}
{\cal L}_{VS_iS_j,VVh^0}&=&(-i e A^\mu + i e\frac{(c_W^{2}-s_W^{2})}{2c_W s_W} Z^\mu) H^+\stackrel{\leftrightarrow}{\partial}_\mu H^-\nonumber\\
&&+ \frac{e}{2c_W s_W} Z^\mu H^0\stackrel{\leftrightarrow}{\partial}_\mu A^0 + i \frac{e m_Z}{c_W s_W} h^0
Z^\mu Z_\mu,
\label{eq:lag}
\end{eqnarray}
where $c_W=\cos\,\theta_W$,
$s_W = \sin\, \theta_W$. We stress first that all the above couplings are fixed just in terms of gauge coupling.  \\
For  $e^+e^-\to Zh^0$, it is easy to compute the differential cross section which is given by \cite{Djouadi:2005gj}:
\begin{eqnarray}
\frac{d\sigma^0(Zh^0)}{d\cos\theta}&=& \frac{\alpha^2 \pi }{256  s_W^4 c_W^4 s}\biggl(1+(1-4 s_W^2)^2\biggr) \kappa_{Zh^0} \frac{8 m_Z^2/s  + \kappa_{Zh^0}^2\sin^2\theta}{(1-m_Z^2/s)^2} \label{eeZhcos} 
\end{eqnarray}
where 
\begin{equation}
\kappa_{ij}^2=\biggl(1-\frac{(m_i+m_j)^2}{s}\biggr)\biggl(1-\frac{(m_i-m_j)^2}{s}\biggr) 
\label{eq.phase}
\end{equation}
The total cross section is obtained after integration over the scattering angle. 
The analytical result can be found in \cite{Djouadi:2005gj}:

\begin{eqnarray}
\sigma^0(Zh^0)&=& \frac{\alpha^2 \pi}{192  s_W^4 c_W^4 s}\biggl(1+(1-4 s_W^2)^2\biggr) \kappa_{Zh^0} \frac{12m_Z^2/s+ \kappa_{Zh^0}^2}{(1-m_Z^2/s)^2} \label{eeZh} 
\end{eqnarray}
The total cross section for the associate production $e^+e^-\to H^0A^0$ 
is given by:
\begin{eqnarray}
\sigma^0(H^0A^0)&=&  
\frac{\alpha^2 \pi }{192  s_W^4 c_W^4 s} \biggl(1+(1-4 s_W^2)^2\biggr) \frac{\kappa_{A^0H^0}^3 }{(1-m_Z^2/s)^2}
\label{eeHA} 
\end{eqnarray}
Because of the presence of two scalars in the final state, $e^+e^-\to H^0 A^0$ process
has the suppression factor $\kappa_{A^0H^0}^3$ which reduces tremendously 
the cross section.

In the Higgstrahlung process $e^+e^-\rightarrow Zh^0$, since there is no mixing between the two CP even neutral Higgs bosons, the cross-section  is the same as in the SM. It is clear that $\sigma^0(Zh^0)$ scales like $1/s$ and  is significant only 
at low energy just after the production threshold $\sqrt{s}\approx m_{h^0} +m_Z$.  We stress that 
the term $\kappa_{Zh^0}^2$ in $\sigma^0(Zh^0)$ originates from  the longitudinal component of the Z,
therefore one could conclude that at high energy the cross section is dominated by the longitudinally polarized Z boson.
On the other hand, the production cross section for $\sigma^0(H^0 A^0)$ drops quickly due to the phase space suppression factor  
$\kappa_{H^0A^0}^3$ (see Fig.~(\ref{fig:eeHA-0})).

\subsection{$e^+ e^- \to Zh^0/H^0A^0$ at one loop}
\label{sec:renormalization}
\begin{figure}[!ht]
\begin{center}
\framebox[0.8\textwidth]{\includegraphics[width=0.8\textwidth]{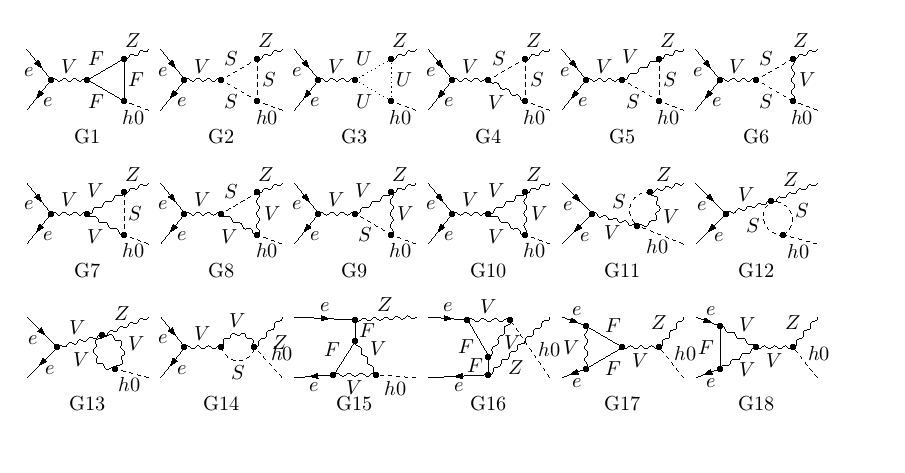}}\\
\framebox[0.8\textwidth]{\includegraphics[width=0.8\textwidth]{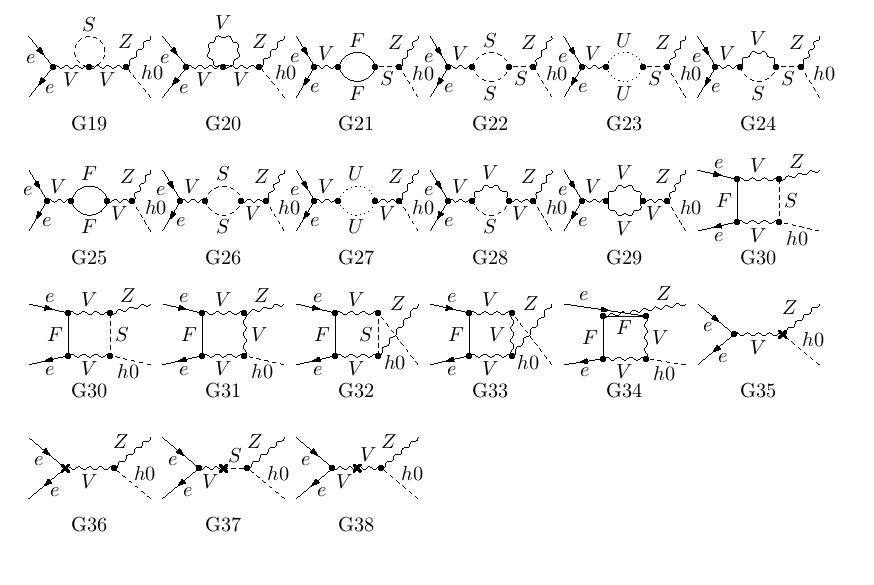}}
\caption{Generic one-loop Feynman diagrams for $e^+e ^- \to Zh^0$
 where $F$ stands for SM fermions, $V$ stands for generic vector boson which could be $\gamma$, $Z$ or $W^\pm$ and 
 $S$ could be either a Goldstone $G^0$, $G^\pm$ or a Higgs boson $h^0,H^0,A^0$ or $H^\pm$.}
\label{figure-zh}
\end{center}
\end{figure}

For all the above processes introduced in the previous section, we have evaluated both the weak corrections as well as the  virtual photons ones in the 'tHooft-Feynman gauge.
The generic Feynman diagrams for $e^+e^- \to Zh^0$ are drawn in Fig.~(\ref{figure-zh}). These comprise: 
1) one-loop corrections to the vertices $V Zh^0$ ($V=\gamma, Z$), $G_1$ to $G_{14}$; 
2) t-channel diagram with one-loop correction to $e^+e^- h^0$ vertices, $G_{15}$  and $G_{16}$; 
3) one-loop corrections to the initial state vertices $Ve^+e^-$ ($V=\gamma, Z$) which is purely SM, $G_{17}$  and $G_{18}$; 
4) one-loop corrections to Z boson and photon propagators as well as $\gamma$-$Z$ and $Z$-$G^0$  mixings, $G_{19}$ to $G_{29}$; 
5) box contributions, $G_{30}$  to $G_{34}$. 
The various counter-terms for initial and final states and also the $\gamma$ and $Z$
propagators, $\gamma$-Z and Z-$G^0$ mixings are also depicted in $G_{35}$ to $G_{38}$.

\begin{figure}[!ht]
\begin{center}
\framebox[0.9\textwidth]{\includegraphics[width=0.9\textwidth]{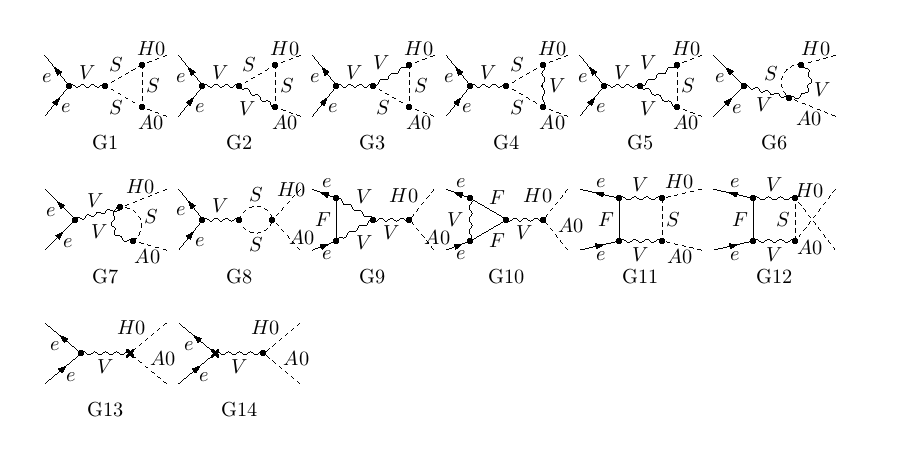}}
\caption{Generic one-loop Feynman diagrams for 
$e^+ e^- \to H^0 A^0$ where $F$ stands for SM fermions, $V$ stands for generic vector boson which could be  $\gamma$, $Z$ or $W^\pm$ and $S$ could be either a Goldstone $G^0$, $G^\pm$ or a Higgs boson $h^0,H^0,A^0$ or $H^\pm$.}
\label{diagramsHA}
\end{center}
\end{figure}

Similarly, we draw in Fig.~(\ref{diagramsHA})
 the generic Feynman diagrams for  
$e^+e^- \to H^0A^0$ and the corresponding 
counter-terms.  For this process, the self energies are not drawn, they are similar to the previous process 
$e^+ e^- \to Zh^0$. 

Evaluation of the one-loop corrections will lead to ultra-violet (UV)  as well as IR divergences. The UV singularities are regularized with dimensional regularization and treated in the on-shell renormalization scheme while the IR ones are regularized with a small fictitious photon mass $\lambda$ and cancelled with  real photon emissions.

For both processes, owing to Lorentz invariance, 
the mixing $Z^\mu$-$G^0$ is proportional to $(p_1^\mu+p_2^\mu)$ where $p_1$ and $p_2$ are the momentum of the external electron and positron. After contracting $Z^\mu$-$G^0$ mixing with the initial state vertices $e^+e^-Z$ and using Dirac equation the amplitude would be proportional to the electron mass which is neglected in the present calculation.

The presence of $\zz_{2}$ symmetry forbids the mixing between the SM doublet $H_1$ and the inert doublet $H_2$ which tremendously eases the renormalization of the IHDM. 
The full renormalization of the IHDM has been presented recently  in~\cite{Banerjee:2019luv}. In our study, we will use the on-shell scheme developed first for the SM in \cite{Bohm:1986rj,Hollik:1988ii,Denner:1991kt} for all SM parameters supplemented by an on-shell renormalization for the extra-inert Higgs fields and their masses.  
Concerning the renormalization of the SM parameter and fields 
such as: the electric charge, the on-shell definition of the W and Z masses, $\gamma$-$Z$ mixing and  Weinberg angle, 
 we refer  to \cite{Denner:1991kt}.
For the renormalization of the inert Higgses,  we use similar  approach as in~\cite{Denner:1991kt}. 
Because of $\zz_2$ symmetry, there is no mixing between $h^0$-$H^0$, 
$G^0$-$A^0$ and $Z$-$A^0$. This simplifies the renormalization of the Higgs fields. Let us redefine the new Higgs fields and masses as follows:

\begin{eqnarray}
&&\Phi \to Z_\Phi^{1/2}\Phi = (1 + \frac{1}{2} \delta Z_\Phi )\Phi \nonumber\\
&&m_\Phi^2 \to m_\Phi^2 + \delta m_\Phi^2, \quad \Phi=h^0, H^{0}, A^0
\end{eqnarray}

Inserting these redefinitions  into the above Lagrangian in Eq.~(\ref{eq:lag}), we find the following counter terms: 
\begin{eqnarray}
&&\delta {\mathcal{L}}_{Z H^0 A^{0}}=\frac{e}{2c_W s_W}(\delta Z_{e}+\frac{\delta Z_{H^0}}{2}+\frac{\delta Z_{ZZ}}{2} -\frac{\delta{s_W}(c_W^{2}-s_W^{2})}{c_W^{2} s_W}+\frac{\delta Z_{A^{0}}}{2}) Z^\mu  H^0\stackrel{\leftrightarrow}{\partial}_\mu A^0 
\nonumber\\
&&\delta {\mathcal{L}}_{Z Z h^0}=i \frac{e m_W}{s_W c_W^2}(\delta Z_{e}+\frac{\delta Z_{H^0}}{2}+\delta Z_{ZZ} -\frac{\delta{s_W}(c_W^{2}-2s_W^{2})}{c_W^{2} s_W}+\frac{\delta m_W^{2}}{2 m_{W}^2})Z^{\mu} Z_{\mu} h^0 
\label{eq:CT}
\end{eqnarray}
For the counter terms of the initial state vertices $e^+e^-\gamma$ and  $e^+e^-Z$, counter terms of the Z boson, the photon propagators and their mixing, they are exactly the same as in the SM and can be found in \cite{Denner:1991kt}. 

We first stress that the
	 counter-term for $e^+e^-h^0$  would be proportional to the electron mass and then vanishes for $m_e\rightarrow 0$. Second, in the case when both $e^-$and $e^+$ 
are on shell, the one loop contributions coming from 
the $e^+e^-h^0$ vertex are vanishing in the limit of zero electron mass. However, from diagrams G15 and G16, one of the 
fermions ($e^-$ or $e^+$) coupled to the Higgs boson, is off shell. Consequently, for vanishing $m_e$ the correction
to the vertex $e^+e^-h^0$ are UV-finite but non-vanishing (because, in this case there are contributions where
the suppression factor ${m_e}/{m_W}$ is absent). We have checked analytically this feature for this process. The remaining part, is of the same order as the other Feynman diagrams and should be included in the computation.

The Higgs wave function renormalization constants and  mass counter-terms are fixed by the on-shell conditions for the masses and the Higgs fields and also by requiring the residue =1 for the Higgses.
These requirements will lead to:
\begin{eqnarray}
&&\textrm{Re} \frac{\partial\hat{\Sigma}^{\Phi\Phi}}{\partial k^{2}}|_{k^2=m_\Phi^2} = 0 \nonumber\\
&& \textrm{Re} \hat{\Sigma}^{\Phi\Phi}(m_\Phi^2) = 0 \quad, \quad \Phi=h^0, H^{0}, A^0
\end{eqnarray}
However, all the counter terms for the gauge boson masses and wave function renormalization and their mixing as well as Weinberg angle are fixed following  Ref.~\cite{Denner:1991kt}. 

The electric charge renormalization constant $\delta Z_e$  is fixed from the $e^+e^- \gamma$ vertex. We require that the renormalized
three point function $\hat{\Gamma}_{e^+e^- \gamma}^\mu$
 satisfies at the Thomson limit:
$$\hat{\Gamma}^\mu_{e^+e^- \gamma} (\not{p}_1 =
\not{p}_2 =m_e, q^2 = 0) = ie\gamma^\mu, $$
and the renormalization constant for electric charge $\delta Z_e$ is obtained as~\cite{Bohm:1986rj,Hollik:1988ii,Denner:1991kt}
\begin{equation}
\delta Z_e=-\dfrac{1}{2}\delta Z_{AA}-\dfrac{s_W}{c_W}\dfrac{1}{2}\delta Z_{ZA}
=\dfrac{1}{2}\Pi^{AA}(0)-\dfrac{s_W}{c_W}\dfrac{\sum^{AZ}_T(0)}{m_Z^2}
\end{equation}
with
\begin{equation}
\Pi^{AA} (0)\equiv\dfrac{\partial\sum^{AA}_T(s)}{\partial s} |_{ s=0}
\end{equation}

There is no reliable theoretical predictions  
available to extract $\Pi^{AA}_{\mathrm{hadron}}(0)$, but this quantity 
can be extracted  from the  experimental data. A non-perturbative parameter $\Delta\alpha^{(5)}_{\mathrm{hadron}}(m_Z)$ is used to absorb the hadronic contribution, namely $\delta Z_e$ is rewritten as
\begin{equation}
\delta Z_e|_{\alpha(0)}=\dfrac{1}{2}\mathrm{Re}\Pi^{AA(5)}_{\mathrm{hadron}}(m_Z^2)+\dfrac{1}{2}\Delta\alpha^{(5)}_{\mathrm{hadron}}(m_Z)+\dfrac{1}{2}\Pi^{AA(5)}_{\mathrm{remaining}}(0)-\dfrac{s_W}{c_W}\dfrac{\sum^{AZ}_T(0)}{m_Z^2}
\end{equation}

Another popular scheme, $\alpha(m_Z)$ scheme, is more preferred, in which the large logarithm from leptons are also absorbed into the redefinition of running coupling constant~\cite{Denner:1991kt,Sun:2016bel,Xie:2018yiv}. 
The corresponding renormalization constant can be converted from $\alpha(0)$ scheme as:
\begin{equation}
\delta Z_e|_{\alpha(m_Z)}=\delta Z_e|_{\alpha(0)}-\dfrac{1}{2}\Delta\alpha(m_Z)
\end{equation}
with
\begin{equation}
\Delta\alpha(m_Z)=\Pi^{AA}_{f\neq \mathrm{top}}(0)-\mathrm{Re}\Pi^{AA}_{f\neq \mathrm{top}}(m_Z^2),
\end{equation}
and the running coupling constant is replaced with
\begin{equation}
\alpha(m_Z)=\dfrac{\alpha(0)}{1-\Delta\alpha(m_Z)}.
\end{equation}

Let us now discuss the treatment of the IR divergences.
In fact, the IR divergences are present in two sources: i) wave function renormalization of charged particles such as electrons; 
ii) vertex corrections to $e^+e^-\gamma$  and $e^+e^- Z$:   Fig.~(\ref{figure-zh})-$G_{17}$ and Fig.~(\ref{diagramsHA})-$G_{10}$ 
with $V=\gamma$,  where incoming electron and positron exchange an virtual photon.

As mentioned before,
in our calculation, to deal with the IR divergences, a small fictitious photon mass $\lambda$ is introduced to regularize the soft and the virtual emission of the photon. 
Meanwhile, two cutoffs, $\Delta E$ and $\Delta\theta$, are introduced to deal with the IR singularities in real photon emission process. $\Delta E=\delta_s\sqrt{s}/{2}$ 
defines the soft photon energy cut-off for the bremsstrahlung process. It can be viewed as the photon energy cut 
that separates the soft from the hard radiation. 
The angle $\Delta \theta$ cut defining between photon and the beam $\theta_\gamma$ is used to separate hard radiation into hard collinear and hard noncollinear parts.

With $\lambda$, $\Delta E$ and $\Delta\theta$, 
the next-leading-order (NLO) corrections are decomposed into the virtual (V), soft (S), hard collinear (HC), and hard non-collinear ($H\overline{C}$) parts as follows:
\begin{equation}
d\sigma^{1}=d\sigma_V(\lambda)+d\sigma_S(\lambda,\Delta E)+d\sigma_{HC+CT}(\Delta E,\Delta\theta)+d\sigma_{H\overline{C}}(\Delta E,\Delta\theta) 
\label{fullcor}
\end{equation}
Here $V$ denotes the virtual correction including loop diagrams and counter terms from renormalization. $CT$ denotes the ``counter term'' from electron structure function, originated from the 2nd term in Eq.~(\ref{eqn:ff}). 
 
Notice that the soft bremsstrahlung for  $e^+e^- \to Zh^0$ and $e^+e^-\to H^0A^0$ processes  can be found in the literature. For completeness, we give the analytical expressions in the Appendix A, as well as many other details. 
The $\lambda$ independence can be checked when combining 
the soft bremsstrahlung (S part) with the virtual one-loop  
QED contribution (V part) and this has been verified numerically with a good precision. 
Notice that we found a good agreement  when we compare the result
from FDC and FormCalc. Moreover, we have numerically checked that 
our results do not depend on $\Delta E$, $\Delta\theta$ and $\log(m_e)$, as shown in Fig.~(\ref{figcheck}).

The total cross section at NLO, $\sigma^{NLO}$, is the sum of LO cross section $\sigma^{0}$, and NLO corrections $\sigma^{1}$, namely
\begin{eqnarray}
\sigma^{NLO}&=&\sigma^{0} + \sigma^{1} \equiv \sigma^0(1+\Delta)\, ,
\label{sig1}
\end{eqnarray}
where $\Delta$ is the relative correction.  Thus $\Delta$ can be decomposed 
 into two gauge-invariant parts,
\begin{eqnarray}
\Delta =\Delta_{\mbox{weak}}+\Delta_{\mbox{QED}}\label{split}
\end{eqnarray}

In the calculation of $d\sigma_V$, computation of all the one-loop amplitudes and
counter-terms is done with the help of FeynArts and FormCalc~\cite{Hahn:2000kx,Hahn:1998yk,Hahn:2006qw}
packages. Numerical evaluations of the scalar integrals are done with
LoopTools~\cite{Hahn:1999mt,Hahn:2010zi} and we have  also tested the cancellation of
UV divergences both analytically and numerically. The soft part $d\sigma_S$ is also done with FormCalc, while $d\sigma_{HC+CT}$ and $d\sigma_{H\overline{C}}$ are obtained with the help of FDC~\cite{Wang:2004du}.

\section{Numerical results}
\label{sec:results}
In this section, we present our numerical results for the two processes introduced above. 
We adopt the following numerical values of the physical parameters from PDG \cite{Tanabashi:2018oca}:
\begin{enumerate}
\item the fine structure constant: $\alpha(0)=1/{137.036}$, $\alpha(m_Z)={1}/{128.943}$ with $\Delta\alpha^{(5)}_{\mathrm{hadron}}(m_Z)=0.02764$
\item the gauge boson masses: $m_{W}=80.379$ GeV and $m_{Z}=91.1876$ GeV
\item the fermion masses: $m_{e}=0.511$ MeV, $m_{\mu}=0.106$ GeV, $m_{\tau}=1.78$ GeV, $m_{t}=173.0$ GeV while the masses of light quarks are set as $m_u=m_d=3.45$ MeV, $m_s=0.095$ GeV, $m_c=1.275$ GeV and $m_b=4.66$ GeV. 
\end{enumerate}

In the IHDM, the CP even Higgs boson $h^0$ is identified as the Higgs like particle observed  by the LHC collaborations and we use $m_{h^0}=125.18$ GeV. For the other IHDM parameters, we perform a systematic scan which include the physical masses $m_{H^0}$, $m_{A^0}$ and $m_{H^\pm}$, $\lambda_2$ and $\mu_{2}^2$ parameters. We take into account all theoretical requirements given in the subsection \ref{sec:thexcon} as well as  all experimental constraints  given in the subsection \ref{sec:expxcon}.
It is found that our numerical results are almost independent of $\lambda_2$. Therefore, in the following part, we will fix $\lambda_2=2$.

In what follows, we will use the $\alpha(m_Z)$ scheme described before to present our numerical results.

\subsection{Higgs-strahlung: $e^+e^- \to Zh^0$}
\label{sec:higgsstrahlung}
Radiative corrections to $e^+e^-\to Zh^0$ in the SM are well known since 
long time \cite{Fleischer:1982af, Kniehl:1991hk, Denner:1992bc}. 
Here we investigate them in the proposed scenarios given in Table (\ref{tab:scenarios}).

\begin{figure}[H]
\begin{center}
\includegraphics[width=0.32\textwidth]{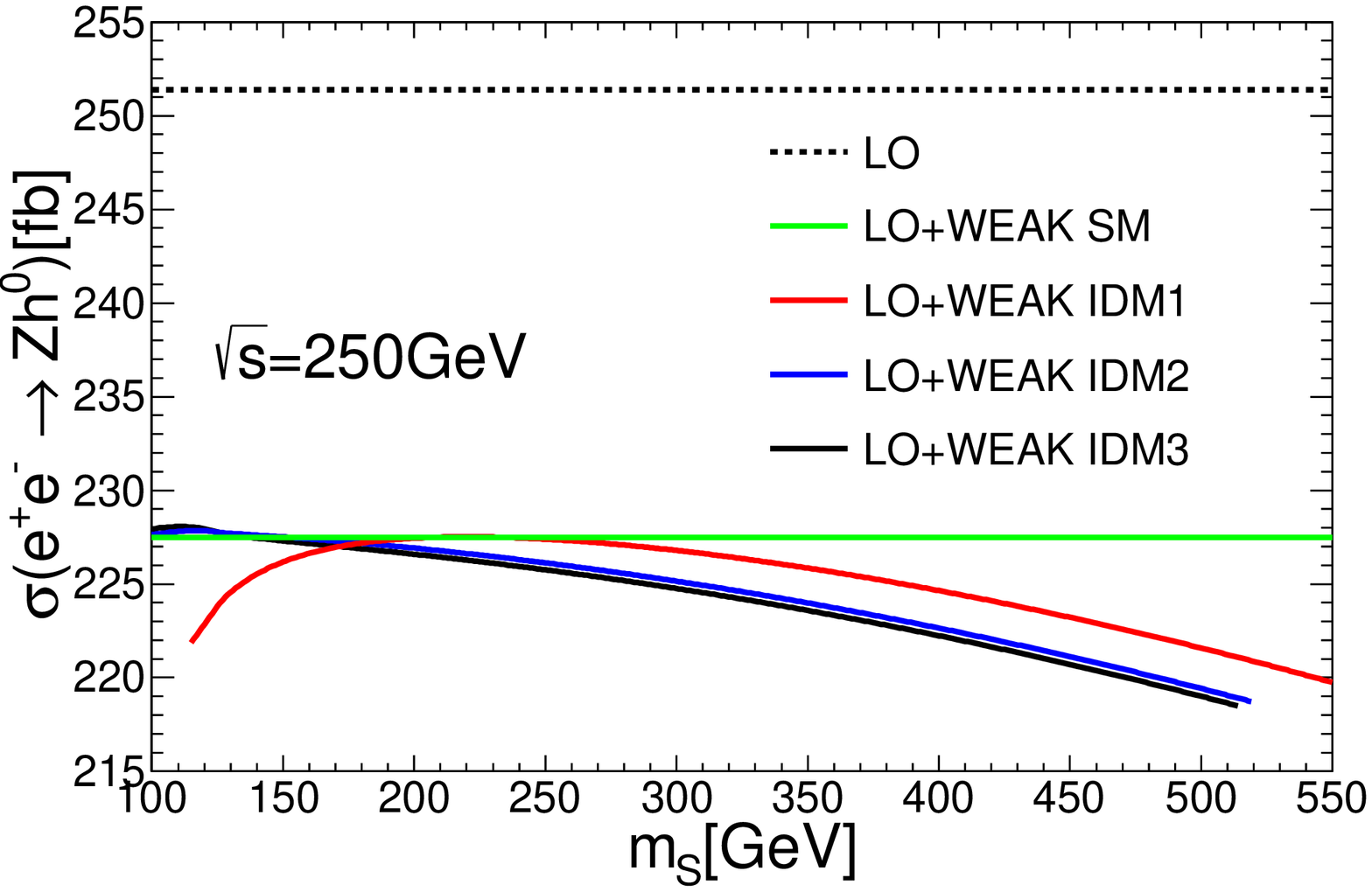}
\includegraphics[width=0.32\textwidth]{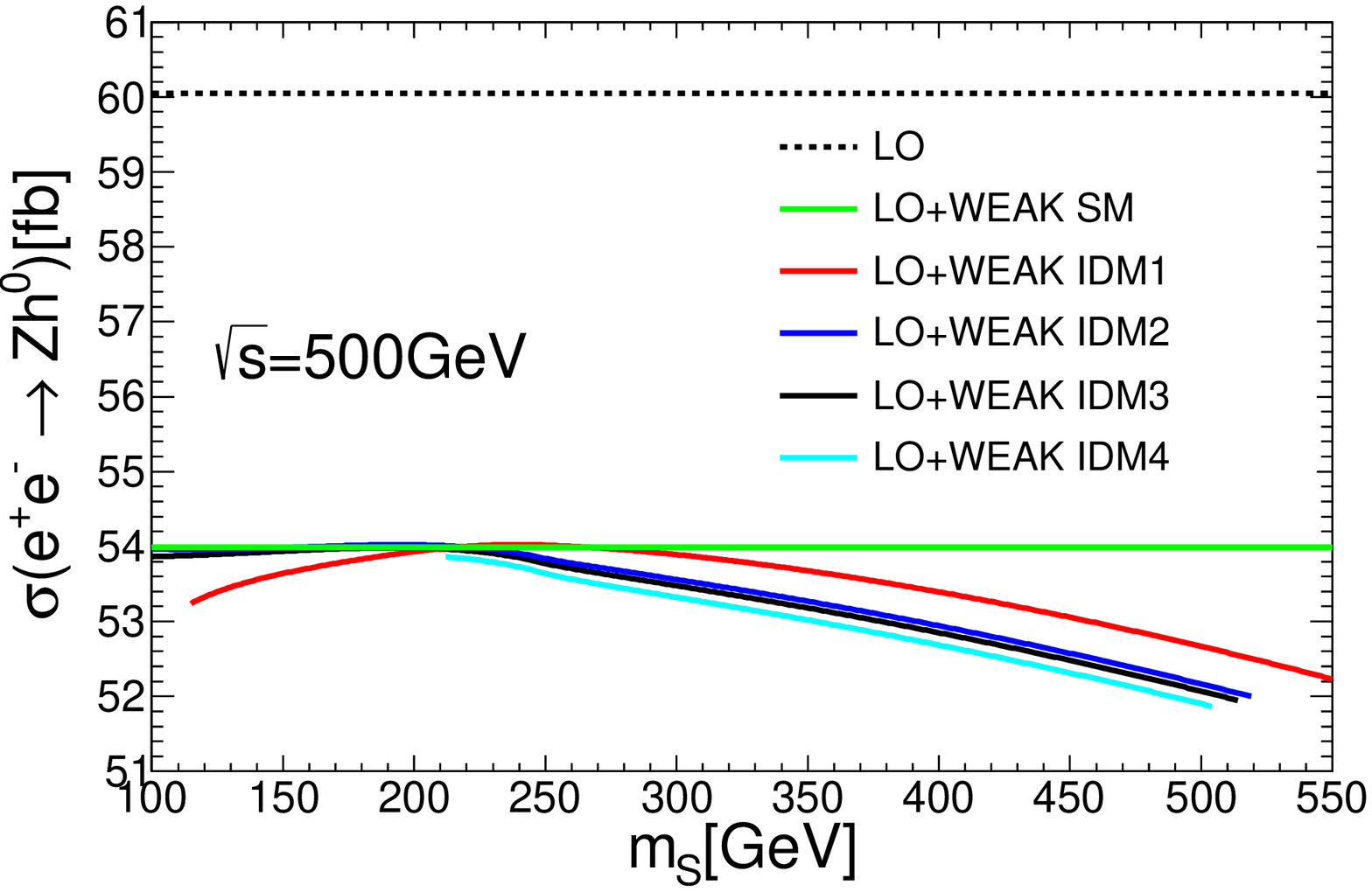}
\includegraphics[width=0.32\textwidth]{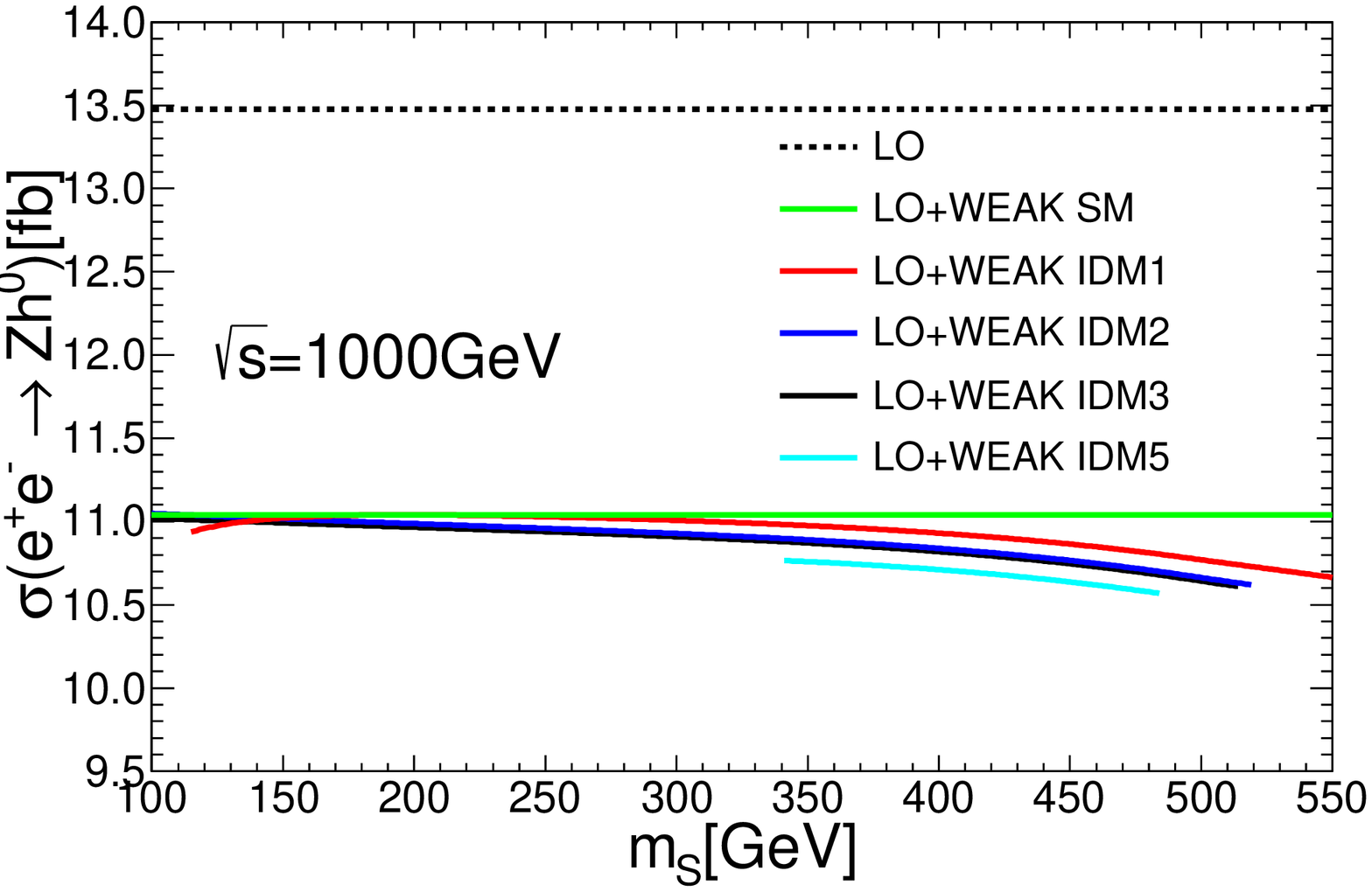}\\
\includegraphics[width=0.32\textwidth]{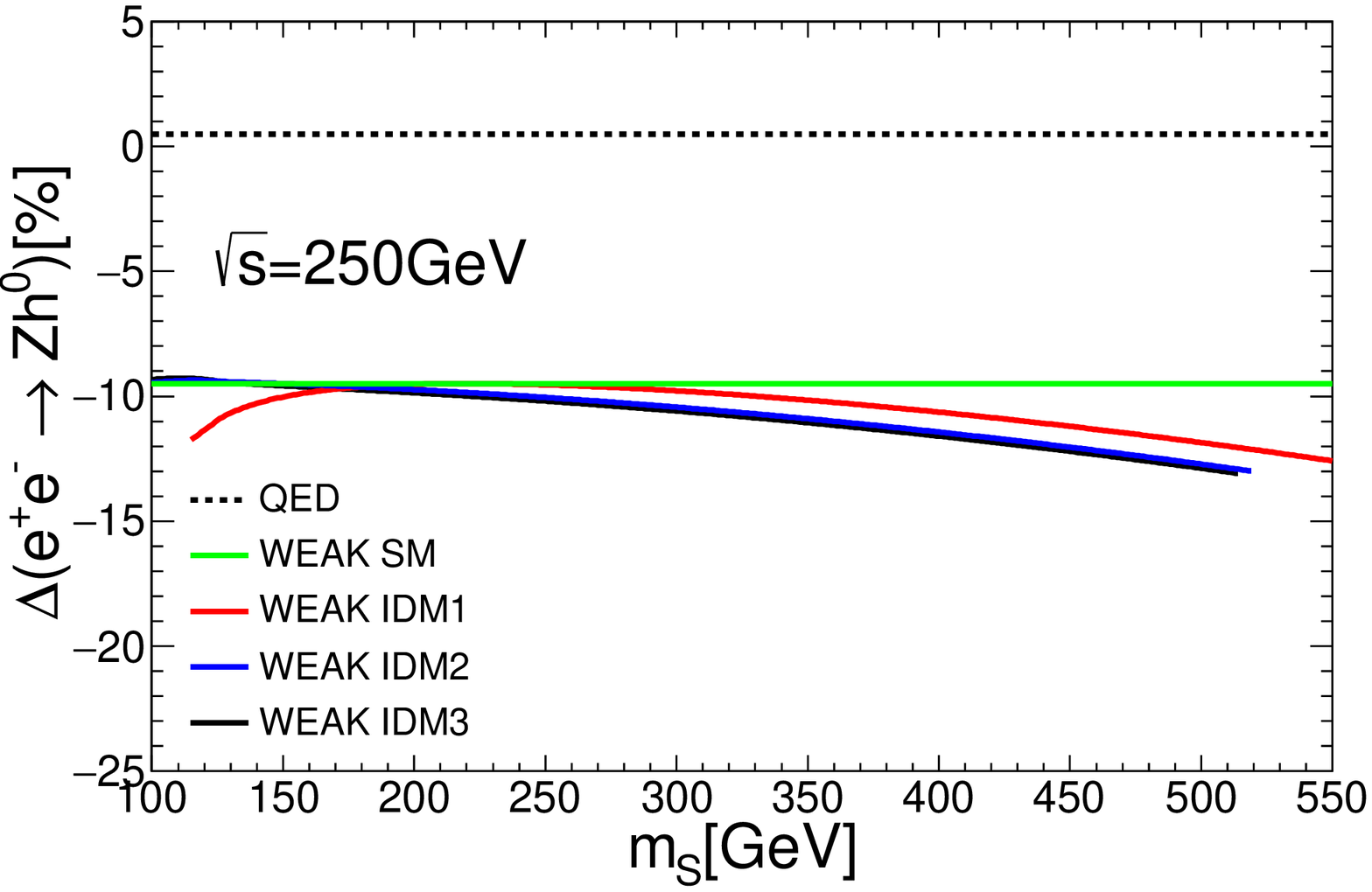}
\includegraphics[width=0.32\textwidth]{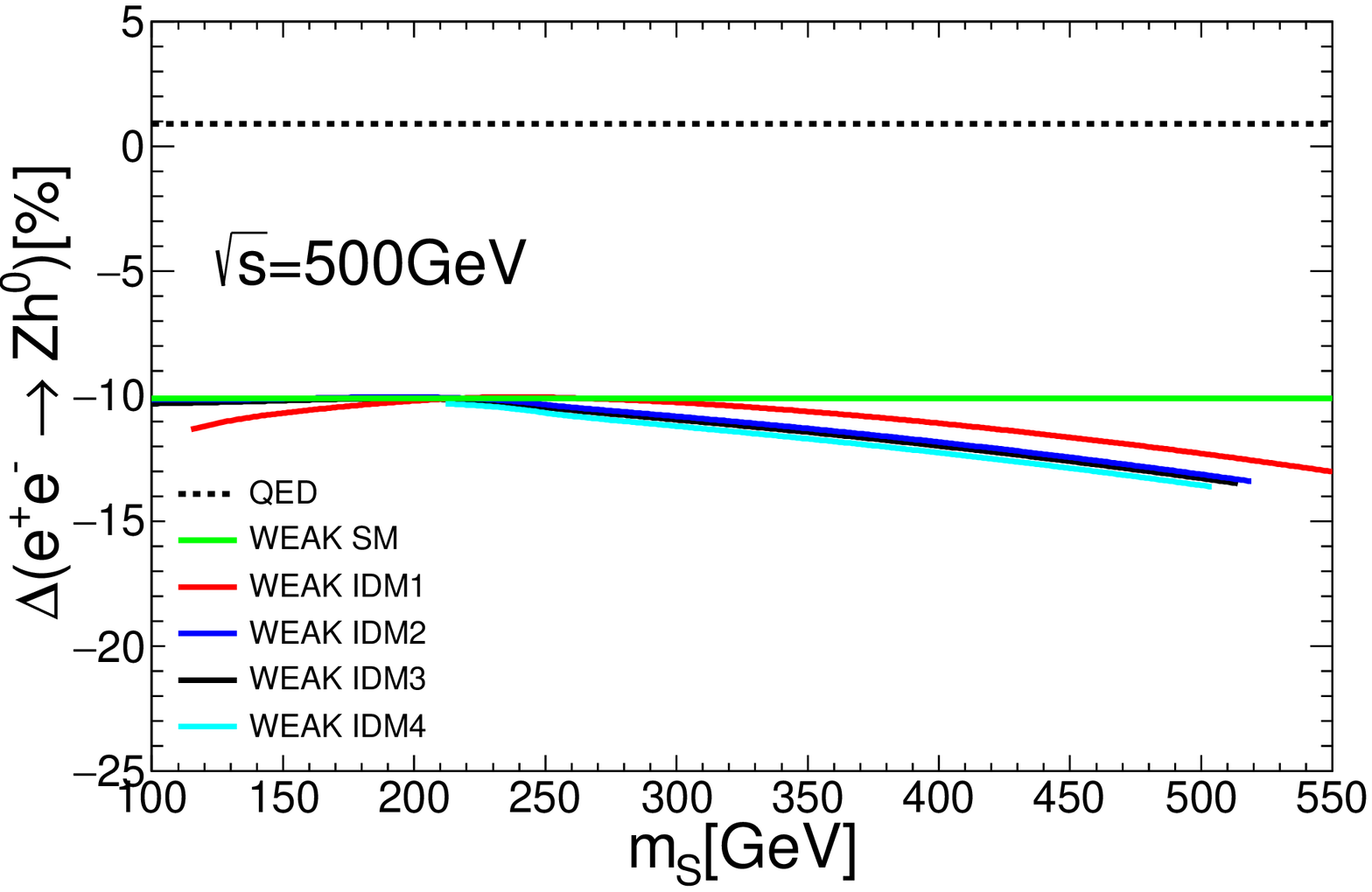}
\includegraphics[width=0.32\textwidth]{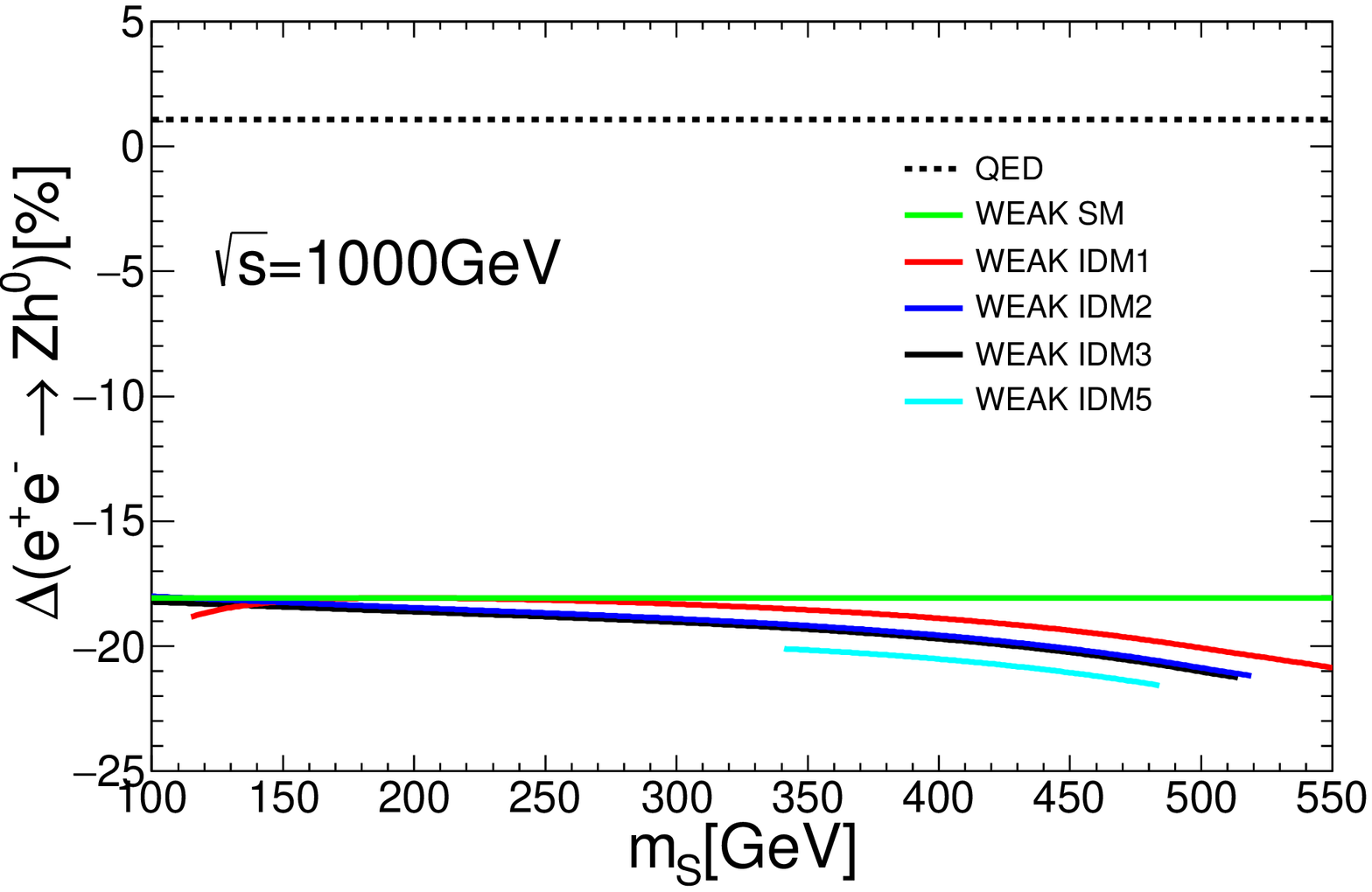}
\caption{In Scenario I, total cross section and relative corrections for $e^+\ e^- \to Z h^0$ 
as a function of the Higgs masses $m_S$ with three collision energies $\sqrt{s}=250$ GeV,  $500$ GeV and $1000$ GeV. Corresponding values of $\mu^2_2$ are given in Table~\ref{tab:idms}.}
\label{eezh-degen}
\end{center}
\end{figure}

In Fig.~(\ref{eezh-degen}), total cross section and relative corrections in Scenario I are shown. 
Three typical collision energies of future electron-positron colliders, namely: $\sqrt{s}=250$ GeV, $\sqrt{s}=500$ GeV and $\sqrt{s}=1000$ GeV are chosen to present the results. Once the mass of charged Higgs boson is fixed, the triple Higgs boson couplings are simply determined by the parameter $\mu_2^2$, as given in Eq. (\ref{eq:lams}). Accordingly, Scenario I also means that $\lambda_3$ can be non-vanishing while $\lambda_4 =\lambda_5=0$.

\begin{figure}[H]\centering
\includegraphics[width=0.30\textwidth]{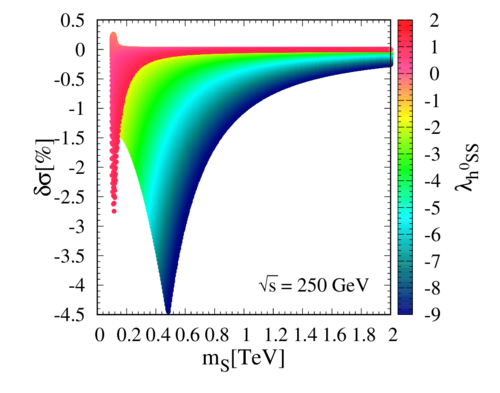}
\includegraphics[width=0.30\textwidth]{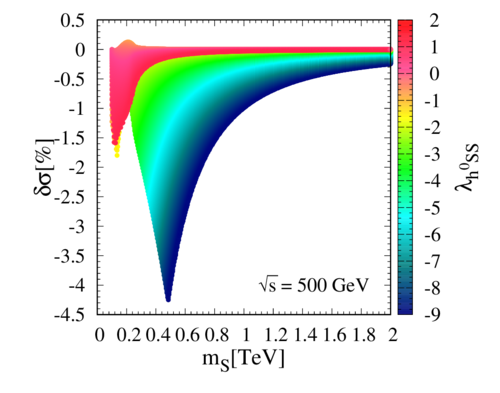}
\includegraphics[width=0.30\textwidth]{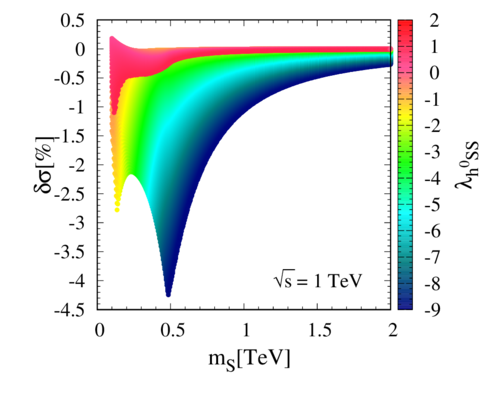}\\
\includegraphics[width=0.30\textwidth]{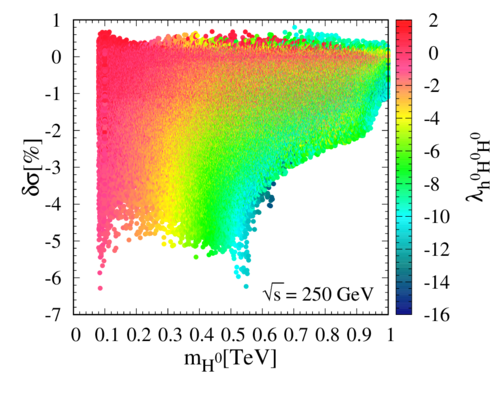}
\includegraphics[width=0.30\textwidth]{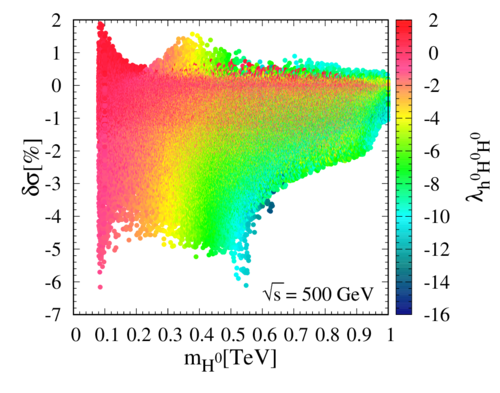}
\includegraphics[width=0.30\textwidth]{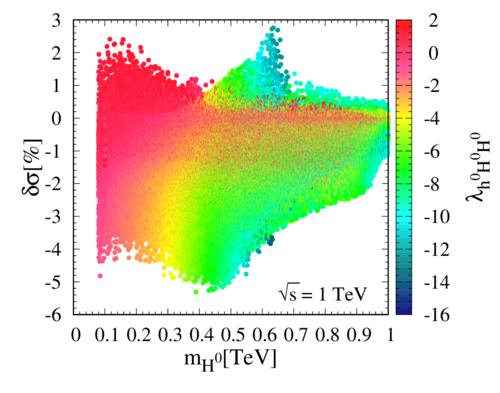}\\\includegraphics[width=0.30\textwidth]{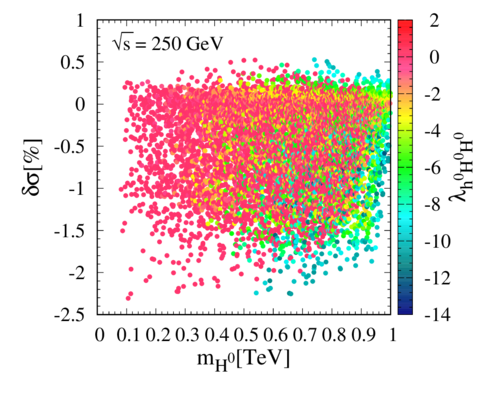}
\includegraphics[width=0.30\textwidth]{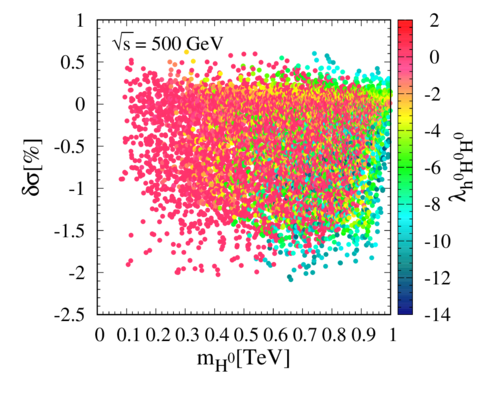}
\includegraphics[width=0.30\textwidth]{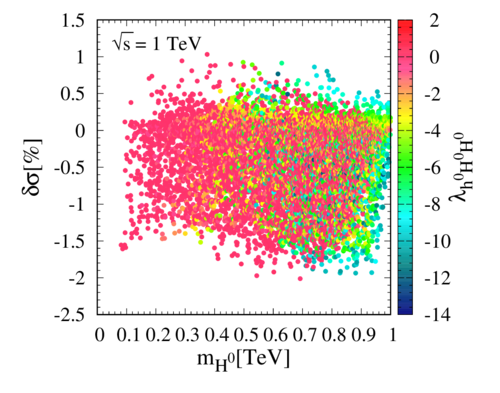}
\caption{IHDM corrections to  $e^+ e^- \to Zh^0$ as a function of the Higgs masses with triple Higgs couplings $\lambda_{h^0SS}$ normalized to the VEV on the vertical color bar  for collision energy 250 GeV, 500 GeV and 1000 GeV, respectively. From upper to lower panels, results for Scenario I, II, and III, 
 are shown, respectively. }
\label{fig:eezh-deg}
\end{figure}

Therefore, in the upper panels of Fig.~(\ref{eezh-degen}), the effect of  the triple Higgs coupling $\lambda_3$ on the cross section is examined by varying the parameter $\mu^2_2$ as shown in Table~\ref{tab:idms}.
Several typical values of $\mu_2^2$ are chosen in the allowed parameter space, and the corresponding results in the IHDM are marked with IDM1-5 in the figure. The values of $\mu_2^2$ are given in Table~\ref{tab:idms}.
 It is observed that weak corrections in the IHDM are typically negative and can reach $9\%-14\%$ at $\sqrt{s}=250$ GeV and become $18\%-23\%$ for $\sqrt{s}=1000$ GeV. 
When the mass of Higgs boson gradually increases to 500 GeV or so, the triple Higgs couplings proportional to $\lambda_3$ also increases, which leads to an increased new physics contribution to the total cross section.  This seems to be against the decoupling limit, but it can be seen from Fig.~(\ref{fig:eezh-deg}) that the decoupling limit is reached at an even higher scale, around 1$\sim$ 2 TeV.
The enhancement is mainly due to the triple scalar couplings $h^0SS$, $S=A^0, H^0, H^\pm$, which are proportional to  $\lambda_3$ in Scenario I. Such triple couplings contribute into the corrections both linearly through the virtual corrections and also quadratically through the wave function renormalization of $h^0$.
A careful reader can also find that the starting points of the Higgs boson mass are also different for different values of $\mu^2_2$, which can be attributed to the theoretical and experimental constraints.

\begin{table}[!htb]
\centering
\begin{tabular}{|c|c|c|c|c|c|}
\hline
& IDM1  & IDM2 & IDM3 & IDM4 & IDM5  \\
\hline
$\mu_2^2$(GeV$^2$)  & 40000 & 6000  & 0  & -10000 & -30000 \\
\hline
\end{tabular}
\caption{In Scenario I, typical values of $\mu_2^2$ labelled as IDM1-5 are shown.}
\label{tab:idms}
\end{table}

Another comment is about the real emission, which can be clearly seen from the 
lower panels in Fig.~(\ref{eezh-degen}). At the $O(\alpha(m_Z))$ order, it is found that the real 
emission contribution is independent of the new physics parameter and can be $0.5\%$, $1\%$ and $1\%$ for three 
collision energies, respectively.\\
In order to illustrate the effect of radiative corrections in the IHDM and to avoid counting the pure SM effects, 
we define the following  ratio given as:
\begin{eqnarray}
\delta =\frac{\sigma_{Zh^0}^{IHDM} -\sigma_{Zh^0}^{SM}}{\sigma_{Zh^0}^{SM}}\,,
\label{eq:sub}
\end{eqnarray}
where $\sigma_{Zh^0}^{IHDM}$ and $\sigma_{Zh^0}^{SM} $ denote the one-loop total 
cross section in the IHDM and the SM, respectively. This ratio is useful for this process. 
We emphasize that there are only the contributions of new physics in the IHDM survived in numerator
of the quantity $\delta$ while the full SM one loop effect has been subtracted. In terms of Feynman diagrams, 
the QED corrections as well as corrections to the initial state vertex $e^+e^-Z$  Fig.~(\ref{figure-zh})-$G_{17,18}$, 
correction to $e^+e^-h^0$ vertex Fig.~(\ref{figure-zh})-$G_{15,16}$ and box contributions will cancel out in
the numerator during the subtraction Eq.~(\ref{eq:sub}). Obviously, the QED and pure SM effects are still present
in the denominator of Eq.~(\ref{eq:sub}). 

In Fig.~(\ref{fig:eezh-deg}), we illustrate $\delta$ as a function of the Higgs masses as well as the triple Higgs couplings $\lambda_{h^0SS}$ with $S=H^0, A^0, H^\pm$ in the color bar for Scenario I, II, and III. Only the results as a function of the mass of $H^0$ are displayed in Scenario II and III, as those of other two mass parameters of $A^0$ and $H^\pm$ are similar and are not shown. 
Same three values of collision energy ($250$ GeV, $500$ GeV, and $1000$ GeV) are chosen for all Scenarios. 
It is noticed that the results for CLIC energy $\sqrt{s}=350$ GeV are rather similar to 250 GeV case and thus they are not shown here.

In Scenario I, from upper panels of Fig.~(\ref{fig:eezh-deg}), one can read that the ratio of new physics with respect to the SM can change from $0.25\%$ to $-4.5\%$ with $\sqrt{s}=250$ GeV, while they are similar for $\sqrt{s}=500$ GeV and $\sqrt{s}=1000$ GeV. 
Generally speaking, the contributions of new physics in the IHDM contributions reduce the production cross section, except from a tiny bump near the region with $m_S=100$ GeV. 

In Scenario II, the contributions can change from $-6\%$ to $2\%$ with $\sqrt{s}=500$ GeV and $-5.5\%$ to $3\%$ with $\sqrt{s}=1000$ GeV. Depending upon model parameters, the contributions can either increase or decrease the production cross section. 

The different behavior of Scenario I and II can be attributed to the different features of triple Higgs couplings. In Scenario I, according to Eq.~(\ref{eq:lams}), it is clear that both $\lambda_4$ and $\lambda_5$ should vanish.
Then the triple Higgs couplings $\lambda_{h^0SS}$ depends only on $\lambda_3$ which is severely constrained from 
di-photon signal strength limit (see discussion in section \ref{sec:hpp}). 
In such a case, as shown in the first row of Fig.~(\ref{fig:eezh-deg}), there is clear dip near the region $m_S=500$ GeV where the ratio can reach from $-1.5\%$ to $-4\%$. This occurs because the triple Higgs coupling $\lambda_{h^0SS}=-\lambda_3 v$ which is driven solely by $\lambda_3$ becomes large in magnitude (say $-8 \sim -9$). 
In fact, there are terms in  $\delta$ which are proportional to the triple Higgs couplings, linear and quadratic.
The linear terms can come from Feynman diagrams Fig.(\ref{figure-zh})-$G_5$ while the quadratic terms come from the wave function renormalization of $h^0$ which contributes to the counter-term of $ZZh^0$ vertex in Eq.~(\ref{eq:CT}). In contrast, in Scenario II, all $\lambda_{3,4,5}$ can contribute to the triple Higgs couplings $\lambda_{h^0SS}$, which can lead to more complicated interferences for each of the terms and can change the signs of new physics contributions.

\begin{figure}[H]\centering
	\includegraphics[width=0.3\textwidth]{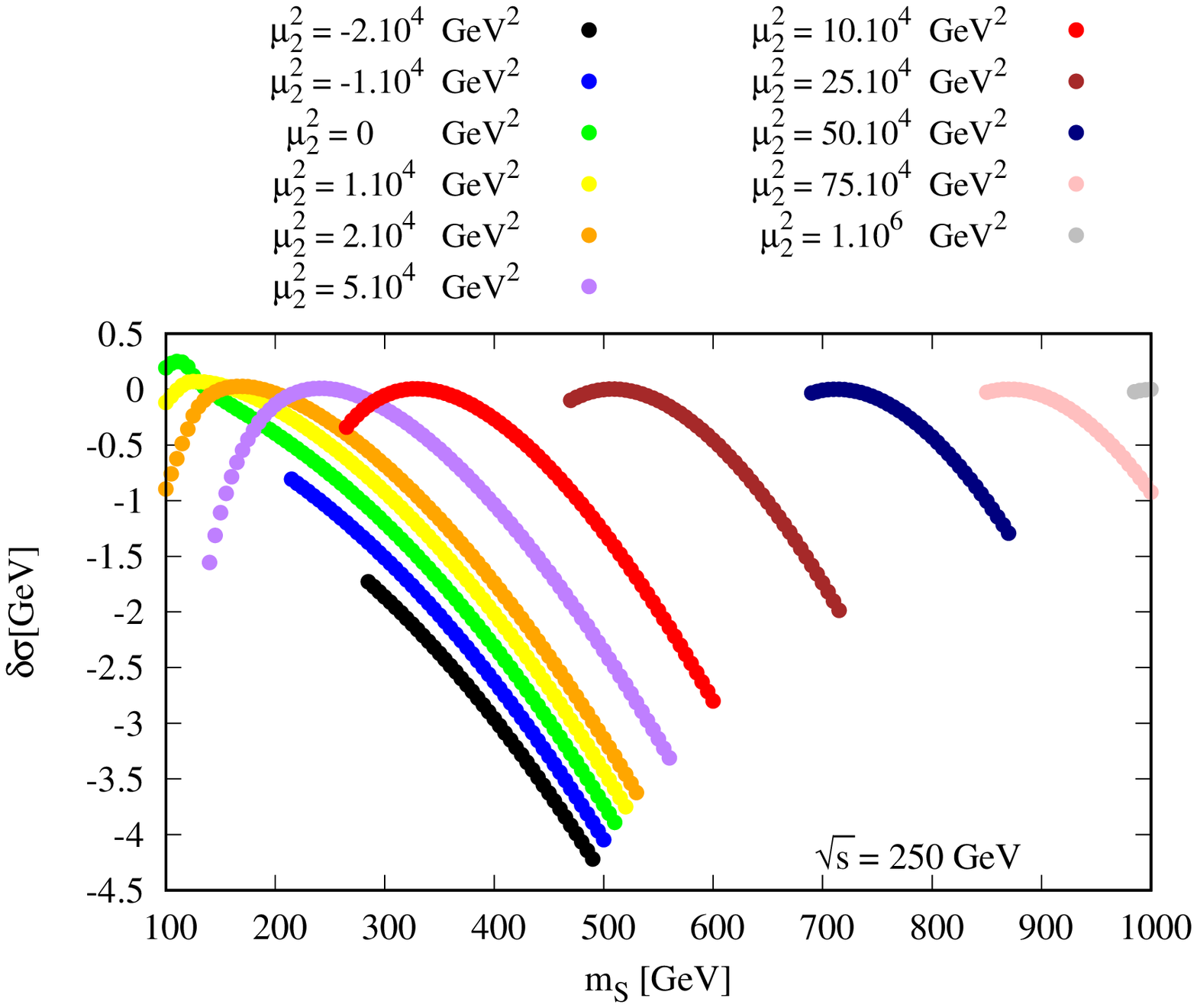}
	\includegraphics[width=0.3\textwidth]{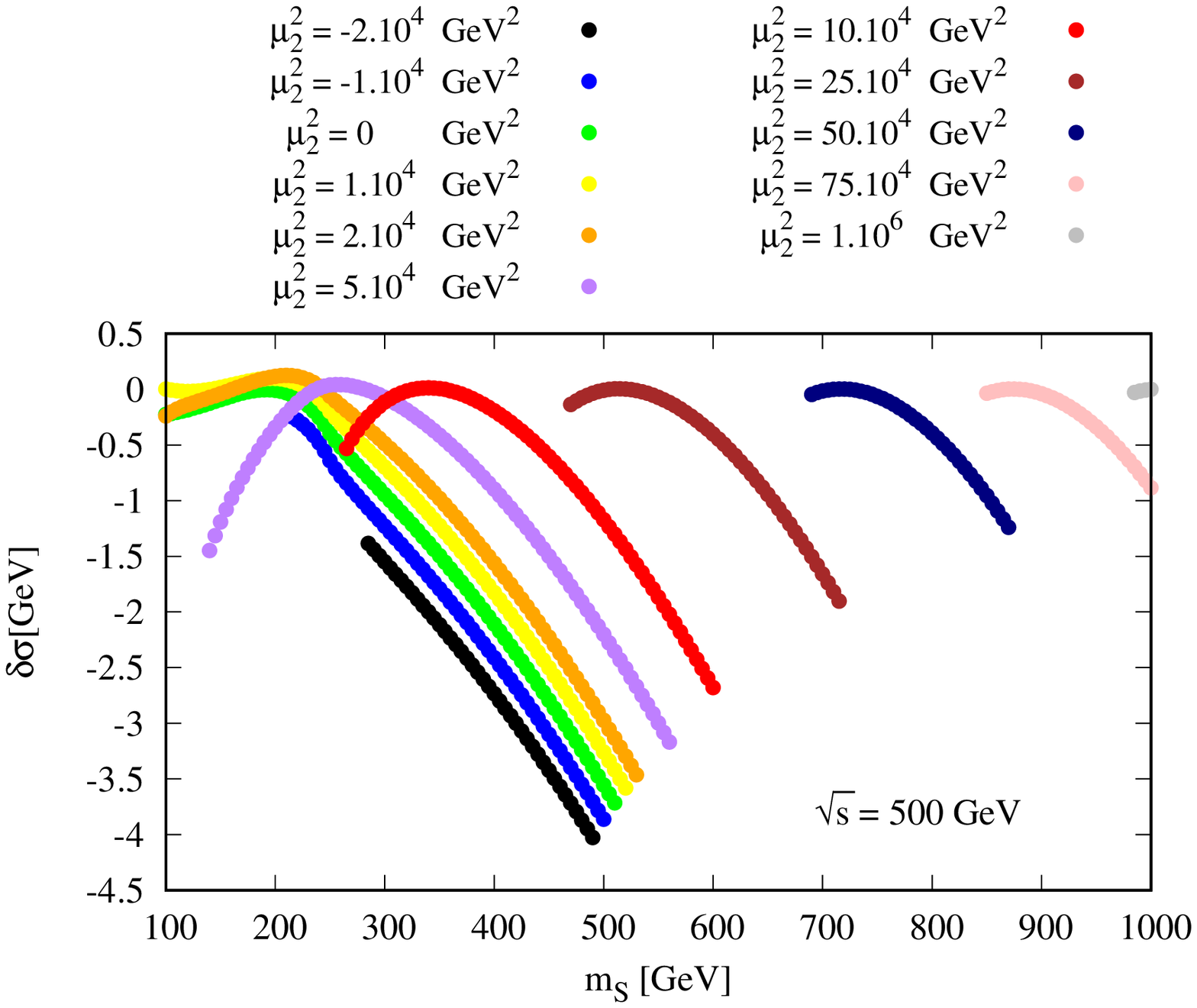}
	\includegraphics[width=0.3\textwidth]{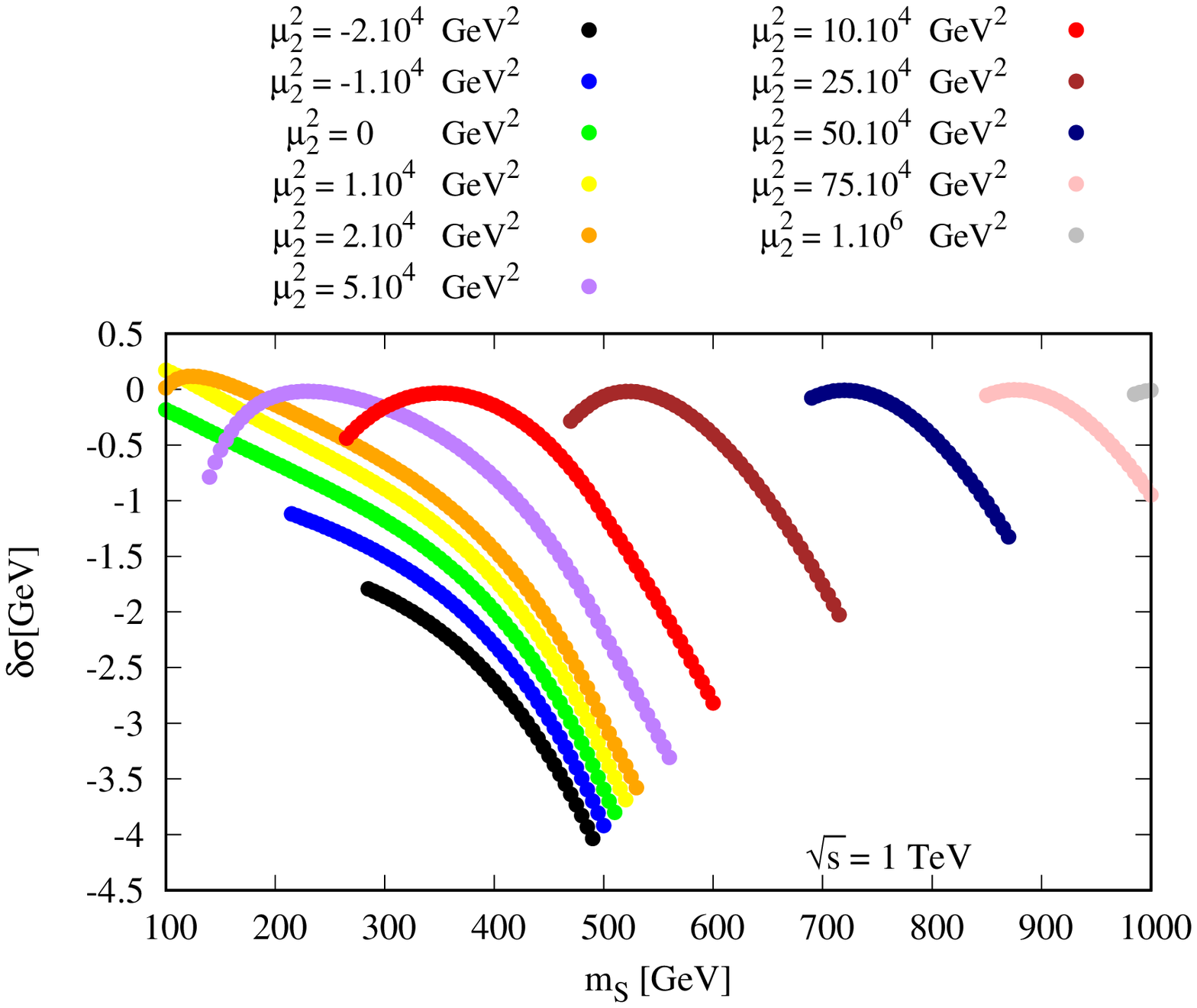}
	\caption{In the degenerate scenario after imposing all theoretical constraints for $\sqrt{s}=250$ GeV,  $500$ GeV and $1$ TeV, 
	the relative corrections of $e^+\ e^- \to Z h^0$ are shown 
		as a function of the Higgs masses $m_S$ for several fixed value of $\mu^2_2$ from $ - 2 \times 10^4$ to $1 \times 10^6$ GeV$^2$.}
	\label{fig:rep19}
\end{figure}

It is interesting to notice that as shown in the upper panel of  Fig.~(\ref{fig:eezh-deg}), there exist a large upside down peak near the region $m_S \sim 500 GeV$ and a tiny bump around $m_S = 150$ GeV. When we look at these structures closely, we find that they are not caused by the mass of any particles nor by threshold effects. Instead, they are caused by the theoretical constraints such as unitarity, vacuum stability, no charged minima, etc. As shown in Fig.~(\ref{fig:rep19}), for a given $\mu^2_2$ (say the case $\mu^2_2=0$ with $\sqrt{s}=250$ GeV), the allowed region of $m_S$ by all theses theoretical constraints produces a tiny bump near 150 GeV and ends at 500 GeV. Similarly with $\sqrt{s}=500$ GeV and $1$ TeV cases, such structures appear. In other words, these accidental structures are produced by the cuts on $m_S$. 

It is necessary to point out that Fig.~(\ref{fig:eezh-deg}) also illustrates the 
decoupling behavior of IHDM in the process $e^+e^-\to Zh^0$. 
To demonstrate this, we 
include points with the Higgs masses $m_S$ up to 1 or 2 TeV 
and also take  $\mu_2^2$ in a wide 
range in order to satisfy theoretical constraints. As expected, 
the radiative corrections become 
smaller when $m_S$ increases, such a decoupling behavior can be clearly seen in Fig.~(\ref{fig:eezh-deg}) in 
the region with a large $m_S$ (say $m_S=1\sim 2$ TeV). 

\begin{figure}[H]\centering
\includegraphics[width=0.32\textwidth]{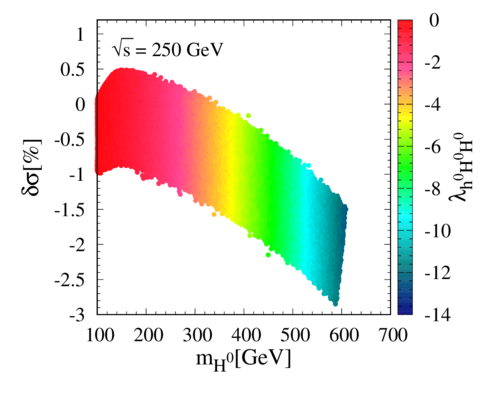}
\includegraphics[width=0.32\textwidth]{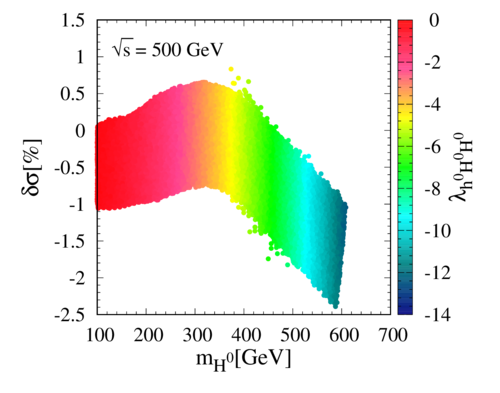}
\includegraphics[width=0.32\textwidth]{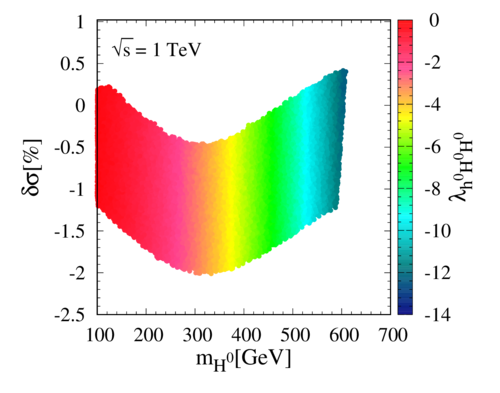}\\
\includegraphics[width=0.32\textwidth]{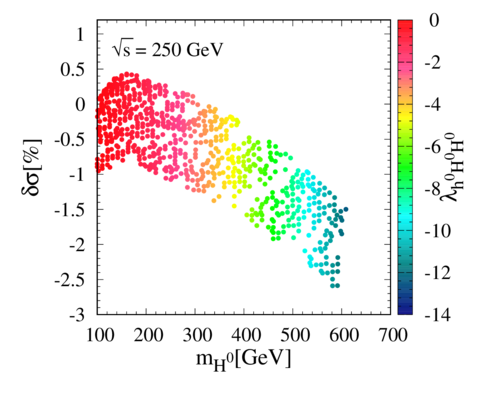}
\includegraphics[width=0.32\textwidth]{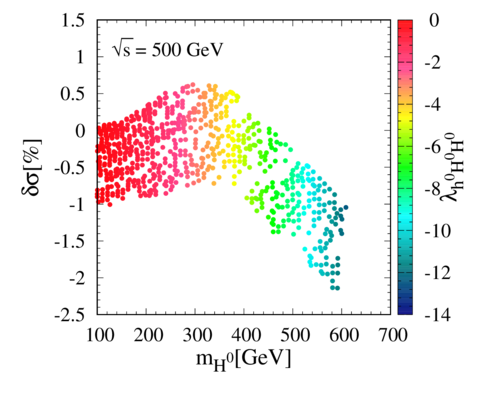}
\includegraphics[width=0.32\textwidth]{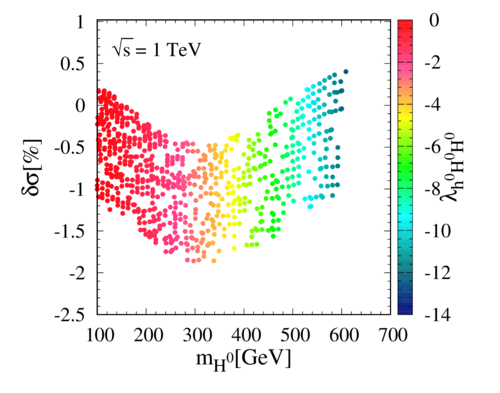}\\
\caption{IHDM corrections to  $e^+ e^- \to Zh^0$ as a function of the Higgs masses are shown for Scenario IV and Scenario V, where the condition of Br$(h^0\to A^0A^0) \le 11 \%$ is imposed. The upper panel is for Scenario IV, and the lower one is for Scenario V.}
\label{fig:eezh-inv}
\end{figure}
In Scenario III, it is noteworthy that the parameter points with large radiative corrections are ruled out by the dark matter constraints, especially by the direct search bounds from XENON1T (Only $1\%$ of points in Scenario II are still survived).  Meanwhile, the radiative corrections can only lead to a change in cross section from $-2.5\%$ to $0.5\%$ for $\sqrt{s}=250$ GeV case. Similar ranges hold for $\sqrt{s}=500$ GeV and $\sqrt{s}=1$ TeV cases. It should be emphasised that dark matter constraints indeed can significantly affect the allowed parameter space and the range of allowed radiative corrections of IHDM.

The results of Scenario IV and V are given in Figure (\ref{fig:eezh-inv}). The band shapes in Figure (\ref{fig:eezh-inv}) are related to the fact that the mass region of dark matter particle is taken from 20 GeV to 62.5 GeV 
 in Scenario IV and from 55 to 65 GeV in Scenario V (the mass range of 20-55 GeV is excluded due to too large relic density). It is observed that only $10\%$ of points from the parameter space of Scenario IV can survive the dark matter searches constraints, which are displayed in Scenario V. 
\begin{figure}[!htb]\centering
\includegraphics[width=0.31\textwidth]{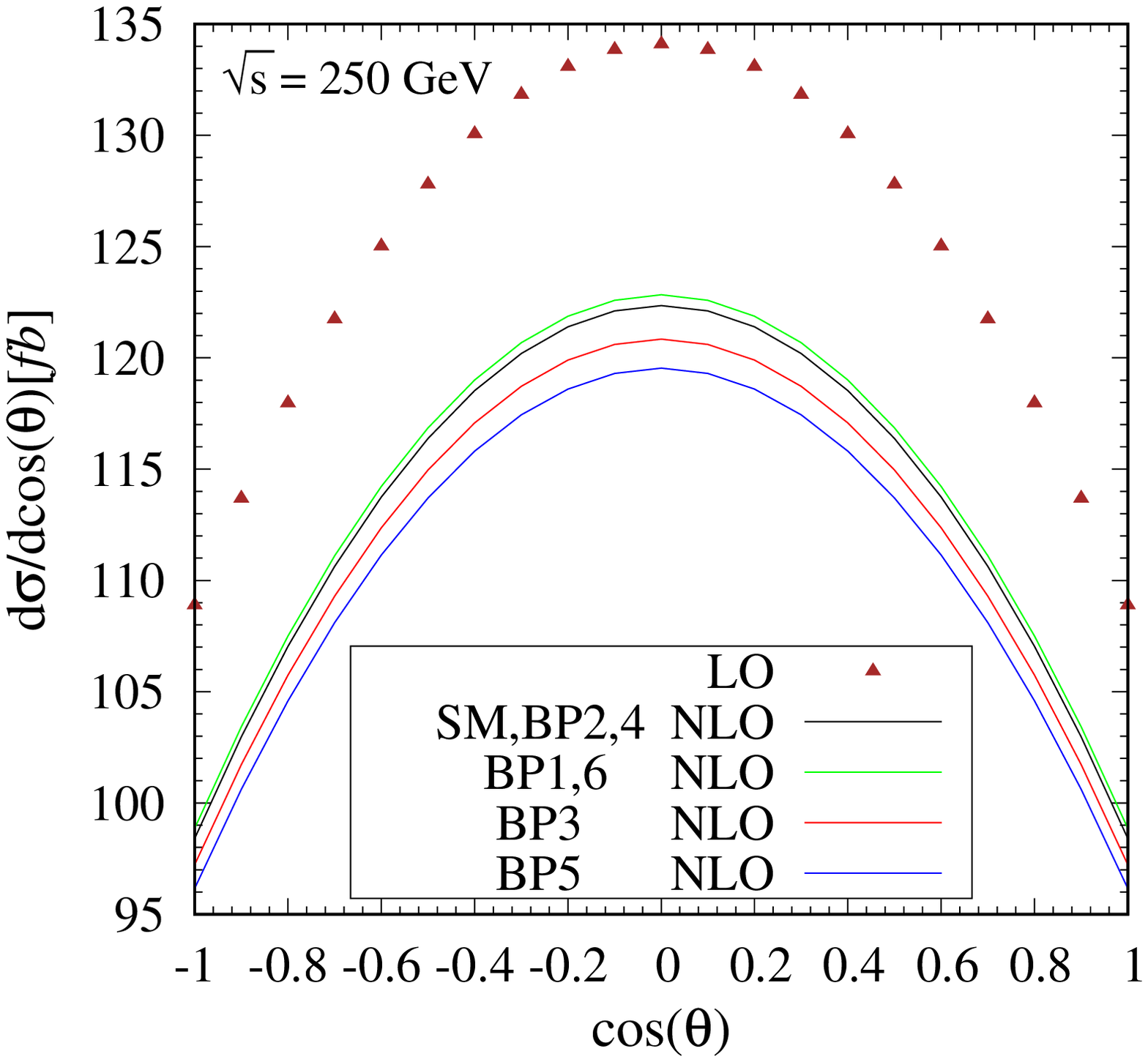}
\includegraphics[width=0.31\textwidth]{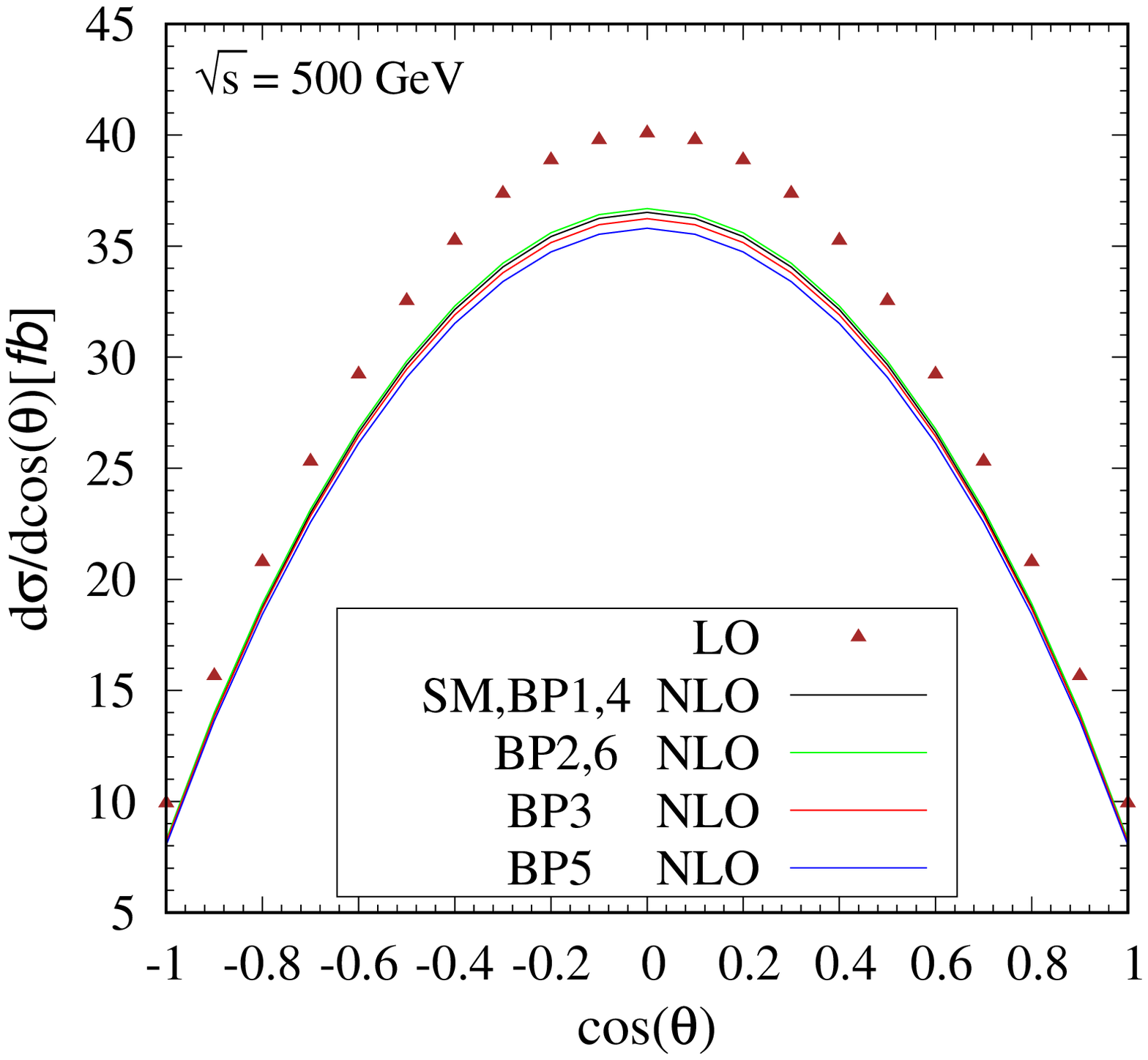}
\includegraphics[width=0.31\textwidth]{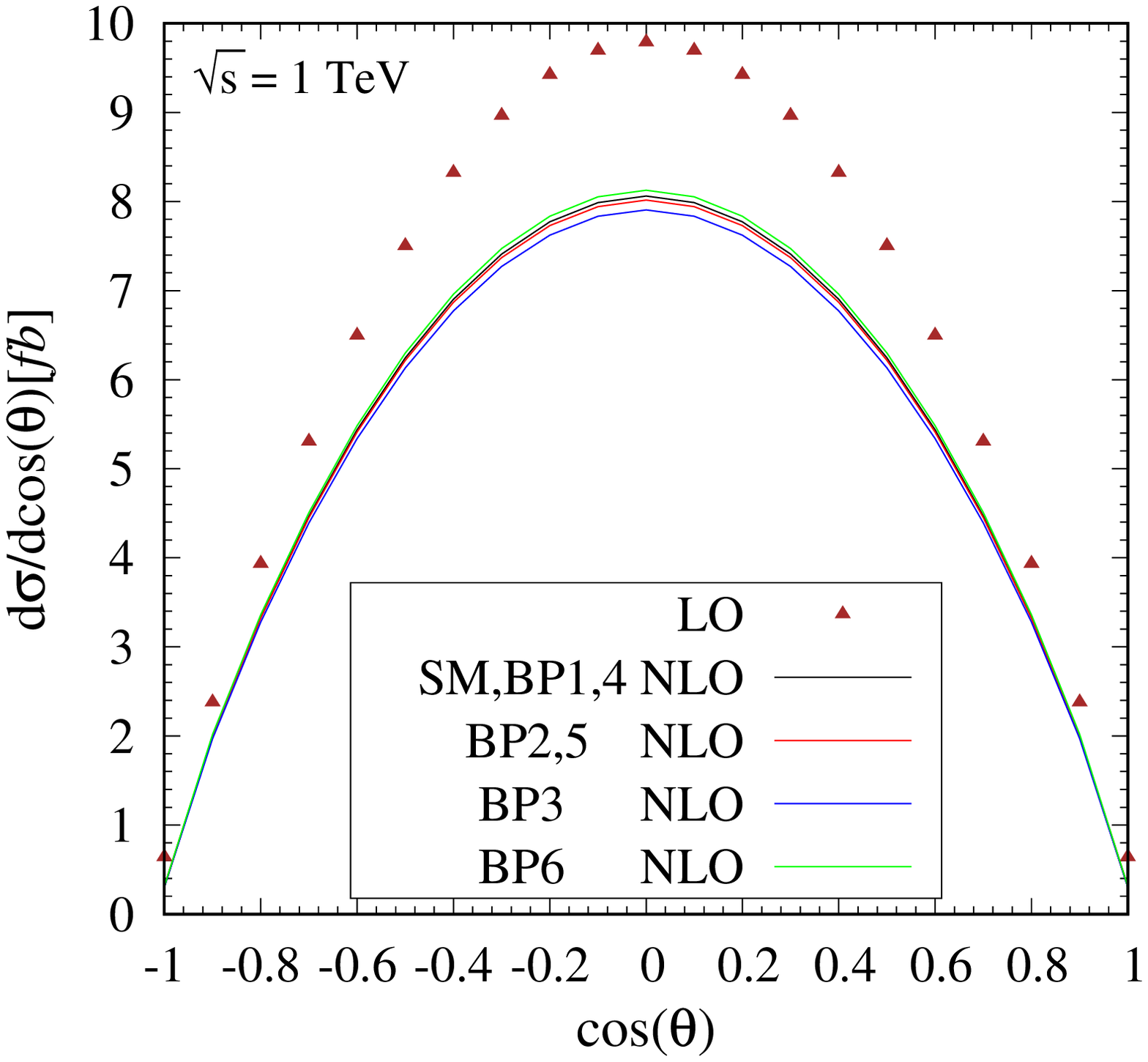}
\\
\caption{Angular distribution  for $e^+e^- \to Zh^0$ with three different collision energies: $\sqrt{s}=250$, 500 GeV and 1 TeV for six benchmark points defined in Table \ref{tab:Benchmark-points}.}
\label{fig:eehZ-dif}
\end{figure}

In Figure (\ref{fig:eehZ-dif}), we illustrate the angular distribution as a function of  $\cos \theta$, where $\theta$ is angle between outgoing Higgs boson and electron beam in the center of mass energy frame. 
At high energy, it is well known that in the SM, the angular distribution behaves like $\sin^2\theta=1-\cos^2 \theta$ \cite{Djouadi:2005gi} and keeps the same shape at the NLO.  This can be seen from Eq.~(\ref{eeZhcos}) where the dominant term is $\kappa_{Zh}^2 \sin^2\theta$.
 In the figure, we illustrate both the LO and NLO distributions in the SM, as well as the angular distributions in the IHDM for six benchmark points, BP1$-$BP6, which are given in Table \ref{tab:Benchmark-points}.  From the plots, for $\sqrt{s}=250$ GeV, one can see that the results of BP1 and BP6 (BP2 and BP4) overlaps 
 and for $\sqrt{s}=500$ GeV, the results of BP1 and BP4 (BP2 and BP6) overlaps. 
 While for $\sqrt{s}=1000$ GeV, the results of BP1 and BP4 (BP2 and BP5) overlaps.
 In all cases, the IHDM contributions reduce the SM differential cross section.
In the SM with 250 GeV, away from the forward and backward direction, the relative correction at NLO is  about $-10\%$ to LO, while in the forward and backward direction, because of the box contributions, the relative correction could be slightly larger depending on the CM energy.
As one can see, the angular distributions in the IHDM have the same shapes as those in the SM.
For the $\sqrt{s}=500$ GeV and 1 TeV cases, the curves of IHDM almost overlap with the one of the NLO SM, i.e. the difference is very subtle and will be challenging to measure. While for  the $\sqrt{s} = 250$ GeV case, sensible deviations from the NLO SM can be observed which are detectable by experiments hopefully.

Before ending this section, we would like to stress that the case of 350 GeV CM energy is quite similar to that of 250 GeV case and is omitted here.

\subsection{$e^+e^- \to H^0 A^0$}
In the general 2HDM or in the MSSM,  there exists a sum rule between the two vertices $Z h^0 A^0$ and $Z H^0 A^0$, which itself simply reflects of the mixing between two CP even Higgs bosons $h^0$ and $H^0$ and the mixing between two CP odd Higgs boson $G^0$ and $A^0$ as well. The sum rule implies that the process $e^+e^- \to H^0 A^0$ and the process $e^+e^- \to h^0 A^0$ could  always happen together for some specific choise of the mixing angles. Thus it is natural to expect that both processes could be detected at the future electron-positron colliders.

In contrast, in the IDHM, due to the fact that there is neither mixing between two CP even Higgs bosons $h^0$ and $H^0$, nor mixing between two CP odd Higgs bosons $G^0$ and $A^0$. A natural consequence from this fact is that the process $e^+e^- \to h^0 A^0$ is forbidden. Therefore, to detect the signature of $e^+e^- \to H^0 A^0$ and to prove that there is no $e^+e^- \to h^0 A^0$ occurred at the same time can help to distinguish the IDHM from other general 2HDM like the MSSM.

It was pointed out in the Refs.~\cite{LopezVal:2009qy,LopezVal:2010vk} that triple Higgs couplings
can greatly enhance the tree level cross section of $e^+e^- \to h^0 A^0 /H^0 A^0$ in the general 2HDM. 
Such processes are supposed to help to probe the structure of Higgs potential of the 2HDM. Below, we examine 
the radiative correction to the cross section of $e^+ e^- \to  H^0 A^0$. In this process, there is no SM results, 
only the ratio $\Delta$ defined in Eq. (\ref{split}) is used.

\begin{figure}[!htb]\centering
\includegraphics[width=0.32\textwidth]{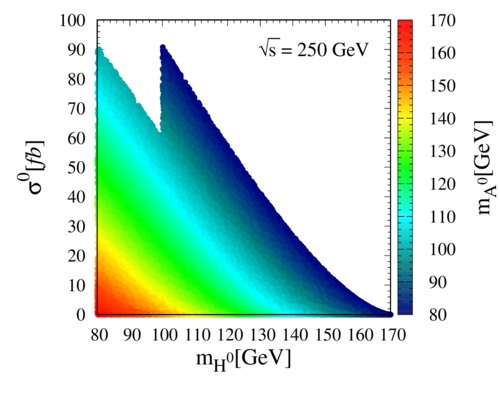}
\includegraphics[width=0.32\textwidth]{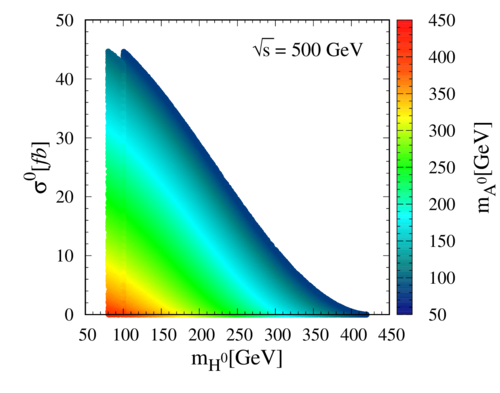}
\includegraphics[width=0.32\textwidth]{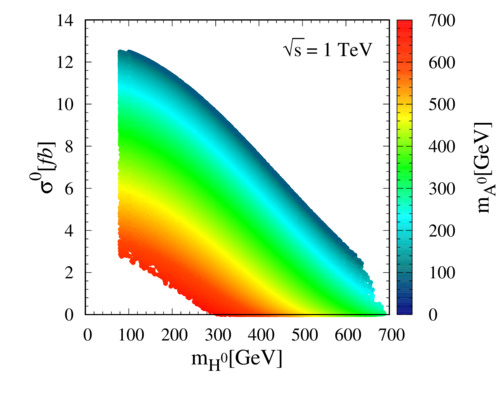}
\caption{In Scenario II, tree level cross sections for $e^+e^-\to H^0 A^0$ are shown as a function of $m_{H^0}$ and $m_{A^0}$ for various CM energy: $\sqrt{s}=250, 500, 1000$ GeV.}
\label{fig:eeHA-0}
\end{figure}
We first give the numerical size for the tree level cross section. 
In Scenario II, Fig.~(\ref{fig:eeHA-0}) is to show the cross section of 
$e^+ e^- \to H^0A^0 $ in a scatter plot as a function of $m_{H^0}$ and $m_{A^0}$. As shown in Eq. 
(\ref{eeHA}), the cross section depends only on the collision energy, the mass of $H^0$ and the mass of $A^0$.


\begin{figure}[!htb]
\begin{center}
\includegraphics[width=0.31\textwidth]{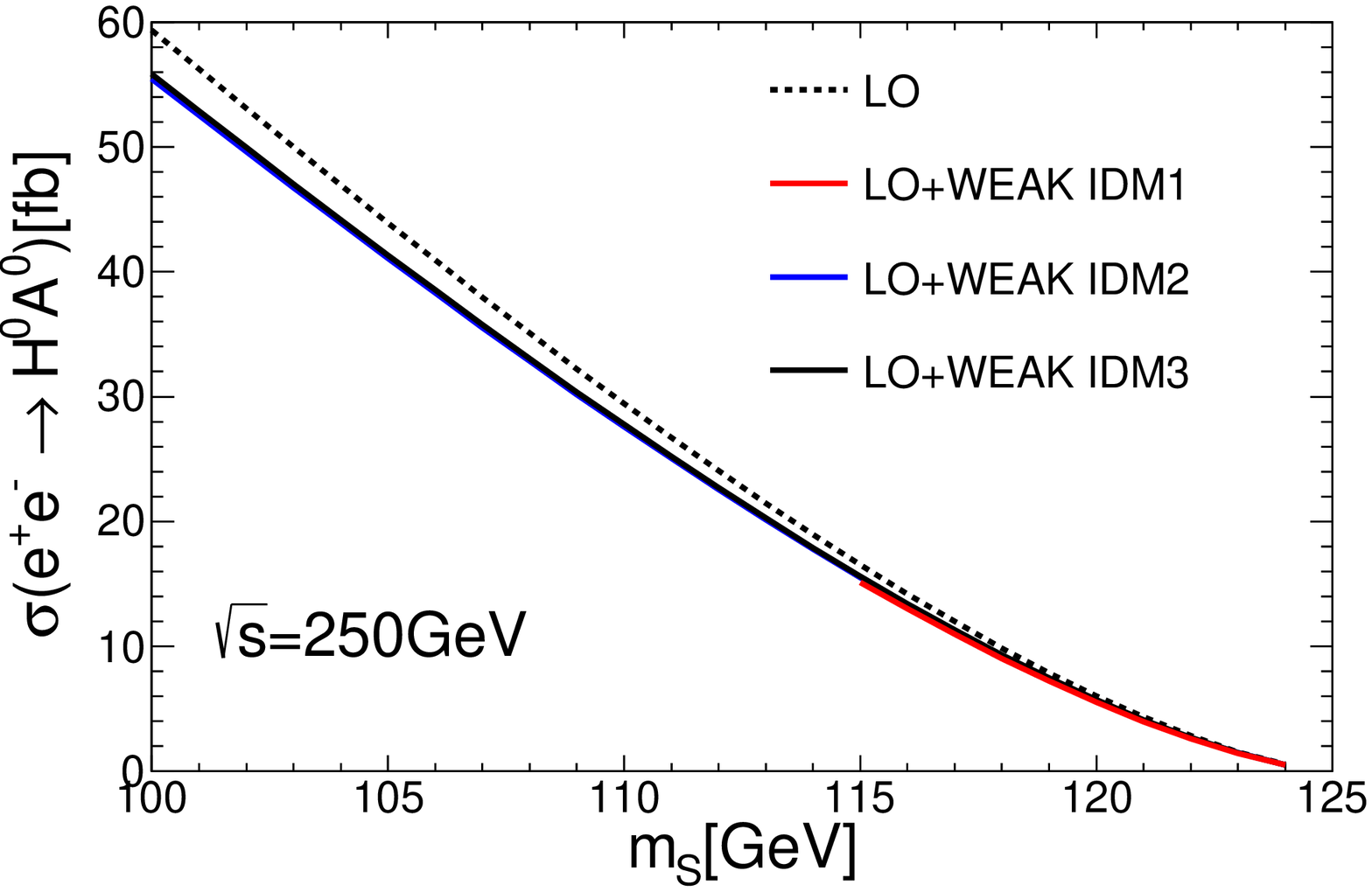}
\includegraphics[width=0.31\textwidth]{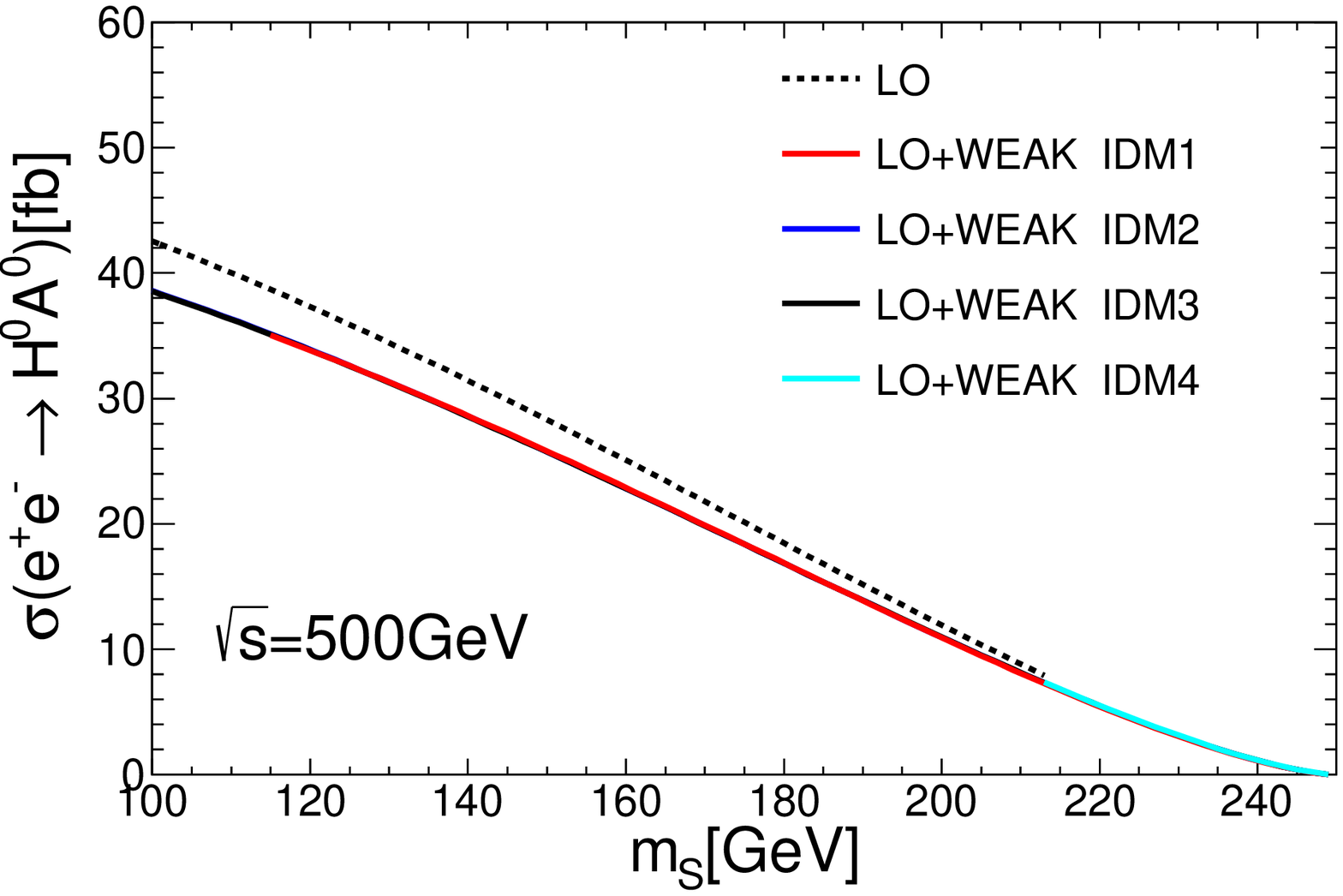}
\includegraphics[width=0.31\textwidth]{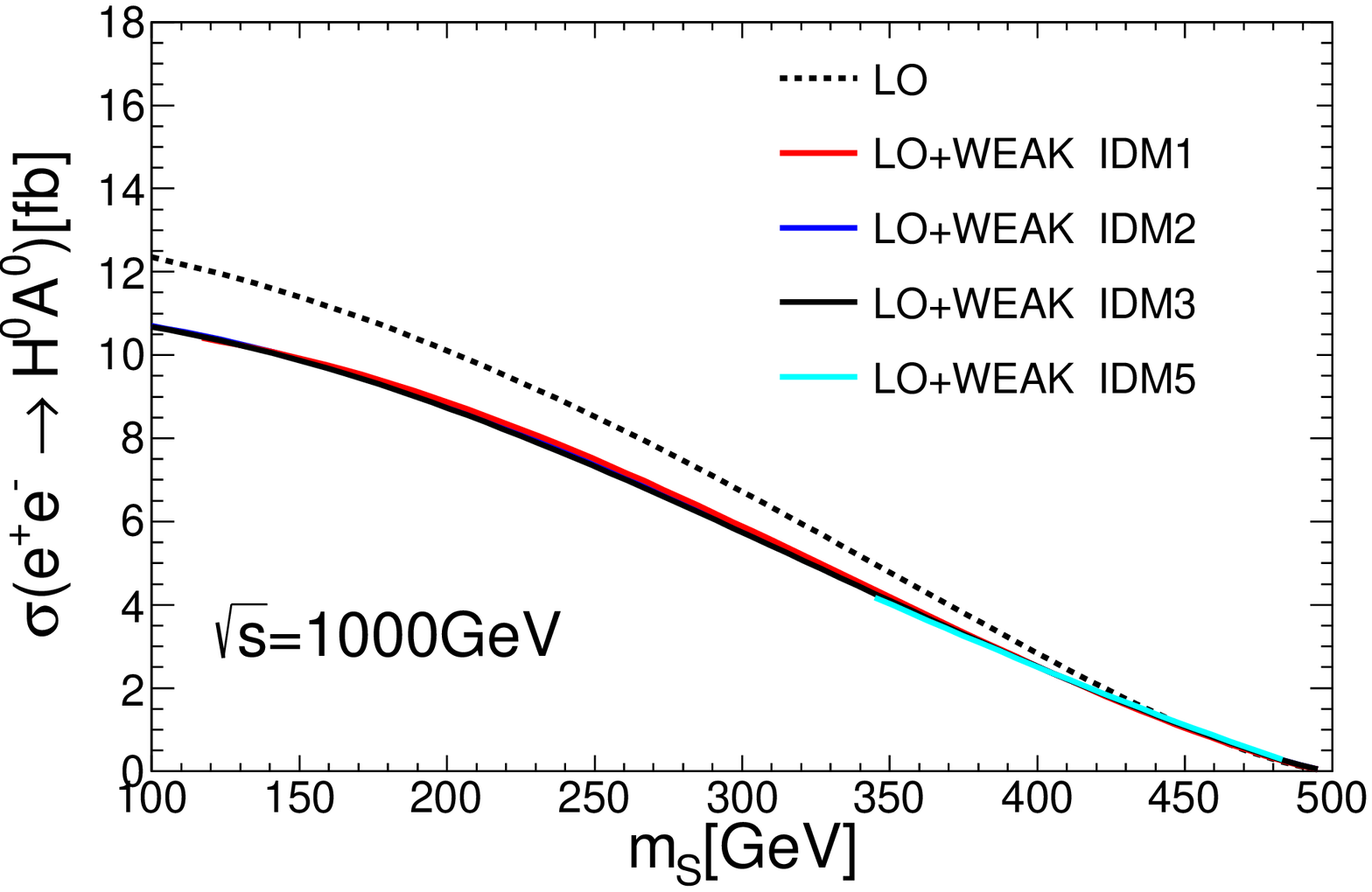}\\
\includegraphics[width=0.32\textwidth]{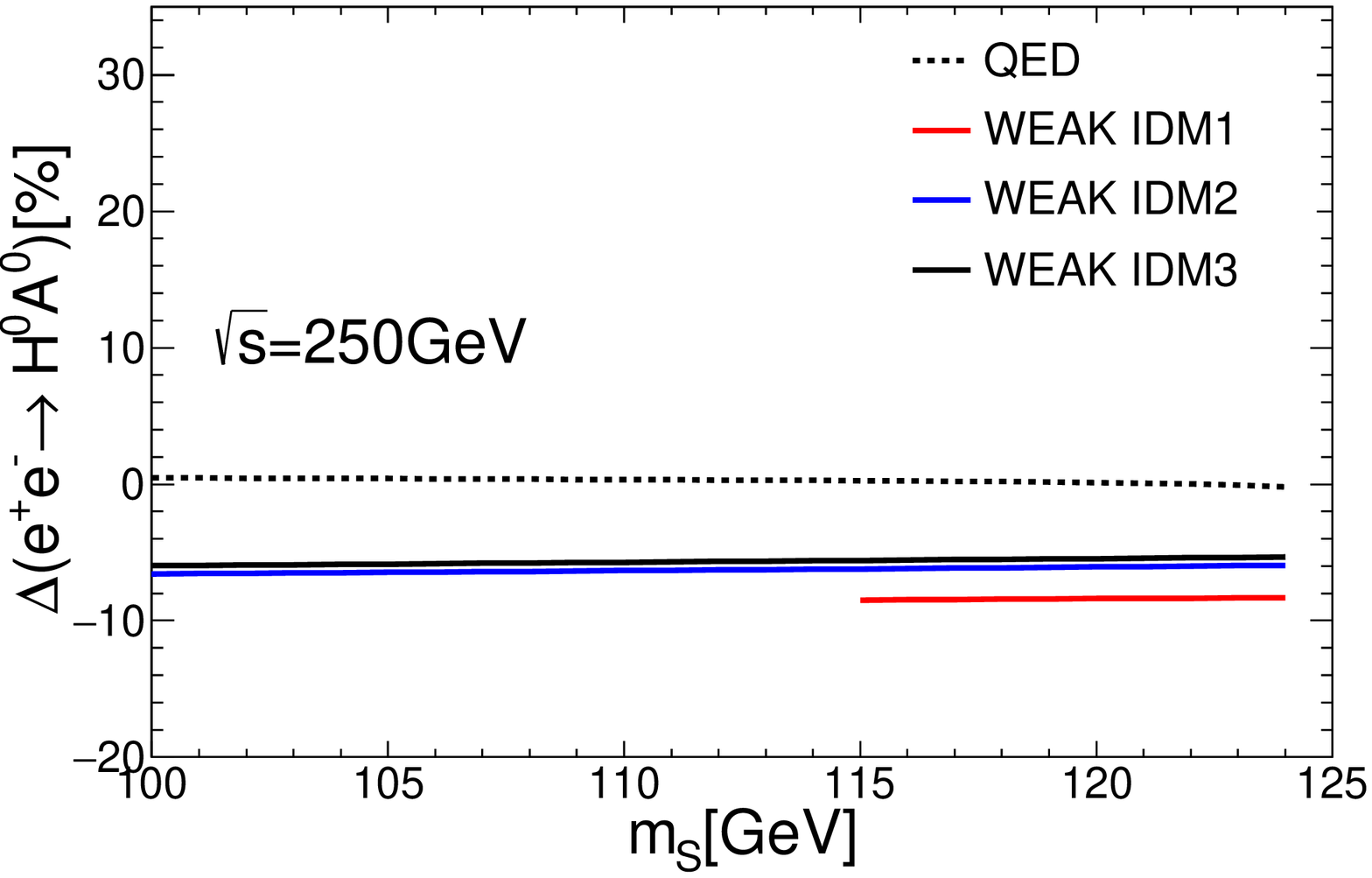}
\includegraphics[width=0.32\textwidth]{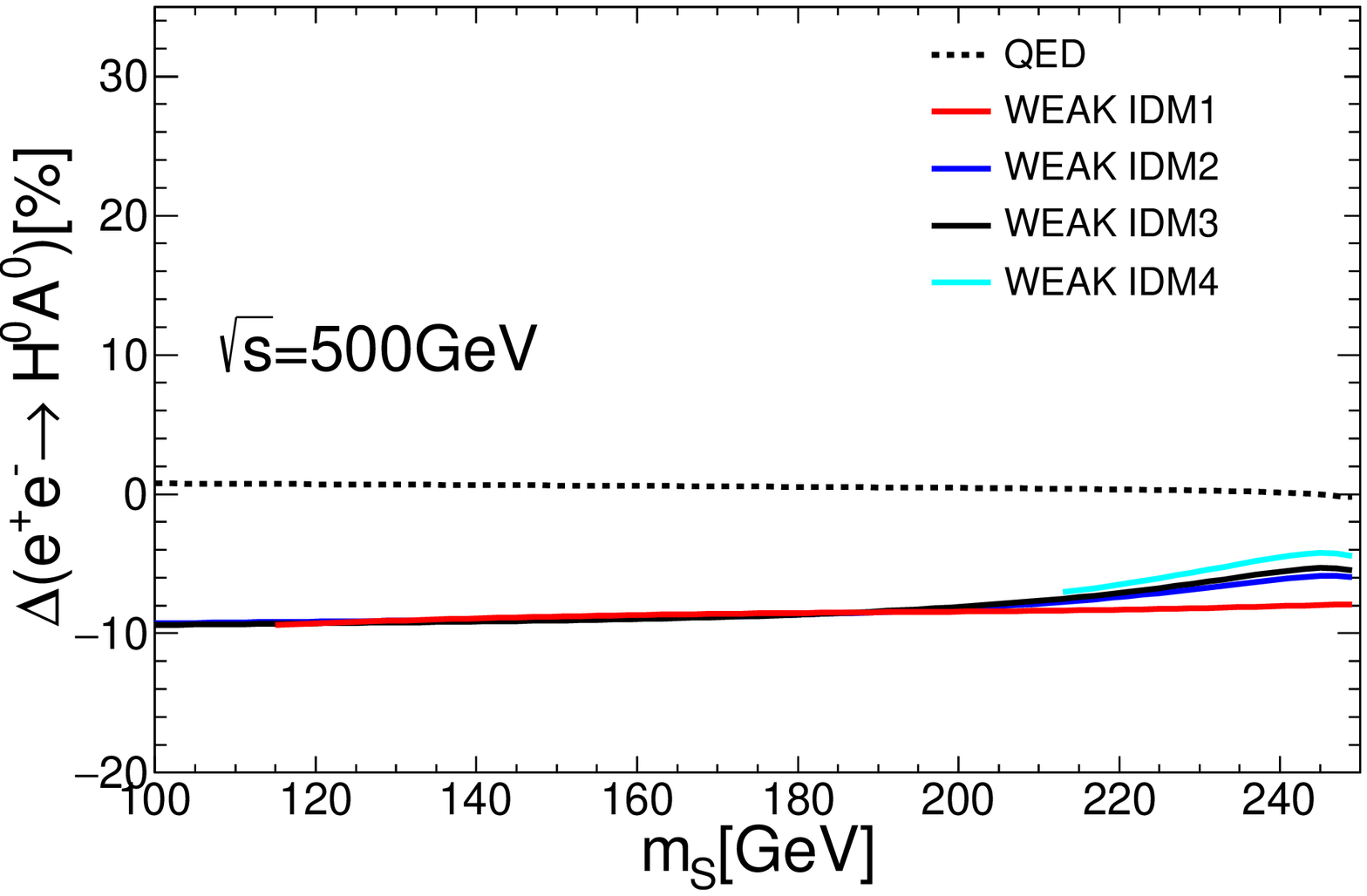}
\includegraphics[width=0.32\textwidth]{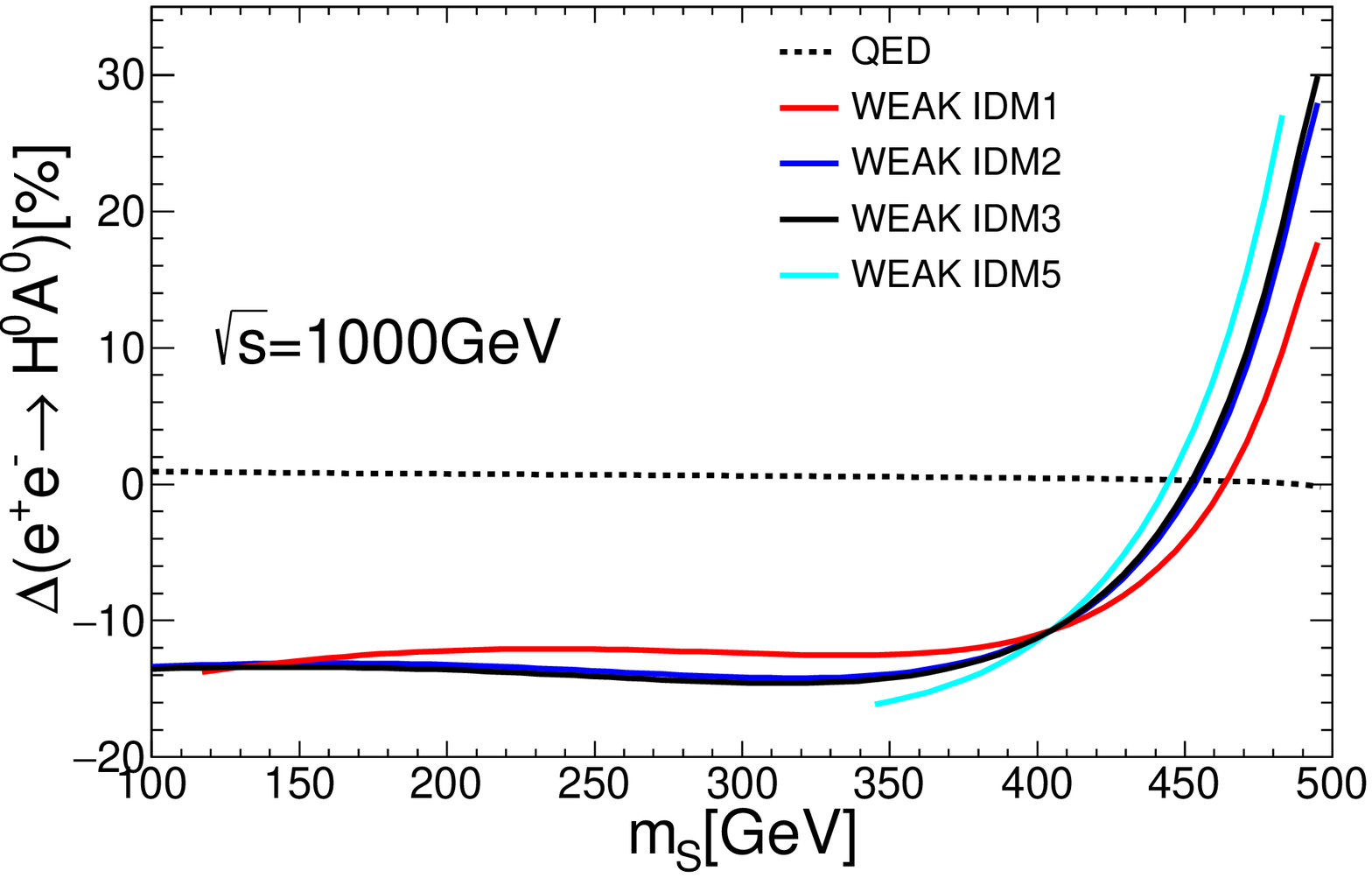}
\caption{In Scenario I, total cross sections and relative corrections as a function of the Higgs mass for $e^+\ e^- \to H^0A^0$ with three CM energies $\sqrt{s}=250$ GeV,  $500$ GeV and $1000$ GeV, where the corresponding values of $\mu^2_2$ are given in Table~\ref{tab:idms}.}
\label{fig:HAdeg}
\end{center}
\end{figure}
For Scenario I, we show the total cross section of $e^+ e^- \to H^0 A^0 $ in the IHDM with three typical collision energies in Fig.~(\ref{fig:HAdeg}).
From the upper panels, it is observed that when the collision energy is fixed, the total cross sections decreases with the increase of the Higgs masses.  
The same values of $\mu_2^2$ given in Table \ref{tab:idms} are chosen to show the effect of triple Higgs couplings.
It can also be observed in the figure that the allowed ranges of $m_{S}$ are different for different values of $\mu^2_2$ due to theoretical and experimental constraints. 

From the left plots of lower panel, it is found that for CM energy $\sqrt{s}=250$ GeV, the weak corrections can be $-5.8\%$, $-6\%$ and $-8\%$ 
when $\mu^2_2$ are chosen $0$, $4\times 10^4$ GeV$^2$ and  $6\times 10^3 $ GeV$^2$, respectively. For  
$\sqrt{s}=500$ and $1000$ GeV cases, radiative corrections become larger and change dramatically near the
threshold regions, where $m_S \sim \sqrt{s}/2$. For instance, in the case when $\sqrt{s}=1000$ GeV and $\mu^2_2=0$,
the radiative corrections change from $-12\%$ to $30\%$. The change occurs in the mass range from 400 GeV to 500 GeV. 
Similar thing happens to the cases of other two $\mu^2_2$. 
\begin{figure}[!htb]
\centering
\includegraphics[width=0.31\textwidth]{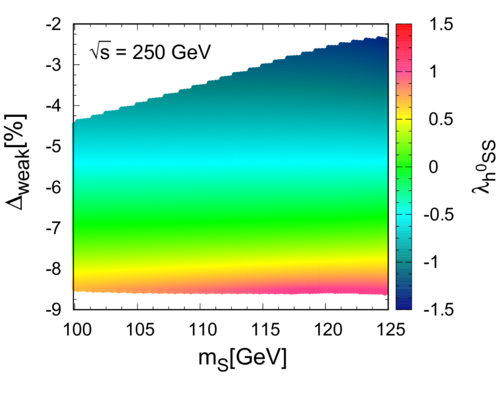}
\includegraphics[width=0.31\textwidth]{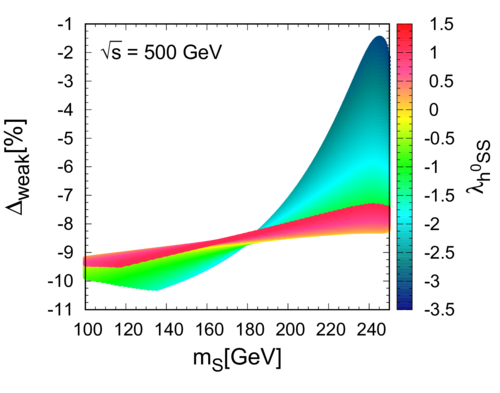}
\includegraphics[width=0.31\textwidth]{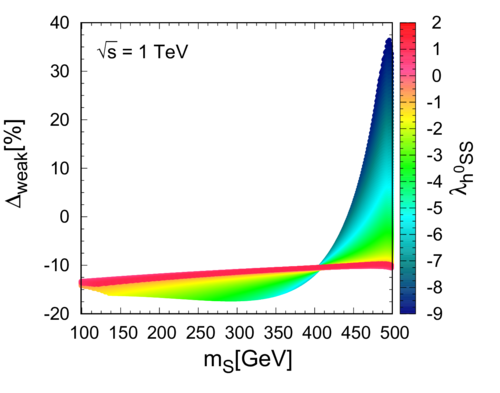}\\
\includegraphics[width=0.31\textwidth]{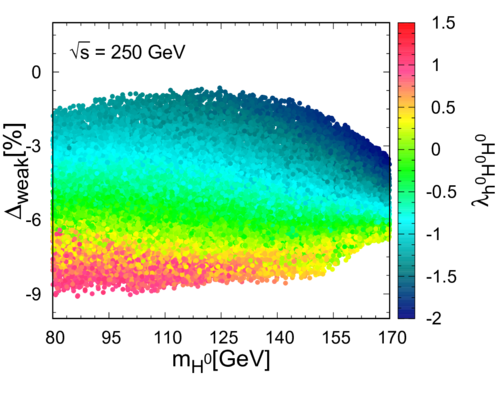}
\includegraphics[width=0.31\textwidth]{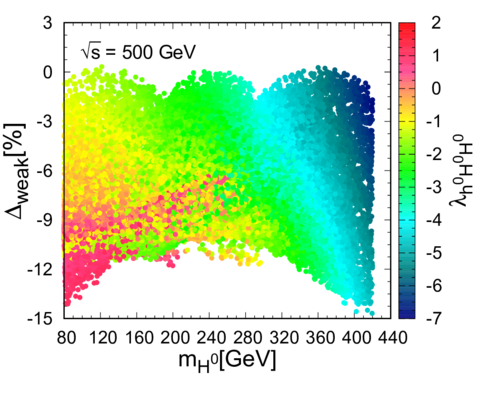}
\includegraphics[width=0.31\textwidth]{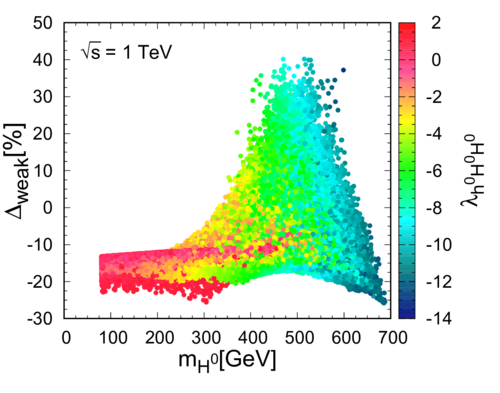}\\
\includegraphics[width=0.31\textwidth]{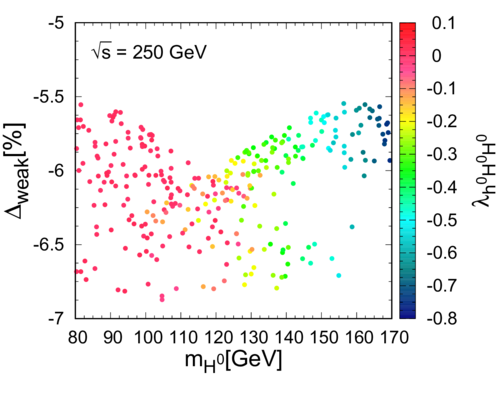}
\includegraphics[width=0.31\textwidth]{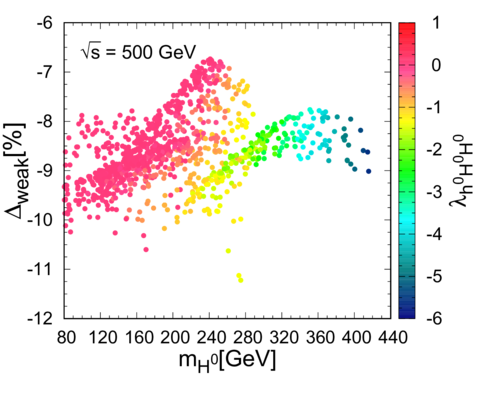}
\includegraphics[width=0.31\textwidth]{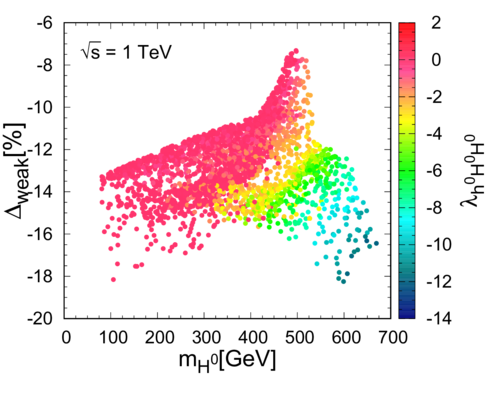}
\caption{Relative weak corrections to $e^+e^- \to H^0A^0$ are shown for $\sqrt{s}=250, 500, 1000$ GeV. From upper to lower panels, Scenario I, II and III are displayed, respectively.}
\label{fig:eeHA-non-deg}
\end{figure}
A careful reader might notice that the behaviour of endpoints looks different for the three CM  energies. As a matter of fact, the different behaviour near endpoints in Fig. (\ref{fig:HAdeg}) are simply determined by the collision energy and the mass range of the new particles. Obviously, with a higher CM energy, the $e^+ e^-$ machine can cover a larger region of parameter space in IHDM. 

Therefore, the CM energy $\sqrt{s}=1000$ GeV can cover the largest region of parameter space than those of $\sqrt{s}=500$ and $\sqrt{s}=250$ GeV cases. Although the CM energy changes the loop integral functions and causes its values to change, it is observed that when $m_S$ is near $120$ GeV, $\Delta(e^+ e^- \to H^0 A^0)$ is similar for these three cases of CM energies. 

Similarly when $m_S=240$ GeV, which is not reachable for $\sqrt{s}=250$ GeV case, the behaviour of $\Delta(e^+ e^- \to H^0 A^0)$ near the endpoint is similar for both $\sqrt{s}=500$ GeV and $\sqrt{s}=1000$ GeV. 

In the case $\sqrt{s}=1000$ GeV near the region $m_S=400$ GeV, it is noticed that there exists a fixed point where new particle's contribution is independent of other parameters of the IHDM. When $m_S$ becomes larger than 400 GeV, the triple Higgs couplings become larger and sizeable, and the contribution of the IHDM can even produce a positive value near the threshold. Such a behaviour can also be observed in Figure (\ref{fig:eeHA-non-deg}) for Scenario I with $\sqrt{s}=500$ GeV at the region $m_S=180$ GeV.

It is also worthy to mention that as shown in Fig.~(\ref{fig:HAdeg}), the ratio of QED corrections is quite small and almost independent of the Higgs mass while the ratio of weak corrections depend significantly on the scalar masses and could be quite large.

For Scenario I, II and III, we show the ratio of the weak corrections in the whole allowed parameter space with corresponding triple Higgs couplings in the color bar in Fig.~(\ref{fig:eeHA-non-deg}). 

For Scenario I, in the case with $\sqrt{s}=500$ GeV, 
the corrections start from $-9.5\%$ or so when $m_S=100$ GeV. When $m_S$ increases from $100$ GeV to $240$ GeV, the ratio keeps increasing and can reach  $-1.5\%$. In the case $\sqrt{s}=1000$ GeV, 
the corrections start from $-12\%$ or so when $m_S=100$ GeV. When $m_S$ increases from $100$ GeV to $490$ GeV, the ratio keeps increasing and can reach  $35\%$. 

For Scenario III, less than $1\%$ of points from Scenario II can survive when the constraints of dark matter are implemented, and the corrections can only be negative. It should be also emphasised that the range of the radiative corrections for those allowed points is shrunk by the dark matter constraints significantly. For example, for $\sqrt{s}=1$ TeV, before the dark matter constraints, the allowed range of radiative corrections can spread from $-25\%$ to $ 40\%$. But after the dark matter constraints, the allowed range can only change from $-18\%$ to $-8\%$. All points with positive radiative corrections have been killed by the dark matter constraints.


\begin{figure}[!htb]\centering
\includegraphics[width=0.32\textwidth]{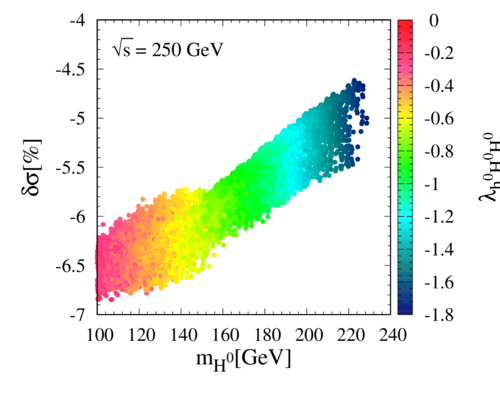}
\includegraphics[width=0.32\textwidth]{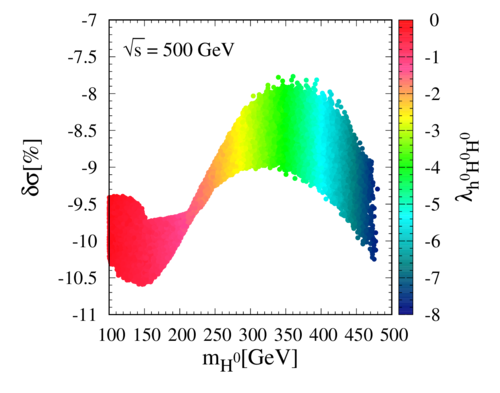}
\includegraphics[width=0.32\textwidth]{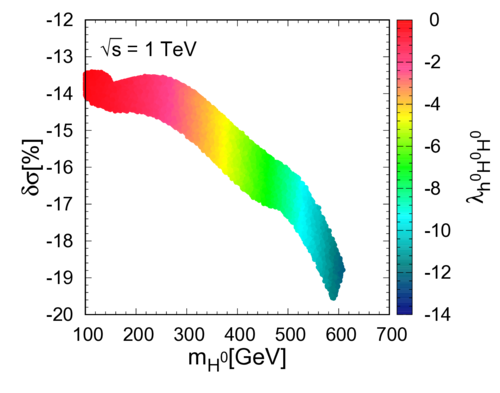}\\
\includegraphics[width=0.32\textwidth]{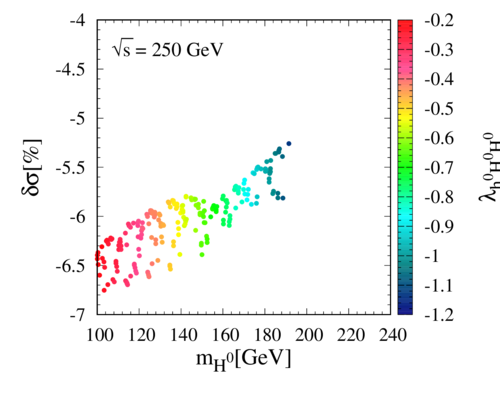}
\includegraphics[width=0.32\textwidth]{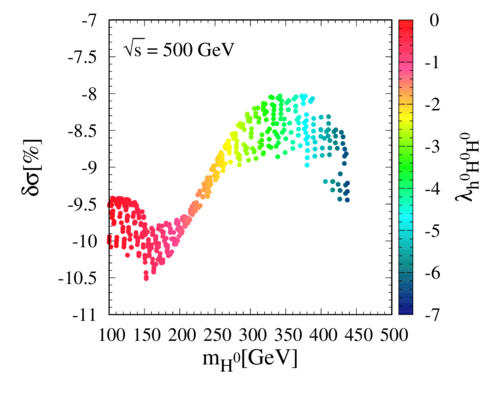}
\includegraphics[width=0.32\textwidth]{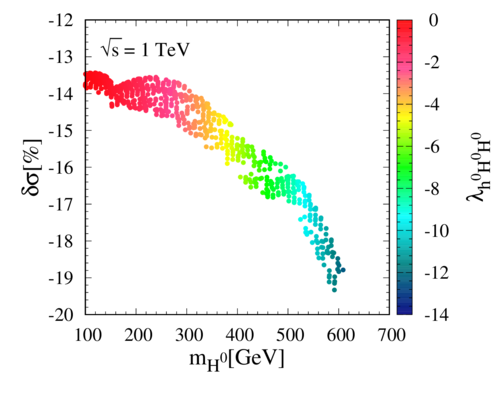}
\caption{ In Scenario IV and V, the IHDM corrections to  $e^+ e^- \to H^0 A^0$ as a function of Higgs masses are shown. }
\label{fig:eeHA-inv}
\end{figure}

There are a couple of comments on  Figure (\ref{fig:eeHA-non-deg}):
\begin{itemize}
\item As shown in Scenario I, the ratio of weak corrections is within the range from $-8.5\%$ to $-2.5\%$ in the $\sqrt{s}=250$ GeV case, $-10.5\%$ to $-2\%$ in the $\sqrt{s}=500$ GeV case, and $-15\%$ to $35\%$ in the $\sqrt{s}=1000$ GeV case. Whether such corrections can be detected at future electron-position colliders is determined by the production cross section. Meanwhile, the increase in the magnitude of the ratio when collision energy increases from 250 GeV to 1000 GeV does not mean the breakdown of the perturbation expansion. As a matter of fact, it more or less demonstrates the decrease of the LO cross section. Moreover, the larger collision energy also means a larger theoretical parameter space can be probed.
\item As shown in Scenario II, in the $\sqrt{s}=250$ GeV case, the ratio is around $-9\%$ when triple Higgs coupling is around $1 \sim 1.5$; while it is around $-1\%$ when the coupling is around $-2\sim-1.5$. For $\sqrt{s}=1000$ GeV case, when $m_{H^0}$ and $m_{A^0}$ are larger than 400 GeV and when the triple Higgs couplings normalized to the vev  become larger than $-14$, the ratio becomes larger than $+40\%$.
\item As shown in Scenario III, only $1\%$ points from the parameter space can survive, which have been displayed. The radiative corrections can vary from $-7\%$ to $-5.5\%$ for $\sqrt{s} = 250$ GeV, from $-11\%$ to $-7\%$ for $\sqrt{s} = 500$ GeV, and from $-18\%$ to $-8\%$ for $\sqrt{s} = 1000$ GeV. Points with positive corrections, like those shown in Scenario II for $\sqrt{s}=1000$ GeV  case have been removed by the dark matter constraints.
\end{itemize}
For Scenario IV and V,  we display the scatter plots of the allowed points from the parameter space in Figure (\ref{fig:eeHA-inv}). Similar to Figure (\ref{fig:eezh-inv}), the band shapes are related to the mass range of the dark matter particles. In Scenario IV and V, the radiative corrections are always negative. It is noteworthy that there is a considerable amount of points that have been ruled out by the mono-jet constraints, which is different from Scenario III where mono-jet constraints has no perceivable affects to the parameter space.

\section{Benchmark points}
\label{bps5}
\begin{table}[!htb]
\centering
\begin{tabular}{|c|c|c|c|c|c|c|}
\hline
Benchmark Points& BP1  & BP2 & BP3 & BP4 & BP5 & BP6 \\
\hline
Scenarios& V & V & V & III & III & III \\
\hline
\hline
$m_{H^0}$ (GeV) &~59.3 &\ 62.0 &\ 57.2  &\ 94.1  & 105.6 & 514.2 \\
\hline
$m_{A^0}$  (GeV)&170.5& 339.3    &     307.9       &     101.9  &   576.9   &   740.3  \\
\hline
$m_{H^\pm}$ (GeV)&145.1& 327.5   &    342.0     &    110.4  &   593.6    &   516.2 \\
\hline
$\mu_2^2$  (GeV$^2$)&3642.7&  3733.9     &    3514.3      & 9059.5 &  10945.8    &  265875.8  \\
\hline
 $\lambda_{L} (10^{-3})$  &-2.055  & 1.824&-4.043   &-3.501   & 3.549 &-23.995  \\
 \hline
 $\lambda_{S}$                &~0.418 & 1.834&~1.503  &~0.022   &5.300  &~~4.646\\
\hline
 $\Omega h^2 (\times 10^{-2})$  &~1.870 & 0.092&~6.490   &~0.298   & 0.089 &~~0.060 \\
\hline
$Br(h^0 \to H^0H^0)$ & $0.6\%$ & $0.2\%$ & $3.1\%$ & - & - &-\\
\hline
$Br(A^0 \to W^{\pm(*)} H^{\mp})$ &  $\sim 0\%$& $\sim 0\%$& - & -& -&$32.4\%$ \\
\hline
$Br(A^0 \to Z^{(*)} H^0)$ &  $\sim 100\%$& $\sim 100\%$& $100\%$ & $100\%$& $100\%$ &$67.6\%$\\
\hline
$Br(H^\pm \to W^{\pm(*)} A^0)$ & - & -& $\sim 0\%$ & $3.9\%$& $\sim 0\%$& -\\
\hline
$Br(H^\pm \to W^{\pm(*)} H^0)$ & $100\%$& $100\%$& $\sim 100\%$ & $96.1\%$& $\sim 100\%$& $100\%$\\
\hline
\end{tabular}
\caption{Benchmark points consistent with collider experiments  and dark matter constraints on the relic density  are proposed. Decay information of $H^0$, $A^0$ and $H^\pm$ are also given.}
\label{tab:Benchmark-points}
\end{table}

\begin{table}[!htb]
\centering
\begin{tabular}{|c|c|r|r|r|r|r|}
\hline\hline
 \multicolumn{6}{|c|}{$e^+e^- \to Zh^0$} \\ \hline
 \multicolumn{1}{|c|}{ }    
& \multicolumn{1}{|c|}{$\sqrt{s}$ (GeV) }                   &\multicolumn{1}{c|}{$\sigma_{\mathrm{SM}}^{0}$}&\multicolumn{1}{c|}{$\sigma^{1,\mathrm{QED}}$ (fb)}&\multicolumn{1}{c|}{$\sigma_{\mathrm{SM}}^{1,\mathrm{weak}}$ (fb)}&\multicolumn{1}{c|}{$\sigma_{\mathrm{SM}}^{\mathrm{NLO}}$(fb) }\\ \hline
 \multirow{4}*{SM}
 &$250$                &251.380 	&1.258 	&-23.890 	&228.748     \\\cline{2-6}
 & $350$            &135.381 	&1.034 	&-13.023 	&123.392  \\\cline{2-6}
  &$500$&60.047 	&0.540 	&-6.059 	&54.528   \\\cline{2-6}
&$1000$   &13.475 	&0.144 	&-2.436 	&11.183\\ \hline\hline
\multicolumn{1}{|c|}{IHDM }
& \multicolumn{1}{|c|}{$\sqrt{s}$ (GeV) }                   &\multicolumn{1}{c|}{$\sigma_{\mathrm{IHDM}}^{1,\mathrm{weak}}$ (fb)}&\multicolumn{1}{c|}{$\sigma_{\mathrm{IHDM}}^{\mathrm{NLO}}$ (fb)}&\multicolumn{1}{c|}{$\Delta$(\%)}&\multicolumn{1}{c|}{$\delta$(\%)}\\ \hline
\multirow{4}*{BP1}
 &$250$            &-22.942     &229.696&-8.626 &0.414     \\\cline{2-6}
 & $350$           &-12.457     &123.958        &-8.438    &0.459  \\\cline{2-6}
  &$500$           &-5.955      &54.632  &-9.018           &0.191    \\\cline{2-6}
&$1000$            &-2.444      &11.175    &-17.069        &-0.072 \\ \hline
\multirow{4}*{BP2}
 &$250$            &-24.103     &228.535        &-9.088 &-0.093    \\\cline{2-6}
 & $350$           &-12.898     &123.517        &-8.763 &0.101   \\\cline{2-6}
  &$500$           &-5.741      &54.846     &-8.661     &0.583    \\\cline{2-6}
&$1000$            &-2.487      &11.132     &-17.387    &-0.456\\ \hline
\multirow{4}*{BP3}
 &$250$   &-26.682      &225.956        &-10.114            &-1.221   \\\cline{2-6}
 & $350$  &-14.249      &122.166        &-9.761     &-0.994   \\\cline{2-6}
  &$500$  &-6.427       &54.160         &-9.804         &-0.675    \\\cline{2-6}
&$1000$   &-2.640       &10.979 &-18.523        &-1.824 \\ \hline
\multirow{4}*{BP4}
 &$250$         &-24.115 &228.523       &-9.093         &-0.098   \\\cline{2-6}
 &$350$         &-13.136 &123.279   &-8.939     &-0.092   \\\cline{2-6}
  &$500$        &-6.084 &54.503     &-9.233     &-0.046    \\\cline{2-6}
 &$1000$    &-2.425     &11.194     &-16.928    &0.098 \\ \hline
\multirow{4}*{BP5}
 &$250$    &-29.138     &223.500        &-11.091        &-2.294    \\\cline{2-6}
 &$350$   &-15.663      &120.752        &-10.806        &-2.140   \\\cline{2-6}
  &$500$   &-7.089      &53.498 &-10.906        &-1.889    \\\cline{2-6}
 &$1000$    &-2.516     &11.103     &-17.603    &-0.715 \\ \hline
 \multirow{4}*{BP6}
  &$250$            &-22.689     &229.949        &-8.525 &0.525   \\\cline{2-6}
 & $350$           &-12.340     &124.075        &-8.351 &0.554   \\\cline{2-6}
  &$500$           &-5.735      &54.852 &-8.652 &0.594    \\\cline{2-6}
&$1000$            &-2.341      &11.278     &-16.304    &0.850
 \\ \hline
\hline
\end{tabular}
\caption{Total cross section for $e^+\ e^-\to Zh^0$ for different CM energy. The IHDM parameters are fixed according to Table \ref{tab:Benchmark-points}.}
\label{Tab:hZBP}
\end{table}
In Table \ref{tab:Benchmark-points}, we propose six benchmark points for the future $e^+ e^-$ collider search. Among them, BP1-BP3 belong to Scenario V and BP4-BP6 belong to Scenario III. BP1-BP3 can also be examined at the LHC or full higher energy $pp$ colliders, via mono-jet measurement as shown in \cite{Belyaev:2018ext} or mono-W/$\gamma$ signal as shown in \cite{Jueid:2020rek}. BP1-BP3 are chosen such that the  LOP is the CP-even Higgs boson $H^0$. It has been checked that we can exchange the mass of $H^0$ with $A^0$. For BP4-BP6, the invisible decay of the SM Higgs is not open.

A few more explanations on these BPs are provided below:
\begin{itemize}
\item In these BPs, BP1 represents a favourable case where the loop correction to the process $e^+ e^- \to Z h^0$ can be detectable and the new particles, rather light, can be directly produced at the future Higgs factories via the process $e^+ e^- \to H^0 A^0$ at all energy cases. 

\item BP4 provides a case where for all the CM energies, the effects of new physics are small and difficult to be detected in the process $e^+ e^- \to Z h^0$. Although there are light particles like $H^0$, $A^0$ and $H^\pm$ for BP4, the contributions of new particles yield a small correction to the cross section of $e^+ e^- \to Z h^0$ due to the smallness triple Higgs couplings. But when we consider the process $e^+ e^- \to H^0 A^0$, it is possible to produce these new particles directly at the future Higgs factories.

\item BP6 provides a case where the mass of new particles is too heavy to be produced directly at future $e^+ e^-$ colliders with CM energy less than $1000$ GeV. But the effects of new physics can be detected via the loop effects in $e^+ e^- \to Z h^0$. The reason for such a sizeable correction can be attributed to the large triple Higgs couplings for this case.

\item BP3 and BP5 represent more complicated cases where new physics  contribution induced via loop to the process $e^+ e^- \to Z h^0$ can be sizeable, but to confirm the case we need future colliders with high CM energies like 500 GeV or higher.

\item BP2 represents the case that a 250 GeV Higgs factory might be difficult to detect the effects of NP, but Higgs factories with a collision energy larger than $350$ GeV might be better. The production of new particles needs a CM energy higher than 500 GeV.
\end{itemize}

 Except the collider searches, these BPs can also be searched by the future dark matter searches, which can be left for  future study.

In Table~\ref{Tab:hZBP}, the total cross section for $e^+e^-\to Zh^0$ in the SM with various center of mass energies are presented, as well as the total cross 
section for the benchmark points in the IHDM.
For the SM results, we present the LO cross section, one-loop QED corrections, one-loop weak corrections and full NLO cross section.
For the IHDM, as the LO cross section and one-loop QED corrections are exactly same as the ones in the SM, only one-loop weak corrections and full NLO cross section are presented, with the addition of $\Delta$ and $\delta$, where $\Delta$ is the relative one-loop corrections to LO and $\delta$ is the relative correction of IHDM to the full NLO SM result defined in Eqs.~(\ref{sig1}) and (\ref{eq:sub}), respectively.

There are a few comments on the results in Table \ref{Tab:hZBP}:
\begin{itemize}
\item As shown in the column $\delta$, when the CM energy $\sqrt{s}=250$ GeV, the ratio of the contribution of new physics in BP1 and BP6 can increase the cross section by a factor $0.414\%$ and $0.525\%$, respectively. In contrast, the contribution of the new physics in BP3 and BP5 decreases the cross section sizeably by a factor $-1.221\%$ and $-2.294\%$, respectively. The contribution of the NP in BP2 and BP4 is $-0.093\%$ and $-0.098\%$, respectively, which is less than the projected precision $0.1\%$ of the future Higgs factories.

\item When the CM energy increases to 350 GeV,  the ratio  in BP1 and BP6 keeps to be positive and increase the cross section slightly. The corrections in BP3 and BP5 decreases the cross section by a factor $-0.994\%$ and $-2.140\%$. While the contribution in BP2 increases the cross section and the effect can reach $0.1\%$. However, the contribution in BP4 is still small and  the effect is $-0.091\%$.

\item When the CM  energy increases to 500 GeV, the corrections in BP1, BP2 and BP6 are positive, and can reach $0.191\%$, $0.583\%$, and $0.594\%$, respectively. The ratio of BP3, BP4 and BP5 are negative, and is $-0.675\%$, $-0.046\%$, $-1.889\%$, respectively.

\item When the CM energy increases to 1000 GeV, the correction in BP1 becomes negative and is given by $-0.071\%$, while in BP6 it remains positive and is $0.85\%$. The corrections of BP3 and BP5 are negative: $-1.824\%$ and $-0.715\%$. While the one  of BP2 and BP4 becomes $-0.456\%$ and $0.098\%$, respectively.
\end{itemize}

\begin{table}[!htb]
\centering
\begin{tabular}{|c|c|r|r|r|r|r|r|}
\hline\hline
 \multicolumn{7}{|c|}{$e^+e^- \to H^0 A^0$} \\ \hline
  \multicolumn{1}{|c|}{IHDM }
& \multicolumn{1}{|c|}{$\sqrt{s}$ (GeV) }                      &\multicolumn{1}{c|}{$\sigma_{\mathrm{IHDM}}^{0}$ (fb)}&\multicolumn{1}{c|}{$\sigma_{\mathrm{IHDM}}^{1,\mathrm{weak}}$ (fb)}&\multicolumn{1}{c|}{$\sigma_{\mathrm{IHDM}}^{1,QED}$ (fb)}&\multicolumn{1}{c|}{$\sigma_{\mathrm{IHDM}}^{\mathrm{NLO}}$ (fb)}&\multicolumn{1}{c|}{$\Delta$(\%)}
\\ \hline
\multirow{4}*{BP1}
&$250$  &12.080         &-0.697         &0.031  &11.414         &-5.513  \\\cline{2-7}
&$350$ &44.391  &-3.613         &0.256  &41.034 &-7.562  \\\cline{2-7}
&$500$&35.880   &-3.513 &0.255  &32.622 &-9.080  \\\cline{2-7}
&$1000$ &11.879 &-1.637 &0.104  &10.346 &-12.905  \\ \hline
\multirow{2}*{BP2}
&$500$  &6.755  &-0.539         &0.028  &6.244  &-7.565 \\\cline{2-7}
&$1000$  &8.947         &-1.274         &0.064  &7.737  &-13.524  \\ \hline
\multirow{2}*{BP3}
&$500$ &11.422&-1.014   &0.055  &10.463 &-8.396  \\\cline{2-7}
&$1000$ &9.611&-1.449   &0.073  &8.235  &-14.317  \\ \hline
\multirow{4}*{BP4}
&$250$  &65.670 &-4.475         &0.313  &61.508         &-6.338  \\\cline{2-7}
&$350$ &68.939  &-5.443         &0.457  &63.953         &-7.232  \\\cline{2-7}
&$500$&43.011   &-3.972     &0.335      &39.374         &-8.456  \\\cline{2-7}
&$1000$ &12.378 &-1.642         &0.116  &10.852     &-12.328   \\ \hline
\multirow{1}*{BP5}
&$1000$  &3.516 &-0.638         &0.018  &2.896          &-17.634\\ \hline
\hline
\end{tabular}
\caption{Total cross section for $e^+\ e^-\to H^0 A^0$ for 
different CM energies.}
\label{Tab:AHBP}
\end{table}

In Table~\ref{Tab:AHBP}, we present the LO and NLO results for 
$e^+e^-\to H^0A^0$ with various center of mass energies. We give the 
weak contributions and  the QED ones as well as the total NLO cross section. 
We also show the relative corrections with respect to the LO results, as demonstrated by the results given in $\Delta$ column. Roughly speaking, the contributions of NP always reduce the cross sections from $-5.5\%$ to $-17\%$ for these BPs. When the precisions of future Higgs factories are considered, obviously such large corrections must be taken into account in any experimental analysis.
For example, in term of the cross section of BP4 and the projected precisions given in Table \ref{tab:HFs}, the CEPC with $\sqrt{s}=250$ GeV can find $3.075 \times 10^5$ raw events of $e^+ e^- \to H^0 A^0$, which corresponds to a precision of $0.18\%$ on the cross section when only statistic errors are taken into account. Instead, the LO result predicts that a total number of events is $3.284 \times 10^5$, which will lead to a deviation of $60 \,\,\sigma$. Meanwhile, such a huge number of signal events can also lead to a precise measurement on the masses of $H^0$ and $A^0$, and impose a strong constraint on the parameters in the loop.


\section{Conclusions and Discussions}
\label{sec:conclusions}
We have studied the one loop radiative corrections to the neutral processes: the Higgsstrahlung $e^+e^- \to Zh^0$ and
the neutral Higgs-pair production $e^+e^-\to H^0A^0$ in the IHDM.  We have evaluated both the QED corrections, 
the soft and hard photon emissions and the full weak corrections. The Feynman diagrams are evaluated using dimensional 
regularization in the Feynman gauge.
The full one loop analysis is done using the on-shell renormalization scheme.  In addition, in the numerical 
analysis, we first  performed a systematic scan over the IHDM parameter space taking into account theoretical 
as well as experimental constraints and localized allowed parameter space.

For $e^+e^- \to Zh^0$ process, we first evaluated the one-loop radiative corrections in the SM and checked 
that they do agree with the existing results in the literature. Next we have evaluated the one-loop corrections
in the IHDM and  the relative corrections with respect to the one-loop SM result. We have shown that the pure IHDM effect could reach about $-4.5\%$ percent in Scenario I and could be slightly larger and 
reach $-6\%$ in Scenario II, and only at most $-2\%$ in Scenario III. 
It is remarkable that for Scenario III (V) after imposing the dark matter constraints, 
the allowed points in the parameter space significantly are reduced when compared with Scenario II (IV). Meanwhile, the range of the radiative corrections for the allowed points is also greatly shrunk by the dark matter constraints, as shown in the lower panel of Figure 4.2 and Figure 4.8.

Results are shown for 250 GeV, 500 GeV and 1 TeV center of mass energy. 
For 350 GeV, the situation is similar to 250 GeV.
Such effect is large enough to be measured in the precise future linear collider program. We have also presented 
one-loop angular distributions for six BPs. In addition, we have  demonstrated
 that for the heavy internal IHDM spectrum, the one-loop corrections decouple for large $\mu_2^2$. It has been 
 demonstrated  that the triple Higgs couplings $h^0SS$ with $S=H^0,A^0, H^\pm$ which contribute into the one-loop
 virtual corrections as well as to the wave function renormalization of $h^0$ could contribute significantly to the relative corrections.  

In the case of the pairwise production of neutral Higgs bosons $e^+e^-\to H^0 A^0$, since two scalars are involved in
the final state, large effect is found mainly coming from triple Higgs couplings: either from the wave function 
renormalization of $H^0$ and $A^0$ or from the virtual correction  Fig.~(\ref{diagramsHA})-$G_1$. 
 In the case of 250 GeV center of mass energy, the mass of $H^0$ and $A^0$ are restricted by $m_{H^0}+m_{A^0}<250$ GeV, 
 the $\lambda_i$ involved in the triple couplings could not be significant, the effect is rather small and could not exceed $-9\%$ both in Scenario I and II. While in the case of 500 GeV CM energy, the effect is
 slightly larger and could reach $-15\%$. In the case of 1 TeV center of mass energy, with this energy one can
 cover a large range for $m_{H^0}$ and $m_{A^0}$. Therefore, the quartic couplings $\lambda_i$ become large than in the previous  cases which would make the triple scalar coupling large and the relative corrections to the tree level result could be of the order $-30 \to + 40\%$. 
 
 In Scenario III, after taking into account the dark matter constraints, the range of weak corrections $\Delta_{weak}$ of allowed points is confined in a narrower range when compared with those of Scenario I and II. For example, for the case $\sqrt{s} =250$ GeV, $\Delta_{weak}$ of allowed points can only be in the range $-6.8\%$ to $-5.5\%$.
 It should be emphasised that in both processes $e^+e^-\to Zh^0/H^0A^0$, as demonstrated by the six BPs, the radiative corrections in the allowed parameter space can be rather large. When the projected precisions 
of future $e^+e^-$ colliders given in Table \ref{tab:HFs} are considered, it is mandatory to take them into account in any realistic experimental measurements and data analysis.
 
Here it would be interesting to explore whether the future $e^+ e^-$ colliders have the potential to distinguish different new physics models, like the IHDM, the general 2HDM, and the MSSM. For obvious reasons (like huge parameter space in MSSM), an exhaustive and thorough comparison is an impossible mission. Even a fair comparison is difficult. Instead, we confine to compare the results of a few scenarios considered in literature at their face values, and compile them in Table \ref{Tab:comparison}.

\begin{itemize}
\item In the scenarios of the general 2HDM, the one loop radiative corrections to $e^+e^- \to Z h^0$ and $e^+e^- \to H^0 A^0$ have been computed in Refs. \cite{LopezVal:2010vk,Xie:2018yiv,LopezVal:2009qy}. We compile the results of scenarios in the recent Reference \cite{Xie:2018yiv} for the purpose of comparison.

\item  In the context of supersymmetric models, the one loop corrections to $e^+e^- \to Z h^0/ H^0 A^0$ have been
	presented within the MSSM and complex MSSM (CMSSM) ~\cite{Cao:2014rma, Heinemeyer:2015qbu,Driesen:1995ew}.  We compare our results with those of scenarios presented in the Reference \cite{Cao:2014rma} the same ratio $\delta$ is defined and used there.

\end{itemize}

\begin{table}[H]
	\centering
	\begin{tabular}{|c|c|r|r|r|r|}
		\hline\hline
		\multicolumn{6}{|c|}{$e^+e^- \to Zh^0$} \\ \hline
		\multicolumn{1}{|c|}{}    
		& \multicolumn{1}{|c|}{$\sqrt{s}$ (GeV) }                   &\multicolumn{1}{c|}{MSSM \cite{Cao:2014rma}}&\multicolumn{1}{c|}{CMSSM \cite{Cao:2014rma}}&\multicolumn{1}{c|}{2HDM \cite{Xie:2018yiv}}&\multicolumn{1}{c|}{ IHDM}\\ \hline
		\multirow{4}*{$\delta [\%]$}
		&$250$& $[-2.50 , -1.00] $ &$[-0.80,\,\, \,\,0.00] $ &
	$[-5.76,-0.02]$&$[-2.60, +0.55]$  \\\cline{2-6}
		&$500$&$[-0.80,+1.30] $	&$[+0.20, +0.60] $ 	&$[-5.51 , +0.01] $ &$[-2.20 ,+0.60]$   \\\cline{2-6}
		&$1000$   & &  &$[-6.63, +0.29]$ 
		&$ [-2.00  , +1.00]$\\ \hline\hline
\end{tabular}
\caption{The ranges of the size of radiative corrections for $e^+\ e^-\to Zh^0$ for different SUSY and non-SUSY models are tabulated.}
\label{Tab:comparison}
\end{table}
 According to the projected precision presented in Table \ref{tab:HFs}, the precision in measuring $\delta g_{ZZh^0}/g_{ZZh^0}$ can reach $0.2\%$ for future FCC-ee with $\sqrt{s}=250$ GeV, which roughly corresponds to a precision in measuring $\delta= \pm0.4\%$. A recent global analysis given in \cite{deBlas:2019wgy} demonstrated an even more aggressive precision $\pm 0.09\%$ might be achievable at CLIC with a combo runs with three CM energy $\sqrt{s}=380/1500/3000$ GeV and polarized beams, which, roughly speaking, means an error in $\delta$ can reach $\pm0.18\%$. In order to address the issue of model discrimination, below we deliberately take an moderate optimistic assumption that a precision $\delta =\pm 0.2\%$ can be achievable.

Obviously, given a possible precision, whether two models can be distinguished is determined by the central value of $\delta^{EXP}$. 
Suppose that the future experiment could determine $\delta^{EXP}= 0.0\%\pm0.2\%$, it could be able to rule out the parameter region of the MSSM analysed in \cite{Cao:2014rma} more than $5 \sigma$, while the parameter space of the  CMSSM analysed in \cite{Cao:2014rma}, the SM, 2HDM and IHDM might still be consistent with experimental bounds. In contrast, suppose the future experiment could determine $\delta^{EXP}=-2.5\%\pm 0.2\%$, then the parameter region of the  MSSM analysed in \cite{Cao:2014rma} and 2HDM and IHDM could interpret the experimental data while the parameter region of the CMSSM analysed in \cite{Cao:2014rma} and the SM could be ruled out in terms of $8 \sigma$ and $12 \sigma$, respectively. While, suppose the future experiments could determine $\delta^{EXP}=-5.5\%\pm 0.2\%$, then only 2HDM can interpret the experimental data while the other models could be ruled out with more than $10 \sigma$. In the case with $\delta^{EXP}=0.55\%\pm 0.2\%$, only IHDM could interpret with the experimental data comfortably and the deviation from the prediction of the SM can reach to $2.5 \sigma$.

For future searches at $e^+e^-$ colliders, we have presented six benchmark points that satisfy dark matter  and LHC constraints. 
For these BPs, we have given the weak and the QED corrections for various center of mass energies. These BPs can be explored
at the future $e^+ e^-$ colliders, the LHC and future proton-proton colliders, and future dark matter experiments.

For example, the discovery channel of $e^+e^- \to H^0 A^0$ can lead to some interesting signatures, as explored in the 
reference \cite{Dolle:2009ft,Kalinowski:2018kdn}.  Moreover, for BP1-3, $H^0$ is the LOP and the final state of  $e^+ e^- \to H^0 A^0$ would lead to $Z H^0 H^0$ final state since $A^0 \to Z H^0$ dominantly,  we expect a signature with dilepton plus missing energy. Since the Z boson is on-shell, then we expect to observe two energetic leptons and a large missing energy in the final state. For BP4, the Z boson is off-shell and two leptons from off-shell Z boson decay in the final state are soft. Then two soft leptons and large missing energy would be the characteristic signature of BP4. In the parameter space, there are some points where $A^0$ can have a tiny decay width and its lifetime can be larger than $10^{-15}$ s, we expect that  the signature could be displaced vertex and large missing energy, as shown in \cite{Cheung:2019qdr}. 

At the LHC, it is believed that radiative corrections to $pp\to Wh^0, Zh^0, H^0A^0$ 
would be, to some extent, similar to our finding for $e^+e^-$ colliders. Therefore, 
 our BPs can also be explored at the $pp$ colliders via the processes $p p \to H^0 H^\pm \to H^0 H^0 W^{\pm(*)}$ for BP1-6. The signatures of these BPs can be a large missing energy plus a $W$ boson, while the $W$ boson can be either on-shell (for BP1-3 and BP5) and off-shell (for BP4 and BP6).  For both BP4 and BP6, the charged Higgs boson is almost degenerate with the LOP $H^0$ and has a large lifetime, which might lead to a signature of charged displaced vertex at the LHC.

\section*{Acknowledgements}
We thank Jianxiong Wang for helpful discussions about FDC program. Q.S.Yan is supported 
by the Natural Science Foundation of China under the grant No. 11475180 and 
No. 11875260. B. Gong is supported by the Natural Science Foundation of China Nos. 11475183 and 11975242.\\
 J. El Falaki would like to thank Pedro M. Ferreira and Sven Heinemeyer for helpful conversations,  and  the Abdus Salam International Centre for Theoretical Physics (ICTP)  for hospitality and financial support where part of this work has been done.
The work of H.A, A.A, R.B and J.E  is supported by the Moroccan Ministry of Higher 
Education and Scientific Research MESRSFC and CNRST: Projet PPR/2015/6.
\appendix

\section{More details about IR divergences}
As mentioned in section \ref{sec:renormalization}, IR  divergences in this paper are regularized with a small fictitious photon mass $\lambda$. Meanwhile, two cutoffs, $\Delta E$ and $\Delta\theta$ are used to separate the phase space of real photon emission. Thus full NLO corrections are separated into four parts, as given in Eq.~(\ref{fullcor}). The hard non-collinear part, $d\sigma_{H\overline{C}}$, is obtained using traditional Monte-Carlo integration techniques. Here we present results for the soft and hard non-collinear parts, as well as the check on the independence of the cutoffs.

For the Higgsstrahlung $e^+e^-\to Zh^0$ and the associate production $e^+e^-\to H^0 A^0$, the real photon is only emitted from the initial state electron and positron. The analytical expression for the soft bremsstrahlung is given by:
\begin{equation}
d\sigma_S=-\dfrac{\alpha}{\pi}d\sigma^0\times\biggl[\log\dfrac{4\Delta E^2}{\lambda^2}  
+ \log\dfrac{4\Delta E^2}{\lambda^2} \log\dfrac{m_e^2}{s}  
+\dfrac{1}{2}\log^2\dfrac{m_e^2}{s}+\log\dfrac{m_e^2}{s}+\dfrac{1}{3}\pi^2\biggr]
\end{equation}
where $\Delta E$ is the cut on the photon energy and  $\lambda$ is a small fictitious mass for the photon. In the  above formula, the IR term $\log\dfrac{4\Delta E^2}{\lambda^2}$
 (resp $\log\dfrac{4\Delta E^2}{\lambda^2} \log\dfrac{m_e^2}{s}$) are respectively canceled by the wave function renormalization constant of the electron and by the virtual photon correction to the one loop $e^+e^-Z$ vertices.
The large Sudakov term $\log^2\dfrac{m_e^2}{s}$ is also canceled by the virtual QED diagram.

One-loop radiation correction includes collinear singularities when $m_e$ goes to zero. In our calculation electron has nonzero mass, but the singularities will  become terms proportional to $\log(m_e)$.
Some of them are cancelled when summing up virtual and real corrections,
and some of them are absorbed into the redefinition of running coupling constant as mentioned above, but some are remained.
To deal with this, we used following fixed order electron structure function which can be derived~\cite{Kuraev:1985hb}
\begin{equation}
\label{eqn:ff}
f_{ee}(x,s)=\delta(1-x)+\dfrac{\alpha}{2\pi}\log\dfrac{s}{4m_e^2}P_{ee}^+(x,0)
\end{equation}
with
\begin{equation}
P_{ee}^+(z,0)=\dfrac{1+z^2}{(1-z)_+}+\dfrac{3}{2}\delta(1-z),
\end{equation}
being the regularized Altarelli-Parisi  splitting function. The 2nd term in Eq.~(\ref{eqn:ff}) gives an additional ``counter term'' which can be combined with hard collinear part.

The $HC+CT$ part is obtained as
\begin{eqnarray}
d\sigma_{HC+CT}&\equiv& d\sigma_{HC+CT}^{*} + d\sigma_{SC} \nonumber \\
d\sigma_{HC+CT}^{*} &=&\dfrac{\alpha}{2\pi}\left[\dfrac{1+z^2}{1-z}\log\Delta\theta^2-\dfrac{2z}{1-z}\right]\times\biggl[d\sigma^0(zk_1)+d\sigma^0(zk_2)\biggr]dz\nonumber\\
d\sigma_{SC} &=&-\dfrac{\alpha}{\pi}\log\dfrac{s}{4m_e^2} \left[\dfrac{3}{2}+2\log\delta_s\right]d\sigma^0,
\label{eqn:ISR}
\end{eqnarray}
where the approximation $\Delta\theta \gg m_e/\sqrt{s}$ has been taken and $\Delta E$ is replaced with a dimensionless parameter $\delta_s=2\Delta E/\sqrt{s}$.

\begin{figure}[!htb]
\begin{center}
\includegraphics[width=0.31\textwidth]{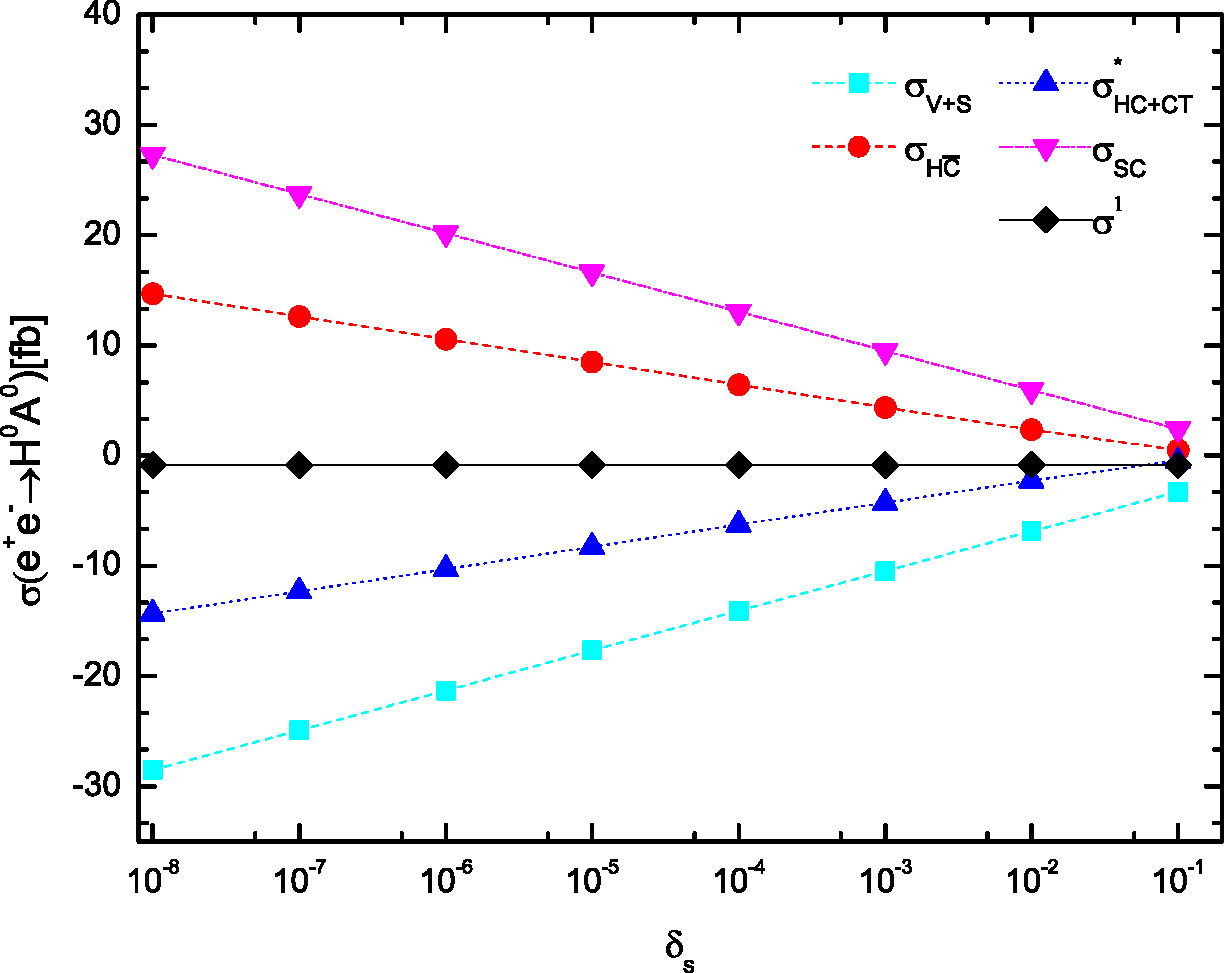}
\includegraphics[width=0.31\textwidth]{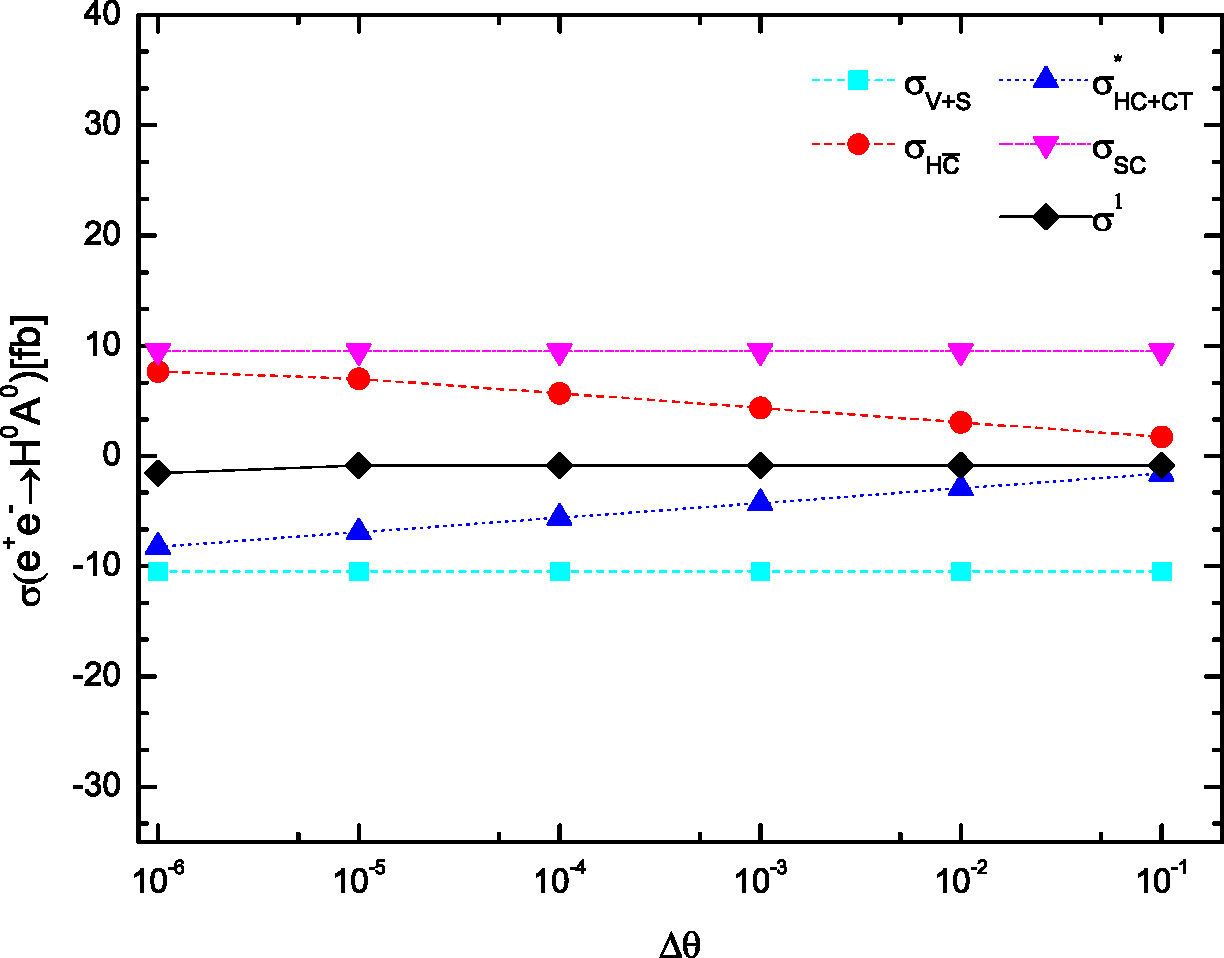}
\includegraphics[width=0.31\textwidth]{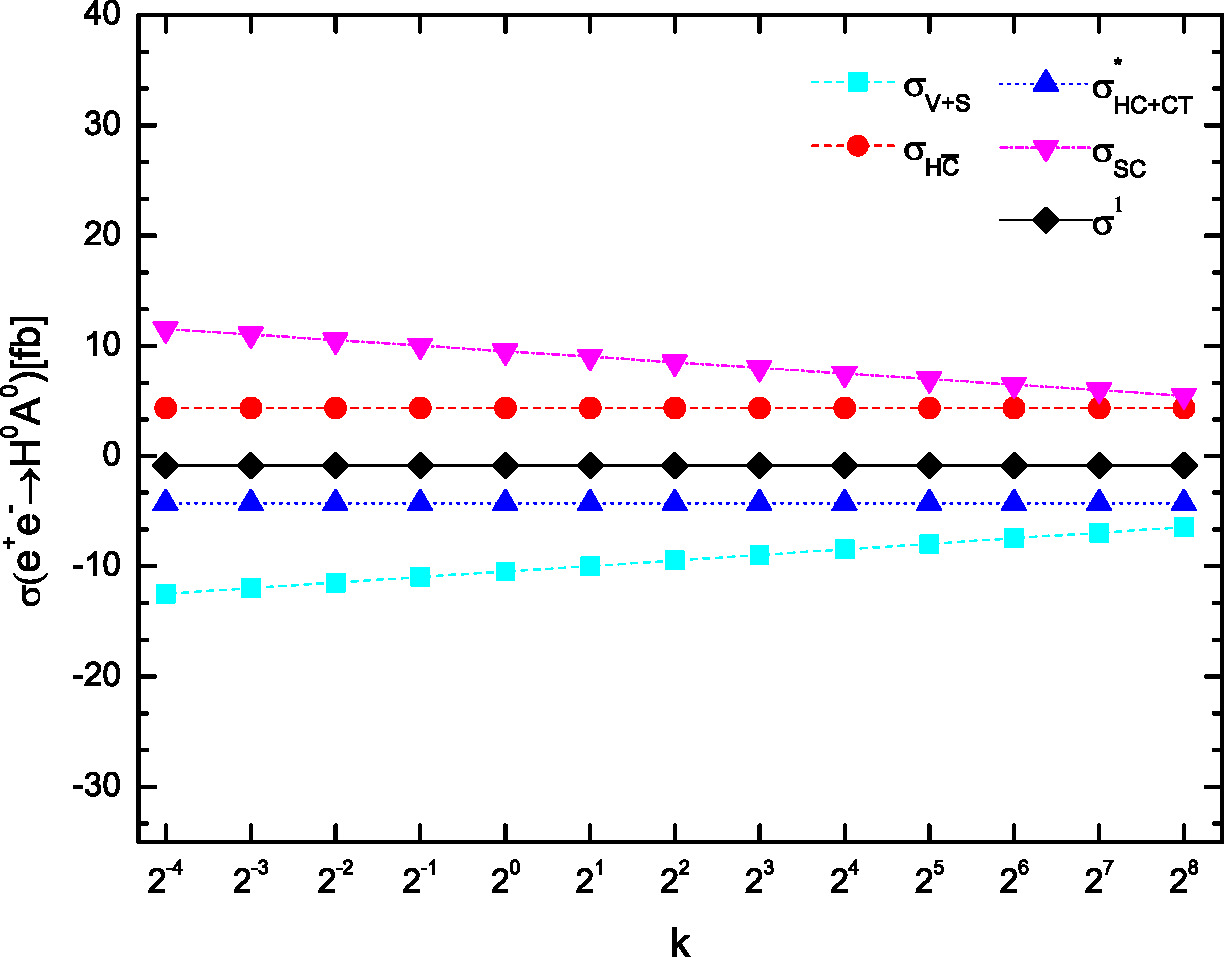}
\caption{One-loop corrections of $e^+e^-\rightarrow H^0A^0$ as functions of $\delta_s$, $\Delta\theta$ and $k$.}
\label{figcheck}
\end{center}
\end{figure}

$\Delta E$ and $\Delta\theta$ are unphysical cutoffs we introduced to deal with IR singularities. Our final results should not depend on them. In Fig.~(\ref{figcheck}), we show our check for this independence, taking one of the processes as an example. From first subfigure, it can be seen that the independence on $\Delta E$ is found in a wide range and we choose $\delta_s=10^{-3}$ as our default choice. In second subfigure, we can see that the result becomes cut dependent when $\Delta\theta$ is smaller than $10^{-4}$. It is because the approximation used in Eq.~(\ref{eqn:ISR}) demands $\Delta\theta\gg m_e/\sqrt{s}\sim 2\times10^{-6}$. Thus we choose $\Delta\theta=10^{-3}$ as our choice. 

Collinear divergences in our calculation appear as terms proportional to $\log(m_e)$. After including the counter term from electron structure function, such divergent terms should vanish in the final result. In order to check this, we vary the mass of electron with  a factor of $k$ from $2^{-4}$ to $2^8$, namely $m_e$ is taken $k\times 0.511$ MeV. The cancellation is shown in last subfigure of Fig.~(\ref{figcheck}), from which we can see that the result remains unchanged when $k$ varies . Also, we can see that singular terms only appear in $\sigma_{V+S}$ and $\sigma_{SC}$ parts.

\bibliographystyle{JHEP}
\bibliography{biblio}

\providecommand{\href}[2]{#2}\begingroup\raggedright\begin{thebibliography}{100}

\bibitem{Aad:2012tfa}
{\bf ATLAS} Collaboration, G.~Aad et~al., {\it {Observation of a new particle
  in the search for the Standard Model Higgs boson with the ATLAS detector at
  the LHC}},  {\em Phys. Lett.} {\bf B716} (2012) 1--29,
  [\href{http://arxiv.org/abs/1207.7214}{{\tt arXiv:1207.7214}}].

\bibitem{Chatrchyan:2012xdj}
{\bf CMS} Collaboration, S.~Chatrchyan et~al., {\it {Observation of a new boson
  at a mass of 125 GeV with the CMS experiment at the LHC}},  {\em Phys. Lett.}
  {\bf B716} (2012) 30--61, [\href{http://arxiv.org/abs/1207.7235}{{\tt
  arXiv:1207.7235}}].

\bibitem{Aaboud:2018urx}
{\bf ATLAS} Collaboration, M.~Aaboud et~al., {\it {Observation of Higgs boson
  production in association with a top quark pair at the LHC with the ATLAS
  detector}},  {\em Phys. Lett.} {\bf B784} (2018) 173--191,
  [\href{http://arxiv.org/abs/1806.00425}{{\tt arXiv:1806.00425}}].

\bibitem{Sirunyan:2018hoz}
{\bf CMS} Collaboration, A.~M. Sirunyan et~al., {\it {Observation of
  $\mathrm{t\overline{t}}$H production}},  {\em Phys. Rev. Lett.} {\bf 120}
  (2018), no.~23 231801, [\href{http://arxiv.org/abs/1804.02610}{{\tt
  arXiv:1804.02610}}].

\bibitem{Aaboud:2018zhk}
{\bf ATLAS} Collaboration, M.~Aaboud et~al., {\it {Observation of $H
  \rightarrow b\bar{b}$ decays and $VH$ production with the ATLAS detector}},
  {\em Phys. Lett.} {\bf B786} (2018) 59--86,
  [\href{http://arxiv.org/abs/1808.08238}{{\tt arXiv:1808.08238}}].

\bibitem{Sirunyan:2018kst}
{\bf CMS} Collaboration, A.~M. Sirunyan et~al., {\it {Observation of Higgs
  boson decay to bottom quarks}},  {\em Phys. Rev. Lett.} {\bf 121} (2018),
  no.~12 121801, [\href{http://arxiv.org/abs/1808.08242}{{\tt
  arXiv:1808.08242}}].

\bibitem{Dawson:2013bba}
S.~Dawson et~al., {\it {Working Group Report: Higgs Boson}},  in {\em
  {Proceedings, 2013 Community Summer Study on the Future of U.S. Particle
  Physics: Snowmass on the Mississippi (CSS2013): Minneapolis, MN, USA, July
  29-August 6, 2013}}, 2013.
\newblock \href{http://arxiv.org/abs/1310.8361}{{\tt arXiv:1310.8361}}.

\bibitem{Zeppenfeld:2000td}
D.~Zeppenfeld, R.~Kinnunen, A.~Nikitenko, and E.~Richter-Was, {\it {Measuring
  Higgs boson couplings at the CERN LHC}},  {\em Phys. Rev.} {\bf D62} (2000)
  013009, [\href{http://arxiv.org/abs/hep-ph/0002036}{{\tt hep-ph/0002036}}].

\bibitem{Gianotti:2000tz}
F.~Gianotti and M.~Pepe-Altarelli, {\it {Precision physics at the LHC}},  {\em
  Nucl. Phys. Proc. Suppl.} {\bf 89} (2000) 177--189,
  [\href{http://arxiv.org/abs/hep-ex/0006016}{{\tt hep-ex/0006016}}].

\bibitem{Cepeda:2019klc}
M.~Cepeda et~al., {\it {Report from Working Group 2}},  {\em CERN Yellow Rep.
  Monogr.} {\bf 7} (2019) 221--584,
  [\href{http://arxiv.org/abs/1902.00134}{{\tt arXiv:1902.00134}}].

\bibitem{deBlas:2019rxi}
J.~de~Blas et~al., {\it {Higgs Boson Studies at Future Particle Colliders}},
  {\em JHEP} {\bf 01} (2020) 139, [\href{http://arxiv.org/abs/1905.03764}{{\tt
  arXiv:1905.03764}}].

\bibitem{DiMicco:2019ngk}
J.~Alison et~al., {\it {Higgs boson potential at colliders: Status and
  perspectives}},  {\em Rev. Phys.} {\bf 5} (2020) 100045,
  [\href{http://arxiv.org/abs/1910.00012}{{\tt arXiv:1910.00012}}].

\bibitem{Moortgat-Picka:2015yla}
A.~Arbey et~al., {\it {Physics at the e+ e- Linear Collider}},  {\em Eur. Phys.
  J.} {\bf C75} (2015), no.~8 371, [\href{http://arxiv.org/abs/1504.01726}{{\tt
  arXiv:1504.01726}}].

\bibitem{Fujii:2015jha}
K.~Fujii et~al., {\it {Physics Case for the International Linear Collider}},
  \href{http://arxiv.org/abs/1506.05992}{{\tt arXiv:1506.05992}}.

\bibitem{Fujii:2017vwa}
K.~Fujii et~al., {\it {Physics Case for the 250 GeV Stage of the International
  Linear Collider}},  \href{http://arxiv.org/abs/1710.07621}{{\tt
  arXiv:1710.07621}}.

\bibitem{CEPC-SPPCStudyGroup:2015csa}
{\bf CEPC-SPPC Study Group} Collaboration, M.~Ahmad et~al., {\it {CEPC-SPPC
  Preliminary Conceptual Design Report. 1. Physics and Detector}},  {\em
  IHEP-CEPC-DR-2015-01, IHEP-TH-2015-01, IHEP-EP-2015-01} (2015).

\bibitem{An:2018dwb}
F.~An et~al., {\it {Precision Higgs physics at the CEPC}},  {\em Chin. Phys. C}
  {\bf 43} (2019), no.~4 043002, [\href{http://arxiv.org/abs/1810.09037}{{\tt
  arXiv:1810.09037}}].

\bibitem{Battaglia:2004mw}
{\bf CLIC Physics Working Group} Collaboration, E.~Accomando et~al., {\it
  {Physics at the CLIC multi-TeV linear collider}},  in {\em {Proceedings, 11th
  International Conference on Hadron spectroscopy (Hadron 2005): Rio de
  Janeiro, Brazil, August 21-26, 2005}}, 2004.
\newblock \href{http://arxiv.org/abs/hep-ph/0412251}{{\tt hep-ph/0412251}}.

\bibitem{Aicheler:2012bya}
M.~Aicheler, P.~Burrows, M.~Draper, T.~Garvey, P.~Lebrun, K.~Peach, N.~Phinney,
  H.~Schmickler, D.~Schulte, and N.~Toge, {\it {A Multi-TeV Linear Collider
  Based on CLIC Technology}}, .

\bibitem{Linssen:2012hp}
L.~Linssen, A.~Miyamoto, M.~Stanitzki, and H.~Weerts, {\it {Physics and
  Detectors at CLIC: CLIC Conceptual Design Report}},
  \href{http://arxiv.org/abs/1202.5940}{{\tt arXiv:1202.5940}}.

\bibitem{Charles:2018vfv}
{\bf CLICdp, CLIC} Collaboration, T.~K. Charles et~al., {\it {The Compact
  Linear Collider (CLIC) - 2018 Summary Report}},
  \href{http://arxiv.org/abs/1812.06018}{{\tt arXiv:1812.06018}}.

\bibitem{deBlas:2018mhx}
R.~Franceschini et~al., {\it {The CLIC Potential for New Physics}},
  \href{http://arxiv.org/abs/1812.02093}{{\tt arXiv:1812.02093}}.

\bibitem{Gomez-Ceballos:2013zzn}
{\bf TLEP Design Study Working Group} Collaboration, M.~Bicer et~al., {\it
  {First Look at the Physics Case of TLEP}},  {\em JHEP} {\bf 01} (2014) 164,
  [\href{http://arxiv.org/abs/1308.6176}{{\tt arXiv:1308.6176}}].

\bibitem{Abada:2019zxq}
{\bf FCC} Collaboration, A.~Abada et~al., {\it {FCC-ee: The Lepton Collider}},
  {\em Eur. Phys. J. ST} {\bf 228} (2019), no.~2 261--623.

\bibitem{Fujii:2019zll}
{\bf LCC Physics Working Group} Collaboration, K.~Fujii et~al., {\it {Tests of
  the Standard Model at the International Linear Collider}},
  \href{http://arxiv.org/abs/1908.11299}{{\tt arXiv:1908.11299}}.

\bibitem{Deshpande:1977rw}
N.~G. Deshpande and E.~Ma, {\it {Pattern of Symmetry Breaking with Two Higgs
  Doublets}},  {\em Phys. Rev.} {\bf D18} (1978) 2574.

\bibitem{Gustafsson:2007pc}
M.~Gustafsson, E.~Lundstrom, L.~Bergstrom, and J.~Edsjo, {\it {Significant
  Gamma Lines from Inert Higgs Dark Matter}},  {\em Phys. Rev. Lett.} {\bf 99}
  (2007) 041301, [\href{http://arxiv.org/abs/astro-ph/0703512}{{\tt
  astro-ph/0703512}}].

\bibitem{Hambye:2007vf}
T.~Hambye and M.~H.~G. Tytgat, {\it {Electroweak symmetry breaking induced by
  dark matter}},  {\em Phys. Lett.} {\bf B659} (2008) 651--655,
  [\href{http://arxiv.org/abs/0707.0633}{{\tt arXiv:0707.0633}}].

\bibitem{Agrawal:2008xz}
P.~Agrawal, E.~M. Dolle, and C.~A. Krenke, {\it {Signals of Inert Doublet Dark
  Matter in Neutrino Telescopes}},  {\em Phys. Rev.} {\bf D79} (2009) 015015,
  [\href{http://arxiv.org/abs/0811.1798}{{\tt arXiv:0811.1798}}].

\bibitem{Dolle:2009fn}
E.~M. Dolle and S.~Su, {\it {The Inert Dark Matter}},  {\em Phys. Rev.} {\bf
  D80} (2009) 055012, [\href{http://arxiv.org/abs/0906.1609}{{\tt
  arXiv:0906.1609}}].

\bibitem{Andreas:2009hj}
S.~Andreas, M.~H.~G. Tytgat, and Q.~Swillens, {\it {Neutrinos from Inert
  Doublet Dark Matter}},  {\em JCAP} {\bf 0904} (2009) 004,
  [\href{http://arxiv.org/abs/0901.1750}{{\tt arXiv:0901.1750}}].

\bibitem{Barbieri:2006dq}
R.~Barbieri, L.~J. Hall, and V.~S. Rychkov, {\it {Improved naturalness with a
  heavy Higgs: An Alternative road to LHC physics}},  {\em Phys. Rev.} {\bf
  D74} (2006) 015007, [\href{http://arxiv.org/abs/hep-ph/0603188}{{\tt
  hep-ph/0603188}}].

\bibitem{Arhrib:2015hoa}
A.~Arhrib, R.~Benbrik, J.~El~Falaki, and A.~Jueid, {\it {Radiative corrections
  to the Triple Higgs Coupling in the Inert Higgs Doublet Model}},  {\em JHEP}
  {\bf 12} (2015) 007, [\href{http://arxiv.org/abs/1507.03630}{{\tt
  arXiv:1507.03630}}].

\bibitem{Banerjee:2019luv}
S.~Banerjee, F.~Boudjema, N.~Chakrabarty, G.~Chalons, and H.~Sun, {\it {Relic
  density of dark matter in the inert doublet model beyond leading order: The
  heavy mass case}},  {\em Phys. Rev.} {\bf D100} (2019), no.~9 095024,
  [\href{http://arxiv.org/abs/1906.11269}{{\tt arXiv:1906.11269}}].

\bibitem{Banerjee:2021oxc}
S.~Banerjee, F.~Boudjema, N.~Chakrabarty, and H.~Sun, {\it {Relic density of
  dark matter in the inert doublet model beyond leading order for the low mass
  region: 1. Renormalisation and constraints}},
  \href{http://arxiv.org/abs/2101.02165}{{\tt arXiv:2101.02165}}.

\bibitem{Banerjee:2021anv}
S.~Banerjee, F.~Boudjema, N.~Chakrabarty, and H.~Sun, {\it {Relic density of
  dark matter in the inert doublet model beyond leading order for the low mass
  region: 2. Co-annihilation}},  \href{http://arxiv.org/abs/2101.02166}{{\tt
  arXiv:2101.02166}}.

\bibitem{Banerjee:2021xdp}
S.~Banerjee, F.~Boudjema, N.~Chakrabarty, and H.~Sun, {\it {Relic density of
  dark matter in the inert doublet model beyond leading order for the low mass
  region: 3. Annihilation in 3-body final state}},
  \href{http://arxiv.org/abs/2101.02167}{{\tt arXiv:2101.02167}}.

\bibitem{Banerjee:2021hal}
S.~Banerjee, F.~Boudjema, N.~Chakrabarty, and H.~Sun, {\it {Relic density of
  dark matter in the inert doublet model beyond leading order for the low mass
  region: 4. The Higgs resonance region}},
  \href{http://arxiv.org/abs/2101.02170}{{\tt arXiv:2101.02170}}.

\bibitem{Braathen:2019pxr}
J.~Braathen and S.~Kanemura, {\it {On two-loop corrections to the Higgs
  trilinear coupling in models with extended scalar sectors}},  {\em Phys.
  Lett.} {\bf B796} (2019) 38--46, [\href{http://arxiv.org/abs/1903.05417}{{\tt
  arXiv:1903.05417}}].

\bibitem{Braathen:2019zoh}
J.~Braathen and S.~Kanemura, {\it {Leading two-loop corrections to the Higgs
  boson self-couplings in models with extended scalar sectors}},  {\em Eur.
  Phys. J.} {\bf C80} (2020), no.~3 227,
  [\href{http://arxiv.org/abs/1911.11507}{{\tt arXiv:1911.11507}}].

\bibitem{Senaha:2018xek}
E.~Senaha, {\it {Radiative Corrections to Triple Higgs Coupling and Electroweak
  Phase Transition: Beyond One-loop Analysis}},  {\em Phys. Rev.} {\bf D100}
  (2019), no.~5 055034, [\href{http://arxiv.org/abs/1811.00336}{{\tt
  arXiv:1811.00336}}].

\bibitem{Dolle:2009ft}
E.~Dolle, X.~Miao, S.~Su, and B.~Thomas, {\it {Dilepton Signals in the Inert
  Doublet Model}},  {\em Phys. Rev.} {\bf D81} (2010) 035003,
  [\href{http://arxiv.org/abs/0909.3094}{{\tt arXiv:0909.3094}}].

\bibitem{Aoki:2013lhm}
M.~Aoki, S.~Kanemura, and H.~Yokoya, {\it {Reconstruction of Inert Doublet
  Scalars at the International Linear Collider}},  {\em Phys. Lett.} {\bf B725}
  (2013) 302--309, [\href{http://arxiv.org/abs/1303.6191}{{\tt
  arXiv:1303.6191}}].

\bibitem{Datta:2016nfz}
A.~Datta, N.~Ganguly, N.~Khan, and S.~Rakshit, {\it {Exploring collider
  signatures of the inert Higgs doublet model}},  {\em Phys. Rev.} {\bf D95}
  (2017), no.~1 015017, [\href{http://arxiv.org/abs/1610.00648}{{\tt
  arXiv:1610.00648}}].

\bibitem{Dutta:2017lny}
B.~Dutta, G.~Palacio, J.~D. Ruiz-Alvarez, and D.~Restrepo, {\it {Vector Boson
  Fusion in the Inert Doublet Model}},  {\em Phys. Rev.} {\bf D97} (2018),
  no.~5 055045, [\href{http://arxiv.org/abs/1709.09796}{{\tt
  arXiv:1709.09796}}].

\bibitem{Kalinowski:2018ylg}
J.~Kalinowski, W.~Kotlarski, T.~Robens, D.~Sokolowska, and A.~F. Zarnecki, {\it
  {Benchmarking the Inert Doublet Model for $e^+ e^-$ colliders}},  {\em JHEP}
  {\bf 12} (2018) 081, [\href{http://arxiv.org/abs/1809.07712}{{\tt
  arXiv:1809.07712}}].

\bibitem{Kalinowski:2018kdn}
J.~Kalinowski, W.~Kotlarski, T.~Robens, D.~Sokolowska, and A.~F. Zarnecki, {\it
  {Exploring Inert Scalars at CLIC}},  {\em JHEP} {\bf 07} (2019) 053,
  [\href{http://arxiv.org/abs/1811.06952}{{\tt arXiv:1811.06952}}].

\bibitem{Guo-He:2020nok}
Y.~Guo-He, S.~Mao, L.~Gang, Z.~Yu, and G.~Jian-You, {\it {Searches for dark
  matter via charged Higgs pair production in the Inert Doublet Model at
  $\gamma \gamma$ collider}},  \href{http://arxiv.org/abs/2006.06216}{{\tt
  arXiv:2006.06216}}.

\bibitem{Yang:2021hcu}
F.-X. Yang, Z.-L. Han, and Y.~Jin, {\it {Same-Sign Dilepton Signature in the
  Inert Doublet Model}},  \href{http://arxiv.org/abs/2101.06862}{{\tt
  arXiv:2101.06862}}.

\bibitem{Kalinowski:2020rmb}
J.~Kalinowski, T.~Robens, D.~Sokolowska, and A.~F. Zarnecki, {\it {IDM
  benchmarks for the LHC and future colliders}},
  \href{http://arxiv.org/abs/2012.14818}{{\tt arXiv:2012.14818}}.

\bibitem{Melfo:2011ie}
A.~Melfo, M.~Nemevsek, F.~Nesti, G.~Senjanovic, and Y.~Zhang, {\it {Inert
  Doublet Dark Matter and Mirror/Extra Families after Xenon100}},  {\em Phys.
  Rev.} {\bf D84} (2011) 034009, [\href{http://arxiv.org/abs/1105.4611}{{\tt
  arXiv:1105.4611}}].

\bibitem{Abercrombie:2015wmb}
D.~Abercrombie et~al., {\it {Dark Matter Benchmark Models for Early LHC Run-2
  Searches: Report of the ATLAS/CMS Dark Matter Forum}},  {\em Phys. Dark
  Univ.} {\bf 27} (2020) 100371, [\href{http://arxiv.org/abs/1507.00966}{{\tt
  arXiv:1507.00966}}].

\bibitem{Ilnicka:2015jba}
A.~Ilnicka, M.~Krawczyk, and T.~Robens, {\it {Inert Doublet Model in light of
  LHC Run I and astrophysical data}},  {\em Phys. Rev.} {\bf D93} (2016), no.~5
  055026, [\href{http://arxiv.org/abs/1508.01671}{{\tt arXiv:1508.01671}}].

\bibitem{Blinov:2015qva}
N.~Blinov, J.~Kozaczuk, D.~E. Morrissey, and A.~de~la Puente, {\it {Compressing
  the Inert Doublet Model}},  {\em Phys. Rev.} {\bf D93} (2016), no.~3 035020,
  [\href{http://arxiv.org/abs/1510.08069}{{\tt arXiv:1510.08069}}].

\bibitem{Poulose:2016lvz}
P.~Poulose, S.~Sahoo, and K.~Sridhar, {\it {Exploring the Inert Doublet Model
  through the dijet plus missing transverse energy channel at the LHC}},  {\em
  Phys. Lett.} {\bf B765} (2017) 300--306,
  [\href{http://arxiv.org/abs/1604.03045}{{\tt arXiv:1604.03045}}].

\bibitem{Hashemi:2016wup}
M.~Hashemi and S.~Najjari, {\it {Observability of Inert Scalars at the LHC}},
  {\em Eur. Phys. J.} {\bf C77} (2017), no.~9 592,
  [\href{http://arxiv.org/abs/1611.07827}{{\tt arXiv:1611.07827}}].

\bibitem{Wan:2018eaz}
N.~Wan, N.~Li, B.~Zhang, H.~Yang, M.-F. Zhao, M.~Song, G.~Li, and J.-Y. Guo,
  {\it {Searches for Dark Matter via Mono-W Production in Inert Doublet Model
  at the LHC}},  {\em Commun. Theor. Phys.} {\bf 69} (2018), no.~5 617.

\bibitem{Belyaev:2018ext}
A.~Belyaev, T.~R. Fernandez Perez~Tomei, P.~G. Mercadante, C.~S. Moon,
  S.~Moretti, S.~F. Novaes, L.~Panizzi, F.~Rojas, and M.~Thomas, {\it
  {Advancing LHC probes of dark matter from the inert two-Higgs-doublet model
  with the monojet signal}},  {\em Phys. Rev.} {\bf D99} (2019), no.~1 015011,
  [\href{http://arxiv.org/abs/1809.00933}{{\tt arXiv:1809.00933}}].

\bibitem{Dercks:2018wch}
D.~Dercks and T.~Robens, {\it {Constraining the Inert Doublet Model using
  Vector Boson Fusion}},  {\em Eur. Phys. J.} {\bf C79} (2019), no.~11 924,
  [\href{http://arxiv.org/abs/1812.07913}{{\tt arXiv:1812.07913}}].

\bibitem{Lu:2019lok}
Y.-L.~S. Tsai, V.~Q. Tran, and C.-T. Lu, {\it {Confronting dark matter
  co-annihilation of Inert two Higgs Doublet Model with a compressed mass
  spectrum}},  {\em JHEP} {\bf 06} (2020) 033,
  [\href{http://arxiv.org/abs/1912.08875}{{\tt arXiv:1912.08875}}].

\bibitem{Gil:2012ya}
G.~Gil, P.~Chankowski, and M.~Krawczyk, {\it {Inert Dark Matter and Strong
  Electroweak Phase Transition}},  {\em Phys. Lett.} {\bf B717} (2012)
  396--402, [\href{http://arxiv.org/abs/1207.0084}{{\tt arXiv:1207.0084}}].

\bibitem{Swiezewska:2015paa}
B.~Swiezewska, {\it {Inert scalars and vacuum metastability around the
  electroweak scale}},  {\em JHEP} {\bf 07} (2015) 118,
  [\href{http://arxiv.org/abs/1503.07078}{{\tt arXiv:1503.07078}}].

\bibitem{Blinov:2015vma}
N.~Blinov, S.~Profumo, and T.~Stefaniak, {\it {The Electroweak Phase Transition
  in the Inert Doublet Model}},  {\em JCAP} {\bf 1507} (2015) 028,
  [\href{http://arxiv.org/abs/1504.05949}{{\tt arXiv:1504.05949}}].

\bibitem{Huang:2017rzf}
F.~P. Huang and J.-H. Yu, {\it {Exploring inert dark matter blind spots with
  gravitational wave signatures}},  {\em Phys. Rev.} {\bf D98} (2018), no.~9
  095022, [\href{http://arxiv.org/abs/1704.04201}{{\tt arXiv:1704.04201}}].

\bibitem{Keus:2014jha}
V.~Keus, S.~F. King, S.~Moretti, and D.~Sokolowska, {\it {Dark Matter with Two
  Inert Doublets plus One Higgs Doublet}},  {\em JHEP} {\bf 11} (2014) 016,
  [\href{http://arxiv.org/abs/1407.7859}{{\tt arXiv:1407.7859}}].

\bibitem{Arcadi:2019lka}
G.~Arcadi, A.~Djouadi, and M.~Raidal, {\it {Dark Matter through the Higgs
  portal}},  {\em Phys. Rept.} {\bf 842} (2020) 1--180,
  [\href{http://arxiv.org/abs/1903.03616}{{\tt arXiv:1903.03616}}].

\bibitem{Belyaev:2016lok}
A.~Belyaev, G.~Cacciapaglia, I.~P. Ivanov, F.~Rojas-Abatte, and M.~Thomas, {\it
  {Anatomy of the Inert Two Higgs Doublet Model in the light of the LHC and
  non-LHC Dark Matter Searches}},  {\em Phys. Rev.} {\bf D97} (2018), no.~3
  035011, [\href{http://arxiv.org/abs/1612.00511}{{\tt arXiv:1612.00511}}].

\bibitem{Arhrib:2013ela}
A.~Arhrib, Y.-L.~S. Tsai, Q.~Yuan, and T.-C. Yuan, {\it {An Updated Analysis of
  Inert Higgs Doublet Model in light of the Recent Results from LUX, PLANCK,
  AMS-02 and LHC}},  {\em JCAP} {\bf 1406} (2014) 030,
  [\href{http://arxiv.org/abs/1310.0358}{{\tt arXiv:1310.0358}}].

\bibitem{Eiteneuer:2017hoh}
B.~Eiteneuer, A.~Goudelis, and J.~Heisig, {\it {The inert doublet model in the
  light of Fermi-LAT gamma-ray data: a global fit analysis}},  {\em Eur. Phys.
  J.} {\bf C77} (2017), no.~9 624, [\href{http://arxiv.org/abs/1705.01458}{{\tt
  arXiv:1705.01458}}].

\bibitem{Fleischer:1982af}
J.~Fleischer and F.~Jegerlehner, {\it {Radiative Corrections to Higgs
  Production by $e^+ e^- \to Z H$ in the {Weinberg-Salam} Model}},  {\em Nucl.
  Phys.} {\bf B216} (1983) 469--492.

\bibitem{Kniehl:1991hk}
B.~A. Kniehl, {\it {Radiative corrections for associated $Z H$ production at
  future $e^{+} e^{-}$ colliders}},  {\em Z. Phys.} {\bf C55} (1992) 605--618.

\bibitem{Denner:1992bc}
A.~Denner, J.~Kublbeck, R.~Mertig, and M.~Bohm, {\it {Electroweak radiative
  corrections to $e^+ e^- \to H Z$}},  {\em Z. Phys.} {\bf C56} (1992)
  261--272.

\bibitem{Sun:2016bel}
Q.-F. Sun, F.~Feng, Y.~Jia, and W.-L. Sang, {\it {Mixed electroweak-QCD
  corrections to e+e-$\rightarrow$HZ at Higgs factories}},  {\em Phys. Rev. D}
  {\bf 96} (2017), no.~5 051301, [\href{http://arxiv.org/abs/1609.03995}{{\tt
  arXiv:1609.03995}}].

\bibitem{Gong:2016jys}
Y.~Gong, Z.~Li, X.~Xu, L.~L. Yang, and X.~Zhao, {\it {Mixed QCD-EW corrections
  for Higgs boson production at $e^+e^-$ colliders}},  {\em Phys. Rev.} {\bf
  D95} (2017), no.~9 093003, [\href{http://arxiv.org/abs/1609.03955}{{\tt
  arXiv:1609.03955}}].

\bibitem{Kanemura:2015mxa}
S.~Kanemura, M.~Kikuchi, and K.~Yagyu, {\it {Fingerprinting the extended Higgs
  sector using one-loop corrected Higgs boson couplings and future precision
  measurements}},  {\em Nucl. Phys.} {\bf B896} (2015) 80--137,
  [\href{http://arxiv.org/abs/1502.07716}{{\tt arXiv:1502.07716}}].

\bibitem{Kanemura:2016sos}
S.~Kanemura, M.~Kikuchi, and K.~Sakurai, {\it {Testing the dark matter scenario
  in the inert doublet model by future precision measurements of the Higgs
  boson couplings}},  {\em Phys. Rev.} {\bf D94} (2016), no.~11 115011,
  [\href{http://arxiv.org/abs/1605.08520}{{\tt arXiv:1605.08520}}].

\bibitem{Arhrib:2014pva}
A.~Arhrib, R.~Benbrik, and T.-C. Yuan, {\it {Associated Production of Higgs at
  Linear Collider in the Inert Higgs Doublet Model}},  {\em Eur. Phys. J.} {\bf
  C74} (2014) 2892, [\href{http://arxiv.org/abs/1401.6698}{{\tt
  arXiv:1401.6698}}].

\bibitem{Arhrib:2012ia}
A.~Arhrib, R.~Benbrik, and N.~Gaur, {\it {$H\to \gamma \gamma$ in Inert Higgs
  Doublet Model}},  {\em Phys. Rev.} {\bf D85} (2012) 095021,
  [\href{http://arxiv.org/abs/1201.2644}{{\tt arXiv:1201.2644}}].

\bibitem{Swiezewska:2012eh}
B.~Swiezewska and M.~Krawczyk, {\it {Diphoton rate in the inert doublet model
  with a 125 GeV Higgs boson}},  {\em Phys. Rev.} {\bf D88} (2013), no.~3
  035019, [\href{http://arxiv.org/abs/1212.4100}{{\tt arXiv:1212.4100}}].

\bibitem{Krawczyk:2013pea}
M.~Krawczyk, D.~Sokołowska, P.~Swaczyna, and B.~Świeżewska, {\it {Higgs $\to
  \gamma \gamma $, $Z\gamma $ in the Inert Doublet Model}},  {\em Acta Phys.
  Polon.} {\bf B44} (2013), no.~11 2163--2170,
  [\href{http://arxiv.org/abs/1309.7880}{{\tt arXiv:1309.7880}}].

\bibitem{Branco:2011iw}
G.~C. Branco, P.~M. Ferreira, L.~Lavoura, M.~N. Rebelo, M.~Sher, and J.~P.
  Silva, {\it {Theory and phenomenology of two-Higgs-doublet models}},  {\em
  Phys. Rept.} {\bf 516} (2012) 1--102,
  [\href{http://arxiv.org/abs/1106.0034}{{\tt arXiv:1106.0034}}].

\bibitem{Ginzburg:2010wa}
I.~F. Ginzburg, K.~A. Kanishev, M.~Krawczyk, and D.~Sokolowska, {\it {Evolution
  of Universe to the present inert phase}},  {\em Phys. Rev.} {\bf D82} (2010)
  123533, [\href{http://arxiv.org/abs/1009.4593}{{\tt arXiv:1009.4593}}].

\bibitem{Lee:1977eg}
B.~W. Lee, C.~Quigg, and H.~B. Thacker, {\it {Weak Interactions at Very
  High-Energies: The Role of the Higgs Boson Mass}},  {\em Phys. Rev.} {\bf
  D16} (1977) 1519.

\bibitem{Sirunyan:2018owy}
{\bf CMS} Collaboration, A.~M. Sirunyan et~al., {\it {Search for invisible
  decays of a Higgs boson produced through vector boson fusion in proton-proton
  collisions at $\sqrt{s} =$ 13 TeV}},  {\em Phys. Lett.} {\bf B793} (2019)
  520--551, [\href{http://arxiv.org/abs/1809.05937}{{\tt arXiv:1809.05937}}].

\bibitem{Aaboud:2019rtt}
{\bf ATLAS} Collaboration, M.~Aaboud et~al., {\it {Combination of searches for
  invisible Higgs boson decays with the ATLAS experiment}},  {\em Phys. Rev.
  Lett.} {\bf 122} (2019), no.~23 231801,
  [\href{http://arxiv.org/abs/1904.05105}{{\tt arXiv:1904.05105}}].

\bibitem{Aaboud:2018sfi}
{\bf ATLAS} Collaboration, M.~Aaboud et~al., {\it {Search for invisible Higgs
  boson decays in vector boson fusion at $\sqrt{s} = 13$ TeV with the ATLAS
  detector}},  {\em Phys. Lett.} {\bf B793} (2019) 499--519,
  [\href{http://arxiv.org/abs/1809.06682}{{\tt arXiv:1809.06682}}].

\bibitem{Cheung:2018ave}
K.~Cheung, J.~S. Lee, and P.-Y. Tseng, {\it {New Emerging Results in Higgs
  Precision Analysis Updates 2018 after Establishment of Third-Generation
  Yukawa Couplings}},  {\em JHEP} {\bf 09} (2019) 098,
  [\href{http://arxiv.org/abs/1810.02521}{{\tt arXiv:1810.02521}}].

\bibitem{Kraml:2019sis}
S.~Kraml, T.~Q. Loc, D.~T. Nhung, and L.~D. Ninh, {\it {Constraining new
  physics from Higgs measurements with Lilith: update to LHC Run 2 results}},
  {\em SciPost Phys.} {\bf 7} (2019), no.~4 052,
  [\href{http://arxiv.org/abs/1908.03952}{{\tt arXiv:1908.03952}}].

\bibitem{ATLAS:2020kdi}
{\bf ATLAS} Collaboration, {\it {Combination of searches for invisible Higgs
  boson decays with the ATLAS experiment}},
  \href{http://arxiv.org/abs/ATLAS-CONF-2020-052}{{\tt ATLAS-CONF-2020-052}}.

\bibitem{Belanger:2015kga}
G.~Belanger, B.~Dumont, A.~Goudelis, B.~Herrmann, S.~Kraml, and D.~Sengupta,
  {\it {Dilepton constraints in the Inert Doublet Model from Run 1 of the
  LHC}},  {\em Phys. Rev.} {\bf D91} (2015), no.~11 115011,
  [\href{http://arxiv.org/abs/1503.07367}{{\tt arXiv:1503.07367}}].

\bibitem{Lundstrom:2008ai}
E.~Lundstrom, M.~Gustafsson, and J.~Edsjo, {\it {The Inert Doublet Model and
  LEP II Limits}},  {\em Phys. Rev.} {\bf D79} (2009) 035013,
  [\href{http://arxiv.org/abs/0810.3924}{{\tt arXiv:0810.3924}}].

\bibitem{Peskin:1991sw}
M.~E. Peskin and T.~Takeuchi, {\it {Estimation of oblique electroweak
  corrections}},  {\em Phys. Rev.} {\bf D46} (1992) 381--409.

\bibitem{Tanabashi:2018oca}
{\bf Particle Data Group} Collaboration, M.~Tanabashi et~al., {\it {Review of
  Particle Physics}},  {\em Phys. Rev.} {\bf D98} (2018), no.~3 030001.

\bibitem{Zyla:2020zbs}
{\bf Particle Data Group} Collaboration, P.~A. Zyla et~al., {\it {Review of
  Particle Physics}},  {\em PTEP} {\bf 2020} (2020), no.~8 083C01.

\bibitem{Abbott:1982af}
L.~F. Abbott and P.~Sikivie, {\it {A Cosmological Bound on the Invisible
  Axion}},  {\em Phys. Lett.} {\bf 120B} (1983) 133--136.

\bibitem{Dine:1982ah}
M.~Dine and W.~Fischler, {\it {The Not So Harmless Axion}},  {\em Phys. Lett.}
  {\bf 120B} (1983) 137--141.

\bibitem{Preskill:1982cy}
J.~Preskill, M.~B. Wise, and F.~Wilczek, {\it {Cosmology of the Invisible
  Axion}},  {\em Phys. Lett.} {\bf 120B} (1983) 127--132.

\bibitem{Belanger:2020gnr}
G.~Belanger, A.~Mjallal, and A.~Pukhov, {\it {Recasting direct detection limits
  within micrOMEGAs and implication for non-standard Dark Matter scenarios}},
  {\em pre-print} (3, 2020) [\href{http://arxiv.org/abs/2003.08621}{{\tt
  arXiv:2003.08621}}].

\bibitem{Aprile:2018dbl}
{\bf XENON} Collaboration, E.~Aprile et~al., {\it {Dark Matter Search Results
  from a One Ton-Year Exposure of XENON1T}},  {\em Phys. Rev. Lett.} {\bf 121}
  (2018), no.~11 111302, [\href{http://arxiv.org/abs/1805.12562}{{\tt
  arXiv:1805.12562}}].

\bibitem{Amole:2019fdf}
{\bf PICO} Collaboration, C.~Amole et~al., {\it {Dark Matter Search Results
  from the Complete Exposure of the PICO-60 C$_3$F$_8$ Bubble Chamber}},  {\em
  Phys. Rev. D} {\bf 100} (2019), no.~2 022001,
  [\href{http://arxiv.org/abs/1902.04031}{{\tt arXiv:1902.04031}}].

\bibitem{Abdelhameed:2019hmk}
{\bf CRESST} Collaboration, A.~H. Abdelhameed et~al., {\it {First results from
  the CRESST-III low-mass dark matter program}},  {\em Phys. Rev. D} {\bf 100}
  (2019), no.~10 102002, [\href{http://arxiv.org/abs/1904.00498}{{\tt
  arXiv:1904.00498}}].

\bibitem{Agnes:2018ves}
{\bf DarkSide} Collaboration, P.~Agnes et~al., {\it {Low-Mass Dark Matter
  Search with the DarkSide-50 Experiment}},  {\em Phys. Rev. Lett.} {\bf 121}
  (2018), no.~8 081307, [\href{http://arxiv.org/abs/1802.06994}{{\tt
  arXiv:1802.06994}}].

\bibitem{Aaboud:2017phn}
{\bf ATLAS} Collaboration, M.~Aaboud et~al., {\it {Search for dark matter and
  other new phenomena in events with an energetic jet and large missing
  transverse momentum using the ATLAS detector}},  {\em JHEP} {\bf 01} (2018)
  126, [\href{http://arxiv.org/abs/1711.03301}{{\tt arXiv:1711.03301}}].

\bibitem{Sirunyan:2017hci}
{\bf CMS} Collaboration, A.~M. Sirunyan et~al., {\it {Search for dark matter
  produced with an energetic jet or a hadronically decaying W or Z boson at $
  \sqrt{s}=13 $ TeV}},  {\em JHEP} {\bf 07} (2017) 014,
  [\href{http://arxiv.org/abs/1703.01651}{{\tt arXiv:1703.01651}}].

\bibitem{Aaboud:2018xdt}
{\bf ATLAS} Collaboration, M.~Aaboud et~al., {\it {Measurements of Higgs boson
  properties in the diphoton decay channel with 36 fb$^{-1}$ of $pp$ collision
  data at $\sqrt{s} = 13$ TeV with the ATLAS detector}},  {\em Phys. Rev. D}
  {\bf 98} (2018) 052005, [\href{http://arxiv.org/abs/1802.04146}{{\tt
  arXiv:1802.04146}}].

\bibitem{Sirunyan:2018ouh}
{\bf CMS} Collaboration, A.~M. Sirunyan et~al., {\it {Measurements of Higgs
  boson properties in the diphoton decay channel in proton-proton collisions at
  $\sqrt{s} =$ 13 TeV}},  {\em JHEP} {\bf 11} (2018) 185,
  [\href{http://arxiv.org/abs/1804.02716}{{\tt arXiv:1804.02716}}].

\bibitem{Djouadi:2005gj}
A.~Djouadi, {\it {The Anatomy of electro-weak symmetry breaking. II. The Higgs
  bosons in the minimal supersymmetric model}},  {\em Phys. Rept.} {\bf 459}
  (2008) 1--241, [\href{http://arxiv.org/abs/hep-ph/0503173}{{\tt
  hep-ph/0503173}}].

\bibitem{Bohm:1986rj}
M.~Bohm, H.~Spiesberger, and W.~Hollik, {\it {On the One Loop Renormalization
  of the Electroweak Standard Model and Its Application to Leptonic
  Processes}},  {\em Fortsch. Phys.} {\bf 34} (1986) 687--751.

\bibitem{Hollik:1988ii}
W.~F.~L. Hollik, {\it {Radiative Corrections in the Standard Model and their
  Role for Precision Tests of the Electroweak Theory}},  {\em Fortsch. Phys.}
  {\bf 38} (1990) 165--260.

\bibitem{Denner:1991kt}
A.~Denner, {\it {Techniques for calculation of electroweak radiative
  corrections at the one loop level and results for W physics at LEP-200}},
  {\em Fortsch. Phys.} {\bf 41} (1993) 307--420,
  [\href{http://arxiv.org/abs/0709.1075}{{\tt arXiv:0709.1075}}].

\bibitem{Xie:2018yiv}
W.~Xie, R.~Benbrik, A.~Habjia, B.~Gong, and Q.-S. Yan, {\it {Signature of 2HDM
  at Higgs Factories}},  \href{http://arxiv.org/abs/1812.02597}{{\tt
  arXiv:1812.02597}}.

\bibitem{Hahn:2000kx}
T.~Hahn, {\it {Generating Feynman diagrams and amplitudes with FeynArts 3}},
  {\em Comput. Phys. Commun.} {\bf 140} (2001) 418--431,
  [\href{http://arxiv.org/abs/hep-ph/0012260}{{\tt hep-ph/0012260}}].

\bibitem{Hahn:1998yk}
T.~Hahn and M.~Perez-Victoria, {\it {Automatized one loop calculations in
  four-dimensions and D-dimensions}},  {\em Comput. Phys. Commun.} {\bf 118}
  (1999) 153--165, [\href{http://arxiv.org/abs/hep-ph/9807565}{{\tt
  hep-ph/9807565}}].

\bibitem{Hahn:2006qw}
T.~Hahn and M.~Rauch, {\it {News from FormCalc and LoopTools}},  {\em Nucl.
  Phys. Proc. Suppl.} {\bf 157} (2006) 236--240,
  [\href{http://arxiv.org/abs/hep-ph/0601248}{{\tt hep-ph/0601248}}].
  [,236(2006)].

\bibitem{Hahn:1999mt}
T.~Hahn, {\it {Loop calculations with FeynArts, FormCalc, and LoopTools}},
  {\em Acta Phys. Polon.} {\bf B30} (1999) 3469--3475,
  [\href{http://arxiv.org/abs/hep-ph/9910227}{{\tt hep-ph/9910227}}].

\bibitem{Hahn:2010zi}
T.~Hahn, {\it {Feynman Diagram Calculations with FeynArts, FormCalc, and
  LoopTools}},  {\em PoS} {\bf ACAT2010} (2010) 078,
  [\href{http://arxiv.org/abs/1006.2231}{{\tt arXiv:1006.2231}}].

\bibitem{Wang:2004du}
J.-X. Wang, {\it {Progress in FDC project}},  {\em Nucl. Instrum. Meth.} {\bf
  A534} (2004) 241--245, [\href{http://arxiv.org/abs/hep-ph/0407058}{{\tt
  hep-ph/0407058}}].

\bibitem{Djouadi:2005gi}
A.~Djouadi, {\it {The Anatomy of electro-weak symmetry breaking. I: The Higgs
  boson in the standard model}},  {\em Phys. Rept.} {\bf 457} (2008) 1--216,
  [\href{http://arxiv.org/abs/hep-ph/0503172}{{\tt hep-ph/0503172}}].

\bibitem{LopezVal:2009qy}
D.~Lopez-Val and J.~Sola, {\it {Neutral Higgs-pair production at Linear
  Colliders within the general 2HDM: Quantum effects and triple Higgs boson
  self-interactions}},  {\em Phys. Rev.} {\bf D81} (2010) 033003,
  [\href{http://arxiv.org/abs/0908.2898}{{\tt arXiv:0908.2898}}].

\bibitem{LopezVal:2010vk}
D.~Lopez-Val, J.~Sola, and N.~Bernal, {\it {Quantum effects on Higgs-strahlung
  events at Linear Colliders within the general 2HDM}},  {\em Phys. Rev.} {\bf
  D81} (2010) 113005, [\href{http://arxiv.org/abs/1003.4312}{{\tt
  arXiv:1003.4312}}].

\bibitem{Jueid:2020rek}
A.~Jueid, J.~Kim, S.~Lee, S.~Y. Shim, and J.~Song, {\it {Phenomenology of the
  Inert Doublet Model with a global U(1) symmetry}},  {\em Phys. Rev.} {\bf
  D102} (2020), no.~7 075011, [\href{http://arxiv.org/abs/2006.10263}{{\tt
  arXiv:2006.10263}}].

\bibitem{Cao:2014rma}
J.~Cao, C.~Han, J.~Ren, L.~Wu, J.~M. Yang, and Y.~Zhang, {\it {SUSY effects in
  Higgs productions at high energy $e^+e^-$ colliders}},  {\em Chin. Phys.}
  {\bf C40} (2016), no.~11 113104, [\href{http://arxiv.org/abs/1410.1018}{{\tt
  arXiv:1410.1018}}].

\bibitem{Heinemeyer:2015qbu}
S.~Heinemeyer and C.~Schappacher, {\it {Neutral Higgs boson production at
  $e^+e^-$ colliders in the complex MSSM: a full one-loop analysis}},  {\em
  Eur. Phys. J.} {\bf C76} (2016), no.~4 220,
  [\href{http://arxiv.org/abs/1511.06002}{{\tt arXiv:1511.06002}}].

\bibitem{Driesen:1995ew}
V.~Driesen, W.~Hollik, and J.~Rosiek, {\it {Production of neutral MSSM Higgs
  bosons in e+ e- collisions: A Complete one loop calculation}},  {\em Z.
  Phys.} {\bf C71} (1996) 259--266,
  [\href{http://arxiv.org/abs/hep-ph/9512441}{{\tt hep-ph/9512441}}].

\bibitem{deBlas:2019wgy}
J.~De~Blas, G.~Durieux, C.~Grojean, J.~Gu, and A.~Paul, {\it {On the future of
  Higgs, electroweak and diboson measurements at lepton colliders}},  {\em
  JHEP} {\bf 12} (2019) 117, [\href{http://arxiv.org/abs/1907.04311}{{\tt
  arXiv:1907.04311}}].

\bibitem{Cheung:2019qdr}
K.~Cheung and Z.~S. Wang, {\it {Probing Long-lived Particles at Higgs
  Factories}},  {\em Phys. Rev. D} {\bf 101} (2020), no.~3 035003,
  [\href{http://arxiv.org/abs/1911.08721}{{\tt arXiv:1911.08721}}].

\bibitem{Kuraev:1985hb}
E.~A. Kuraev and V.~S. Fadin, {\it {On Radiative Corrections to e+ e- Single
  Photon Annihilation at High-Energy}},  {\em Sov. J. Nucl. Phys.} {\bf 41}
  (1985) 466--472. [Yad. Fiz.41,733(1985)].

\end{thebibliography}\endgroup

\end{document}